\documentclass[structabstract]{aa}  
\usepackage{longtable,lscape,graphicx,float,amsmath,amssymb,mathrsfs,txfonts,color}
\usepackage{multirow,natbib,rotating,ulem} 
\def\msol{\rm{M}$_{\odot}$}
\def\lsol{\rm{L}$_{\odot}$}

\def\arcsec{$^{\prime}$$^{\prime}$}
\def\arcmin{$^{\prime}$}

\def\kms{\,km~s$^{-1}$}

\def\apjs{ApJS}
\def\apjl{ApJL}%
\def\aap{A\&A}%

\newcounter{ppnum4}
\newcounter{ppnum5}
\newcounter{ppnum6}
\begin{document}
\setlength{\tabcolsep}{3pt}
\title{Water in star-forming regions with \textit{Herschel} (WISH)\thanks{\textit{Herschel} is an ESA space observatory with science instruments provided by European-led Principal Investigator consortia and with important participation from NASA.}$^{,}$\thanks{The appendices are available in electronic form at http://www.aanda.org}$^{,}$\thanks{Reduced spectra are available at the CDS via anonymous ftp to http://cdsarc.u-strasbg.fr (ftp://130.79.128.5) or via http://cdsarc.u-strasbg.fr/viz-bin/qcat?J/A+A/}}
\subtitle{V. The physical conditions in low-mass protostellar outflows revealed by multi-transition water observations}

\author{J.~C.~Mottram\inst{1}~\thanks{E-mail:mottram@strw.leidenuniv.nl} 
          \and
          L.~E.~Kristensen\inst{2}
          \and
          E.~F.~van~Dishoeck\inst{1,3}
          \and
          S.~Bruderer\inst{3}
          \and
          I.~San~Jos\'{e}-Garc\'{i}a\inst{1}
          \and
          A.~Karska\inst{3}
          \and
          R.~Visser\inst{4}
          \and
          G.~Santangelo\inst{5,6}
          \and
          A.~O.~Benz\inst{7}
          \and
          E.~A.~Bergin\inst{4}
          \and
          P.~Caselli\inst{8,3}
          \and
          F.~Herpin\inst{9,10}
          \and
          M.~R.~Hogerheijde\inst{1}
          \and
          D.~Johnstone\inst{11,12,13}
          \and
          T.~A.~van~Kempen\inst{1}
          \and
          R.~Liseau\inst{14}
          \and
          B.~Nisini\inst{6}
          \and
          M.~Tafalla\inst{15}
          \and
          F.~F.~S.~van~der~Tak\inst{16,17}
          \and
          F.~Wyrowski\inst{18}
}

\institute{Leiden Observatory, Leiden University, PO Box 9513, 2300 RA Leiden, The Netherlands
  \and
Harvard-Smithsonian Center for Astrophysics, 60 Garden Street, Cambridge, MA 02138, USA
\and
Max Planck Institut f\"{u}r Extraterrestrische Physik, Giessenbachstrasse 1, 85748 Garching, Germany
\and
Department of Astronomy, University of Michigan, 500 Church Street, Ann Arbor, MI 48109-1042, USA
\and
Osservatorio Astrofisico di Arcetri, Largo Enrico Fermi 5, I-50125 Florence, Italy
\and
Osservatorio Astronomico di Roma, via di Frascati 33, 00040 Monteporzio Catone, Italy
\and
Institute for Astronomy, ETH Zurich, 8093 Zurich, Switzerland
\and
School of Physics and Astronomy, University of Leeds, Leeds LS2 9JT, UK
\and
Universit\'e de Bordeaux, Observatoire Aquitain des Sciences de l'Univers, 2 rue de l'Observatoire, BP 89, F-33270 Floirac Cedex, France
\and
CNRS, LAB, UMR 5804, Laboratoire d'Astrophysique de Bordeaux, 2 rue de l'Observatoire, BP 89, F-33270 Floirac Cedex, France
\and
Joint Astronomy Centre, 660 North A’ohoku Place, University Park, Hilo, HI 96720, USA
\and
Department of Physics and Astronomy, University of Victoria, PO Box 3055 STN CSC, Victoria, BC V8W 3P6, Canada
\and
NRC-Herzberg Institute of Astrophysics, 5071 West Saanich Road, Victoria, BC V9E 2E7, Canada
\and
Department of Earth and Space Sciences, Chalmers University of Technology, Onsala Space Observatory, 439 92 Onsala, Sweden
\and
Observatorio Astron\'{o}mico Nacional (IGN), Alfonso XII 3, E-28014 Madrid, Spain
\and
SRON Netherlands Institute for Space Research, PO Box 800, 9700 AV Groningen, The Netherlands
\and
Kapteyn Astronomical Institute, University of Groningen, PO Box 800, 9700 AV Groningen, The Netherlands
\and
Max-Planck-Institut f\"{u}r Radioastronomie, Auf dem H\"{u}gel 69, 53121 Bonn, Germany
}

   \date{Received XXXX; accepted XXXX}
\titlerunning{WISH V. The physical conditions in low-mass protostellar outflows revealed by water}

 
  \abstract
   {Outflows are an important part of the star formation process as both the result of ongoing active accretion and one of the main sources of mechanical feedback on small scales. Water is the ideal tracer of these effects because it is present in high abundance for the conditions expected in various parts of the protostar, particularly the outflow.}
   {To constrain and quantify the physical conditions probed by water in the outflow-jet system for Class 0 and I sources.}
   {We present velocity-resolved \textit{Herschel} HIFI spectra of multiple water-transitions observed towards 29 nearby Class 0/I protostars as part of the WISH Guaranteed Time Key Programme. The lines are decomposed into different Gaussian components, with each component related to one of three parts of the protostellar system; quiescent envelope, cavity shock and spot shocks in the jet and at the base of the outflow. We then use non-LTE \textsc{radex} models to constrain the excitation conditions present in the two outflow-related components.}
   {Water emission at the source position is optically thick but effectively thin, with line ratios that do not vary with velocity, in contrast to CO. The physical conditions of the cavity and spot shocks are similar, with post-shock H$_{2}$ densities of order 10$^{5}-$10$^{8}$\,cm$^{-3}$ and H$_{2}$O column densities of order 10$^{16}-$10$^{18}$\,cm$^{-2}$. H$_{2}$O emission originates in compact emitting regions: for the spot shocks these correspond to point sources with radii of order 10-200\,AU, while for the cavity shocks these come from a thin layer along the outflow cavity wall with thickness of order 1-30\,AU.}
   {Water emission at the source position traces two distinct kinematic components in the outflow; J shocks at the base of the outflow or in the jet, and C shocks in a thin layer in the cavity wall. The similarity of the physical conditions is in contrast to off-source determinations which show similar densities but lower column densities and larger filling factors. We propose that this is due to the differences in shock properties and geometry between these positions. Class I sources have similar excitation conditions to Class 0 sources, but generally smaller line-widths and emitting region sizes. We suggest that it is the velocity of the wind driving the outflow, rather than the decrease in envelope density or mass, that is the cause of the decrease in H$_{2}$O intensity between Class 0 and I sources.}


   \maketitle

%

\section{Introduction}
\label{S:Intro}

Molecular outflows are a ubiquitous and necessary part of the star formation process. They remove angular momentum and material from the protostellar environment in a feedback process which helps the protostar form a disk and gain mass in the short-term while ultimately conspiring with the initial core conditions to starve it in the long term. Thus understanding outflows is at the heart of developing a true law of star formation which can predict the stellar outcome based on initial core properties. 

The classical tracer of such outflows, low-$J$ CO ($J\leq$4), traces material in a mixing layer which has undergone turbulent entrainment from the quiescent envelope \citep[e.g.][]{Canto1991,Raga1995} with gas temperatures of order 50$-$100\,K \citep[e.g.][]{Yildiz2013}. This carries away a significant amount of mass from the envelope, but at relatively low velocities of order 5$-$20\kms{} and likely from material entrained at some distance from the protostar. Therefore, the properties derived from this emission may not accurately reflect the total momentum, angular momentum and kinetic energy transport of the system. In addition, it does not trace the active surface where the envelope is currently being sculpted \citep[e.g.][]{Nisini2010,Santangelo2013} and so does not probe the true feedback conditions.

In contrast, protostellar jets, as traced in atomic gas or shocked H$_{2}$ \citep[e.g.][]{Reipurth2000}, are more directly linked with accretion onto the central protostar \citep[e.g.][]{Pudritz2007,Shang2007}. This material is moving faster \citep[100$-$1000\kms{},][]{Frank2014} and is at higher temperatures but lower H$_{2}$ number densities compared to the outflow \citep[of order 10$^{3}$$-$10$^{4}$\,K and $n_{\mathrm{H}}$$\sim$10$^{3}-$10$^{4}$\,cm$^{-3}$ respectively,][]{Bacciotti1999}, and therefore has higher momentum and kinetic energy but lower mass. So-called 'bullets' can be seen in molecular species such as CO, SiO and more recently H$_{2}$O \citep[][]{Bachiller1990,Bachiller1991,Hirano2006,Santiago-Garcia2009,Kristensen2011}, where material is compressed (and thus cools more efficiently) due to shocks within the jet. However, the pencil-beam nature of the jet means that it is unlikely to be a major factor in the disruption of the envelope as the protostar evolves from Class 0 ($T_{\mathrm{bol}}<$70\,K) to Class I \citep[70$\leq T_{\mathrm{bol}}<$650\,K;][]{Lada1984,Andre1993}.

Between these two extremes are two intermediate regions; the outflow cavity which may be filled with a wind which has a similar or larger density than the jet \citep[][]{Panoglou2012}, and the active cavity shock at the boundary between the cavity and the quiescent envelope \citep[e.g.][]{Velusamy2007,Visser2012}. For the latter, the gas temperature and H$_{2}$ number density are of order 300$-$1000\,K and 10$^{5}-$10$^{7}$\,cm$^{-3}$ respectively, as traced by H$_{2}$O and high-$J$ CO \citep[e.g.][]{Goicoechea2012,Karska2013,Kristensen2013}, while the dust is below 100\,K. Studies based on \textit{Herschel} PACS \citep{Poglitsch2010} and/or SPIRE \citep{Griffin2010} observations see multiple distinct temperature components in CO excitation diagrams: cold emission ($\sim$100\,K) for $J<$14, warm emission ($\sim$300\,K) for 14$\lesssim J \lesssim$24 and sometimes also hot emission ($\sim$750\,K) for $J \gtrsim$24 with the column density decreasing with increasing temperature \citep{Manoj2013,Karska2013,Green2013}. However, these observations are spectrally unresolved, which makes relating these temperature components to physical parts of the protostar more challenging.

Therefore, how the physical and excitation conditions in the different parts of protostellar outflow-jet systems are related, and how these vary both with distance from the central protostar and between sources, is not currently well understood. In addition, while outflow line-width and force decrease on average as the protostar evolves \citep{Bontemps1996}, in particular from Class 0 to Class I, it is not clear if this is because the outflow actually decreases in strength or simply because there is less envelope material available to reveal its presence.

Water is the ideal molecule to resolve these questions. It is the primary ice constituent and oxygen reservoir in protostellar envelopes, sublimates at dust temperatures above $\sim$100\,K and can also be formed efficiently in the gas phase at temperatures above a few 100\,K \citep[see][and references therein]{vanDishoeck2013}. At shock velocities above $\sim$10$-$20\kms{} it can also be sputtered from the grain mantles \citep[see e.g.][]{Jimenez-Serra2008b,VanLoo2013,Neufeld2014,Suutarinen2014}. It is therefore potentially present in relatively high abundance in the gas-phase in the cavity shock, wind and shocks within the jet. Water also has a large dipole moment and Einstein A coefficients, and therefore more intense line emission than species with smaller dipole moments. Even for subthermal excitation, where the number density is well below the critical density, water lines can be more easily detected than emission from species such as CO. The favourable combination of these factors makes water a good tracer of the kinematics of these regions.

The expected kinematic signatures are related to the properties of the shocks in the outflow and jet \citep[see, e.g.][]{Draine1980,Hollenbach1997}. In discontinuous, ``jump'' (J-type) shocks, there is a sharp increase in the acceleration of gas in the shock with respect to the ambient un-shocked material by the passage of the shock front. The line-centre of the emission from these molecules is therefore shifted from the source velocity to some fraction of the shock velocity dependent on the viewing angle. The distribution of velocities in the post-shock material, and thus the FWHM of the emission line, will also be different from that of the ambient material. Alternatively, in ``continuous'' (C-type) shocks, the molecules are smoothly accelerated by the shock and so emission extends from the source velocity to the velocity of the shock. Therefore, for the same shock geometry larger line-widths are expected for C-type shocks. Hybrid C-J-type shocks can be formed if the shock conditions are such that a C-type shock does not have time to reach steady-state; in this case, a J-type front develops at the time the shock is truncated \citep{Chieze1998}. For simplicity, in the remainder of the paper we will only refer to C and J shocks given the time dependant nature of C-J shocks. Multiple discrete shocks with different conditions or orientations with respect to the line of sight will give rise to multiple emission line components. It is also possible that both C and J type shocks exist as part of the same structure \citep[e.g., see Fig.~9 of][]{Suutarinen2014}, in which case the physical conditions will be similar but the two shocks will produce different line profiles.

Strong, broad and complex line profiles have been observed in water towards Class 0 and I protostars \citep[][]{Kristensen2010,Kristensen2012}, most recently using the \textit{Herschel} Space Observatory \citep[][]{Pilbratt2010} as part of the ``Water in star-forming regions with \textit{Herschel}'' \citep[WISH;][]{vanDishoeck2011} Guaranteed Time Key Programme. These have been complemented by spectra at off-source positions along several promenant outflows \citep{Santangelo2012,Vasta2012,Nisini2013,Santangelo2013}. While \citet{Kristensen2012} looked at the dynamical components for one water line, the 1$_{10}-$1$_{01}$ ground-state transition at 557\,GHz, this paper seeks to use multiple water transitions to probe the excitation conditions and water chemistry in these sources. In addition, studying how the excitation of water varies between the different physical components is required to disentangle the temperature components seen in spectrally unresolved PACS/SPIRE observations.

The goals of this paper are therefore: to use the multiple transitions of water observed towards low-mass Class 0/I protostars as part of the WISH survey to identify the physical conditions present where water is emitting within low-mass protostellar outflows, to understand the differences and similarities between the conditions in the various parts of the jet-outflow system, and to explore how this changes with source evolution.

We begin with a brief description of the sample and observations used for this study in Sec.~\ref{S:observations}. Next, we present our results in Sec.~\ref{S:results} and additional analysis in Sec.~\ref{S:analysis}. We then discuss the implications of these results in Sec.~\ref{S:discussion}, both in terms of the different parts of the jet-outflow system (Sec.~\ref{S:discussion_components}), the impact of source evolution (Sec.~\ref{S:discussion_evolution}), and comparison to off-source shocks (Sec.~\ref{S:discussion_on_vs_off}). Finally, we summarise our main findings and reach our conclusions in Sec.~\ref{S:conclusions}.

\section{Observations}
\label{S:observations}

The WISH low-mass sample consists of 15 Class 0 and 14 Class I sources, the properties of which are given in Table~\ref{T:observations_sources}. All sources have been independently verified as truly embedded sources and not edge-on disks. 

This sample was the target of a series of observations of gas-phase water transitions with the Heterodyne Instrument for the Far-Infrared \citep[HIFI;][]{deGraauw2010} on \textit{Herschel} between March 2010 and October 2011. Three of the Class I sources (IRAS3A, RCrA-IRS5A and HH100-IRS) were only observed in the 557\,GHz H$_{2}$O 1$_{10}-$1$_{01}$ line, which was presented for all sources by \citet{Kristensen2012}. All other sources were observed in between four and seven H$_{2}^{16}$O transitions and between one and four H$_{2}^{18}$O transitions. Additional data from two OT2 programmes, OT2\_rvisser\_2 and OT2\_evandish\_4, are also included to augment the WISH data. 

\begin{figure}
\begin{center}
\includegraphics[width=0.47\textwidth]{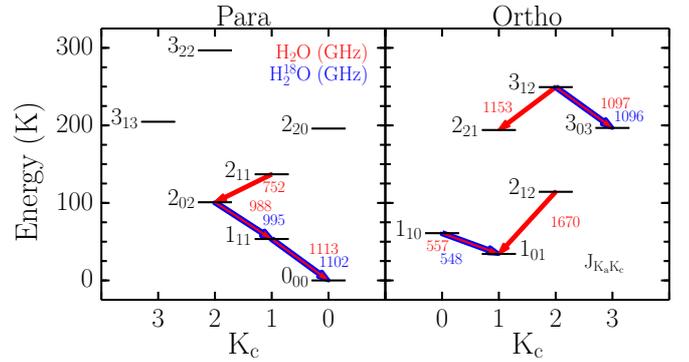}
\caption{Level diagram of the various H$_{2}$O (red) and H$_{2}^{18}$O (blue) transitions observed with HIFI towards the WISH sample of low-mass protostars.}
\label{F:observations_leveldiagram}
\end{center}
\end{figure}

Details of the line frequency, main-beam efficiency, spectral and spatial resolutions, observing time, critical density at 300\,K and upper level energy of the observed transitions are given for all lines in Table~\ref{T:observations_lines}. Settings primarily targeting H$_{2}^{18}$O transitions were only observed towards Class 0 sources where higher line intensities were expected compared to Class I sources. This also motivated the longer integrations in the H$_{2}$O 1$_{11}-$0$_{00}$ transition for Class 0 than Class I sources as that setting also includes the corresponding H$_{2}^{18}$O transition. Longer integrations were performed for Class I sources in the 1$_{10}-$1$_{01}$ transition to ensure detections in at least one line in the maximum number of sources. A level diagram of the various lines is shown in Figure~\ref{F:observations_leveldiagram} and the observations identification numbers of all data used in this paper are given in Table~\ref{T:obsids}.

\begin{table*}
\caption{Source parameters.}
\small
\begin{center}
\begin{tabular}{lcccccccc}
\hline\hline
Source & RA & Dec & $D$\tablefootmark{a} & $\varv_{\rm LSR}$\tablefootmark{b} & $L_{\rm bol}$\tablefootmark{c} & $T_{\rm bol}$\tablefootmark{c} & $M_{\rm env}$\tablefootmark{d} & $F_{\mathrm{CO}}$\tablefootmark{e} \\
       & (h m s) & (\degr\ \arcmin\ \arcsec) & (pc) & (\kms) & (\lsol) & (K) & (\msol) & (\msol{}yr$^{-1}$\kms) \\ \hline
L1448-MM                  & 03 25 38.9 & $+$30 44 05.4 & 235 & \phantom{1}$+$5.2 & \phantom{1}9.0 & \phantom{1}46 &	\phantom{1}3.9 & 3.7$\times$10$^{-3}$ \\
NGC1333-IRAS2A            & 03 28 55.6 & $+$31 14 37.1 & 235 & \phantom{1}$+$7.7 &           35.7 & \phantom{1}50 &	\phantom{1}5.1 & 7.4$\times$10$^{-3}$ \\
NGC1333-IRAS4A            & 03 29 10.5 & $+$31 13 30.9 & 235 & \phantom{1}$+$7.2 & \phantom{1}9.1 & \phantom{1}33 &	\phantom{1}5.2 & 2.1$\times$10$^{-3}$ \\
NGC1333-IRAS4B            & 03 29 12.0 & $+$31 13 08.1 & 235 & \phantom{1}$+$7.4 & \phantom{1}4.4 & \phantom{1}28 &	\phantom{1}3.0 & 2.2$\times$10$^{-4}$ \\
L1527                     & 04 39 53.9 & $+$26 03 09.8 & 140 & \phantom{1}$+$5.9 & \phantom{1}1.9 & \phantom{1}44 &	\phantom{1}0.9 & 4.4$\times$10$^{-4}$ \\
Ced110-IRS4               & 11 06 47.0 & $-$77 22 32.4 & 125 & \phantom{1}$+$4.2 & \phantom{1}0.8 & \phantom{1}56 & \phantom{1}0.2 & $-$ \\
BHR71                     & 12 01 36.3 & $-$65 08 53.0 & 200 & \phantom{1}$-$4.4 &           14.8 & \phantom{1}44 &	\phantom{1}3.1 & $-$ \\
IRAS15398\tablefootmark{f}& 15 43 01.3 & $-$34 09 15.0 & 130 & \phantom{1}$+$5.1 & \phantom{1}1.6 & \phantom{1}52 &	\phantom{1}0.5 & 9.5$\times$10$^{-5}$\\
L483                      & 18 17 29.9 & $-$04 39 39.5 & 200 & \phantom{1}$+$5.2 &           10.2 & \phantom{1}49 &	\phantom{1}4.4 & 5.9$\times$10$^{-4}$\\
Ser-SMM1 & 18 29 49.8 & $+$01 15 20.5 & 415 & \phantom{1}$+$8.5 &           99.0 & \phantom{1}39 &	          52.5 & 3.0$\times$10$^{-3}$\\
Ser-SMM3 & 18 29 59.2 & $+$01 14 00.3 & 415 & \phantom{1}$+$7.6 & \phantom{1}16.6 & \phantom{1}38 &	10.4 & 4.2$\times$10$^{-3}$\\
Ser-SMM4 & 18 29 56.6 & $+$01 13 15.1 & 415 & \phantom{1}$+$8.0 & \phantom{1}6.2 & \phantom{1}26 & \phantom{1}6.9 & 4.8$\times$10$^{-3}$\\
L723                      & 19 17 53.7 & $+$19 12 20.0 & 300 &           $+$11.2 & \phantom{1}3.6 & \phantom{1}39 &	\phantom{1}1.3 & 2.9$\times$10$^{-3}$\\
B335                      & 19 37 00.9 & $+$07 34 09.6 & 250 & \phantom{1}$+$8.4 & \phantom{1}3.3 & \phantom{1}36 &	\phantom{1}1.2 & 6.0$\times$10$^{-4}$\\
L1157                     & 20 39 06.3 & $+$68 02 15.8 & 325 & \phantom{1}$+$2.6 & \phantom{1}4.7 & \phantom{1}46 &	\phantom{1}1.5 & 3.7$\times$10$^{-3}$\\ \hline
NGC1333-IRAS3A            & 03 29 03.8 & $+$31 16 04.0 & 235 & \phantom{1}$+$8.5 &           41.8 &           149 &	\phantom{1}8.6 & $-$\\
L1489                     & 04 04 43.0 & $+$26 18 57.0 & 140 & \phantom{1}$+$7.2 & \phantom{1}3.8 &           200 & \phantom{1}0.2 & 1.6$\times$10$^{-4}$\\
L1551-IRS5                & 04 31 34.1 & $+$18 08 05.0 & 140 & \phantom{1}$+$6.2 &           22.1 & \phantom{1}94 & \phantom{1}2.3 & 5.1$\times$10$^{-4}$\\
TMR1\tablefootmark{f}     & 04 39 13.7 & $+$25 53 21.0 & 140 & \phantom{1}$+$6.3 & \phantom{1}3.8 &           133 &	\phantom{1}0.2 & 2.5$\times$10$^{-5}$\\
TMC1A\tablefootmark{f}    & 04 39 34.9 & $+$25 41 45.0 & 140 & \phantom{1}$+$6.6 & \phantom{1}2.7 &           118 &	\phantom{1}0.3& 1.3$\times$10$^{-4}$ \\
TMC1                      & 04 41 12.4 & $+$25 46 36.0 & 140 & \phantom{1}$+$5.2 & \phantom{1}0.9 &           101 &	\phantom{1}0.2 & 4.5$\times$10$^{-4}$\\
HH46-IRS                  & 08 25 43.9 & $-$51 00 36.0 & 450 & \phantom{1}$+$5.2 &           27.9 &           104 &	\phantom{1}4.4 & 1.1$\times$10$^{-3}$\\
IRAS12496                 & 12 53 17.2 & $-$77 07 10.6 & 178 & \phantom{1}$+$3.1 &           35.4 &           569 &	\phantom{1}0.8 & $-$\\
GSS30-IRS1                & 16 26 21.4 & $-$24 23 04.0 & 125 & \phantom{1}$+$3.5 &           13.9 &           142 & \phantom{1}0.6 & 5.2$\times$10$^{-4}$\\
Elias 29                  & 16 27 09.4 & $-$24 37 19.6 & 125 & \phantom{1}$+$4.3 &           14.1 &           299 & \phantom{1}0.3 & 6.4$\times$10$^{-5}$\\
Oph-IRS63                 & 16 31 35.6 & $-$24 01 29.6 & 125 & \phantom{1}$+$2.8 & \phantom{1}1.0 &           327 & \phantom{1}0.3 & 1.1$\times$10$^{-5}$\\
RNO91                     & 16 34 29.3 & $-$15 47 01.4 & 125 & \phantom{1}$+$0.5 & \phantom{1}2.6 &           340 & \phantom{1}0.5 & 1.0$\times$10$^{-4}$\\
RCrA-IRS5A                & 19 01 48.0 & $-$36 57 21.6 & 130 & \phantom{1}$+$5.7 & \phantom{1}7.1 &           126 & \phantom{1}2.0 & $-$\\
HH100-IRS                 & 19 01 49.1 & $-$36 58 16.0 & 130 & \phantom{1}$+$5.6 &           17.7 &           256 & \phantom{1}8.1 & $-$\\
\hline
\end{tabular}
\tablefoot{Sources above the horizontal line are Class 0, sources below are Class I.
         \tablefoottext{a}{Taken from \citet{vanDishoeck2011} with the exception of sources in Serpens, where we use the distance determined using VLBA observations by \citet{Dzib2010}.}
        \tablefoottext{b}{Obtained from ground-based C$^{18}$O or C$^{17}$O observations \citep{Yildiz2013} with the exception of IRAS4A for which the value from \citet{Kristensen2012} is more consistent with our data.}
	\tablefoottext{c}{Measured using \textit{Herschel}-PACS data from the WISH and DIGIT key programmes \citep[][]{Karska2013}.}
	\tablefoottext{d}{Mass within the 10\,K radius, determined by \citet{Kristensen2012} from \textsc{dusty} modelling of the sources.}
        \tablefoottext{e}{Taken from Yildiz et al., subm. for CO 3$-$2.}
	\tablefoottext{f}{The coordinates used in WISH; more accurate SMA coordinates of the sources are 15$^{\rm h}$43$^{\rm m}$02\fs2, $-$34\degr09\arcmin06\farcs8 (IRAS15398), 04$^{\rm h}$39$^{\rm m}$13\fs9, $+$25\degr53\arcmin20\farcs6 (TMR1) and 04$^{\rm h}$39$^{\rm m}$35\fs2, $+$25\degr41\arcmin44\farcs4 \citep[TMC1A;][]{Jorgensen2009}. For IRAS15398, these coordinates were observed in two settings as part of the OT2 programme OT2\_evandish\_4.}}
\end{center}
\label{T:observations_sources}
\end{table*}

All observations were taken in both horizontal and vertical polarisations with both the Wide Band Spectrometer (WBS) and High Resolution Spectrometer (HRS) backends. Observations were taken as single pointings in dual-beam-switch (DBS) mode with a chop throw of 3\arcmin{}, with the exception of some of the H$_{2}$O 1$_{10}-$1$_{01}$ observations, which were taken in position-switch mode \citep[see][for more details]{Kristensen2012}. The \textit{Herschel} beam ranges from 12.7\arcsec{} to 38.7\arcsec{} over the frequency range of the various water lines, close to the diffraction limit of the primary mirror.

\begin{figure*}
\begin{center}
\includegraphics[width=0.95\textwidth]{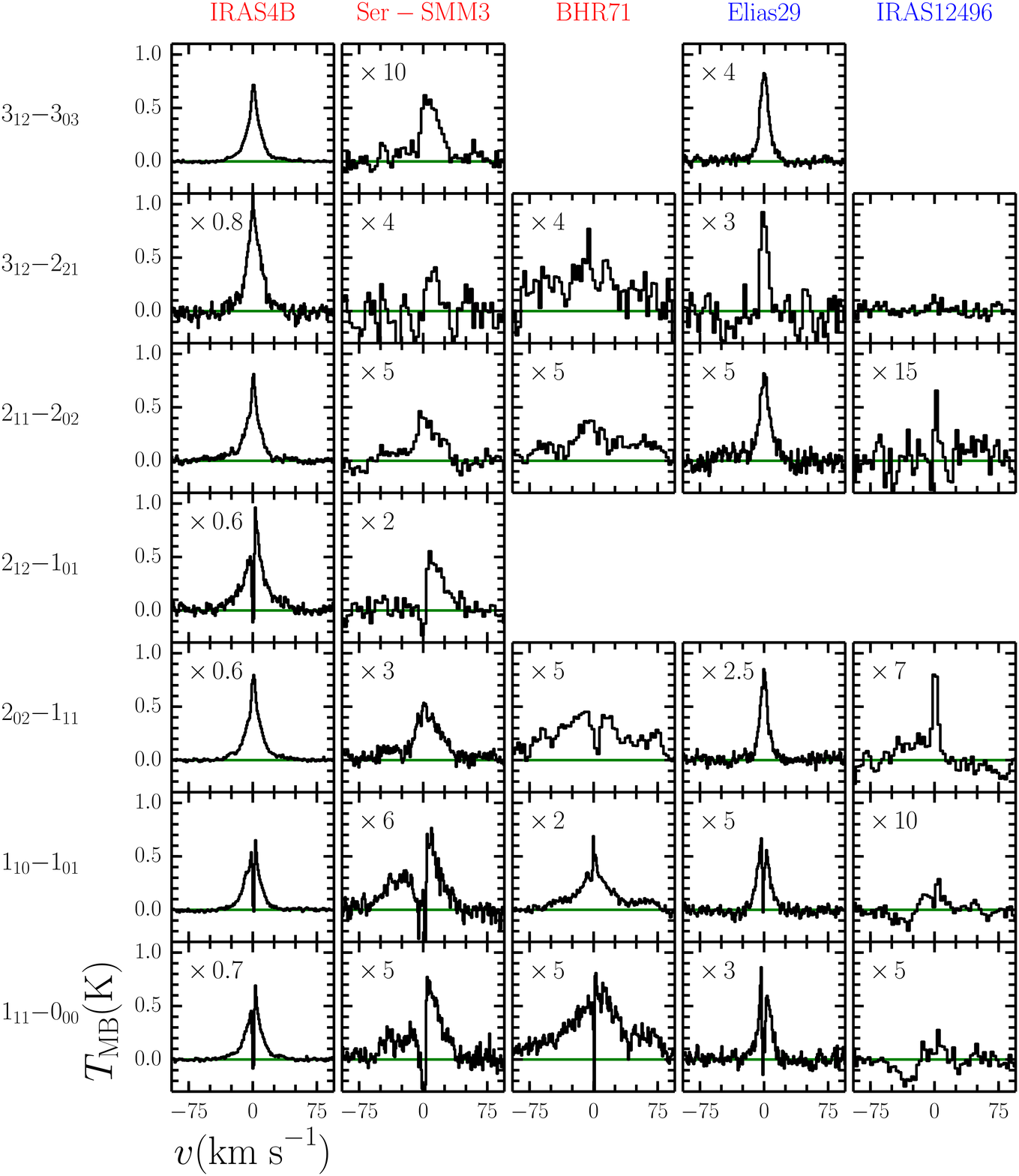}
\caption{Example H$_{2}$O spectra for three Class 0 and two Class I sources (names in red and blue respectively). All spectra have been recentred so that the source velocity is at 0\kms{} and scaled by the number in the top-right corner of each panel. Some spectra have also been resampled to a lower velocity resolution for ease of comparison. The green line indicates the baseline.}
\label{F:results_profiles_basic_comparison}
\end{center}
\end{figure*}

The data were reduced with \textsc{hipe} \citep{Ott2010}. After initial spectrum formation, further processing was also performed using \textsc{hipe}. This began with removal of instrumental standing waves where required, followed by baseline subtraction with a low-order ($\leq$2) polynomial in each sub-band. The fit to the baseline was then used to calculate the continuum level, compensating for the dual-sideband nature of the HIFI detectors i.e. the initial continuum level is the combination of emission from both the upper and lower sideband, which we assumed to be equal. Following this the WBS sub-bands were stitched into a continuous spectrum and all data were converted to the $T_{\mathrm{MB}}$ scale using efficiencies from \citet{Roelfsema2012}. Finally, for ease of analysis all data were converted to FITS format and resampled to 0.3\kms{} spectral resolution on the same velocity grid using bespoke \textsc{python} routines. 

Comparison of the two polarisations for each source revealed insignificant differences, so these were co-added to reduce the noise. Comparison of peak and integrated intensities between the original WISH observations and those obtained as part of OT2\_rvisser\_2 for the same sources suggest that the calibration uncertainty is $\lesssim$10$\%$. For the 2$_{02}-$1$_{11}$ line for BHR71, the off-positions of the DBS mode coincided with outflow emission, resulting in a broad absorption. This is masked out during the analysis so does not impact the results for this source. In addition, as also noted for the 1$_{10}-$1$_{01}$ transition by \citet{Kristensen2012}, observations of the three Serpens sources sometimes show a weak narrow absorption feature at ~$v_{\mathrm{LSR}}$=1\kms{} which probably arrises from emission in the reference position. This does not have any impact on the results derived below and so is ignored.

In five sources the C$^{18}$O $J$=10$-$9 line is detected in the line wing of the H$_{2}$O 3$_{12}-$3$_{03}$ (1097\,GHz) line. Before performing analysis on these data, we remove the C$^{18}$O emission by subtracting a Gaussian with the same FWHM, line-centre and amplitude as obtained by \citet{SanJoseGarcia2013}.

As noted in Table~\ref{T:observations_sources}, the more accurate SMA coordinates for IRAS15398 were observed in two settings, H$_{2}$O 1$_{11}-$0$_{00}$ and H$_{2}^{18}$O 1$_{10}-$1$_{01}$, as part of programme OT2\_evandish\_4. Comparison of these observations with the WISH observations is discussed in Appendix~\ref{S:appendix2_iras15398}. In the rest of this paper we will focus on the WISH observations as these include the most transitions observed towards the same position.

\section{Results}
\label{S:results}

This section begins with presentation of those results that can be obtained simply from the data themselves (Sect.~\ref{S:results_profiles}). The profiles are then fitted with multiple Gaussian components, which are subsequently divided into different physically motivated categories based on their properties (Sect.~\ref{S:results_components}).

\subsection{Line profiles}
\label{S:results_profiles}

\begin{table*}
\begin{center}
\caption[]{Detection statistics, average noise and FWZI for each line and evolutionary stage.}
\begin{tabular}{lccccccccc}
\hline \noalign {\smallskip}
Line & \multicolumn{4}{c}{Class 0} && \multicolumn{4}{c}{Class I}  \\ \cline{2-5} \cline{7-10}
\noalign {\smallskip}
 & D/O.\tablefootmark{a} & $\overline{\sigma_{\mathrm{rms}}}$ & Mean FWZI & Median FWZI && D/O\tablefootmark{a} & $\overline{\sigma_{\mathrm{rms}}}$ & Mean FWZI & Median FWZI \\
 &  & (mK) & (\kms{}) & (\kms{}) && & (mK) & (\kms{}) & (\kms{}) \\
\hline\noalign {\smallskip}
H$_{2}$O 1$_{11}$-0$_{00}$&14/15&\phantom{0}19&\phantom{\tablefootmark{b}}79$\pm$32\tablefootmark{b}&\phantom{\tablefootmark{b}}82\tablefootmark{b}&&\phantom{0}7/11&\phantom{0}24&47$\pm$18&42\\
H$_{2}$O 1$_{10}$-1$_{01}$&15/15&\phantom{0}12&\phantom{\tablefootmark{b}}72$\pm$32\tablefootmark{b}&\phantom{\tablefootmark{b}}69\tablefootmark{b}&&12/14&\phantom{0}10&50$\pm$19&48\\
H$_{2}$O 2$_{12}$-1$_{01}$&\phantom{0}5/\phantom{0}5&123&69$\pm$13&63&&\phantom{0}0/\phantom{0}0&\phantom{0}$-$&$-$&$-$\\
H$_{2}$O 2$_{02}$-1$_{11}$&14/15&\phantom{0}22&\phantom{\tablefootmark{b}}75$\pm$33\tablefootmark{b}&\phantom{\tablefootmark{b}}81\tablefootmark{b}&&\phantom{0}9/11&\phantom{0}22&34$\pm$13&34\\
H$_{2}$O 2$_{11}$-2$_{02}$&12/15&\phantom{0}20&\phantom{\tablefootmark{b}}65$\pm$27\tablefootmark{b}&\phantom{\tablefootmark{b}}62\tablefootmark{b}&&\phantom{0}7/\phantom{0}9&\phantom{0}17&33$\pm$15&33\\
H$_{2}$O 3$_{12}$-2$_{21}$&\phantom{0}7/15&105&\phantom{\tablefootmark{b}}58$\pm$21\tablefootmark{b}&\phantom{\tablefootmark{b}}54\tablefootmark{b}&&\phantom{0}4/11&122&22$\pm$5&22\\
H$_{2}$O 3$_{12}$-3$_{03}$&\phantom{0}8/\phantom{0}8&\phantom{0}17&\phantom{\tablefootmark{c}}81$\pm$26\tablefootmark{c}&\phantom{\tablefootmark{c}}75\tablefootmark{c}&&\phantom{0}2/\phantom{0}2&\phantom{00}9&\phantom{\tablefootmark{d}}42\tablefootmark{d}&\phantom{\tablefootmark{d}}42\tablefootmark{d}\\
\hline\noalign {\smallskip}
H$_{2}^{18}$O 1$_{11}$-0$_{00}$&\phantom{0}1/15&\phantom{0}18&\phantom{\tablefootmark{d}}16\tablefootmark{d}&\phantom{\tablefootmark{d}}16\tablefootmark{d}&&\phantom{0}0/11&\phantom{0}26&$-$&$-$\\
H$_{2}^{18}$O 1$_{10}$-1$_{01}$&\phantom{0}3/13&\phantom{00}4&41$\pm$8&45&&\phantom{0}0/\phantom{0}1&\phantom{00}4&$-$&$-$\\
H$_{2}^{18}$O 2$_{02}$-1$_{11}$&\phantom{0}0/\phantom{0}3&\phantom{0}16&$-$&$-$&&\phantom{0}0/\phantom{0}0&\phantom{0}$-$&$-$&$-$\\
H$_{2}^{18}$O 3$_{12}$-3$_{03}$&\phantom{0}2/\phantom{0}8&\phantom{0}14&\phantom{\tablefootmark{d}}12\tablefootmark{d}&\phantom{ \tablefootmark{d}}12\tablefootmark{d}&&\phantom{0}1/\phantom{0}2&\phantom{00}8&\phantom{\tablefootmark{d}}33\tablefootmark{d}&\phantom{ \tablefootmark{d}}33\tablefootmark{d}\\
\hline\noalign {\smallskip}
\label{T:results_profiles_basic_fwzi}
\end{tabular}
\tablefoot{\tablefoottext{a}{No. of sources with detections out of the total observed in each line.} \tablefoottext{b}{Detections for BHR71 and L1448-MM excluded.} \tablefoottext{c}{Detection for L1448-MM excluded, BHR71 not observed.} \tablefoottext{d}{No standard deviation is given for detections in less than three sources.} }
\end{center}
\end{table*}

All H$_{2}$O spectra for three Class 0 and two Class I sources are shown in Fig.~\ref{F:results_profiles_basic_comparison} as an example, with spectra for all WISH sources presented in Appendix~\ref{S:appendix1} in Figures~\ref{F:H2O_class0_1}~$-$~\ref{F:H2O_classI_3} for all H$_{2}$O transitions. The H$_{2}^{18}$O spectra for sources with at least one detection are shown in Figure~\ref{F:H218O_detections}. 

As can be seen in Figure~\ref{F:results_profiles_basic_comparison}, the water line profiles are often broad and complex, with generally narrower emission towards Class I with respect to Class 0 sources. There is significant variation in line intensity and shape between different sources, which is not particularly surprising given the range the sample covers in terms of luminosity, envelope mass and outflow activity (see further discussion in Sec.~\ref{S:analysis_correlations}).

The basic properties of the spectra; noise level in 0.3\kms{} bins, peak brightness temperature, integrated intensity and full-width at zero intensity (FWZI), are tabulated for all sources and lines in Tables~\ref{T:int_ortho}~$-$~\ref{T:int_h218o}. 

The FWZI is measured on spectra resampled to 3\kms{} to improve the signal-to-noise ratio. First, the furthest points from the source velocity that are above 2$\sigma_{\mathrm{rms}}$ of the resampled spectrum within a window around the line are found. The FWZI is then between the first channel moving away from the source velocity in each direction where the spectrum drops below 1$\sigma_{\mathrm{rms}}$. The integrated intensity is then calculated over the range identified by the FWZI. While this approach is more data than source driven, there is approximately a factor of 10 difference in the noise level between the deepest and shallowest spectra (see Table~\ref{T:results_profiles_basic_fwzi}). Thus using an alternative definition of the FWZI based on a set fraction of the peak in a way that is consistent and comparable between the different transitions would require a high enough threshold that it would not reflect the broadness of the line wings. It could also be skewed in the lower excitation lines by the narrow emission and/or absorption at the source velocity (for example, see IRAS4B in Fig.~\ref{F:results_profiles_basic_comparison}).

Table~\ref{T:results_profiles_basic_fwzi} presents the detection statistics, median noise level, and the mean and median FWZI for all detections separated by the evolutionary stage of the source. For the Class 0 sources, BHR71 and L1448-MM are excluded because they have bullet emission (discussed further in Sect.~\ref{S:results_components_spotshock}) which significantly increases their FWZI compared to other sources but were not observed in all lines. 

The average H$_{2}$O FWZIs (see Table~\ref{T:results_profiles_basic_fwzi}) are remarkably similar for Class 0 sources. There is also little difference between the mean and median values, suggesting that these values are not dominated by a few sources and so are representative of general source properties. Given the order of magnitude difference between the highest and lowest sensitivity observations, this suggests that on average our observations have a high-enough sensitivity to detect the full extent of the line wings. While the Class I sources are fainter and so have a lower signal-to-noise ratio, the transitions also look narrower, so it seems unlikely that higher sensitivity would increase their mean FWZI to the point where it was consistent with the Class 0 sources. Variation in line shape between transitions for a given source is relatively small, particularly in the line wings for the Class 0 sources. In a few cases the FWZI varies between the different transitions for a given source, but in all cases except Ser-SMM3 the results are due to variation in the noise level of the different spectra. The reason that Ser-SMM3 is likely still consistent with the general picture is discussed in Appendix~\ref{S:appendix2_smm3}. 

\begin{figure}
\begin{center}
\includegraphics[width=0.40\textwidth]{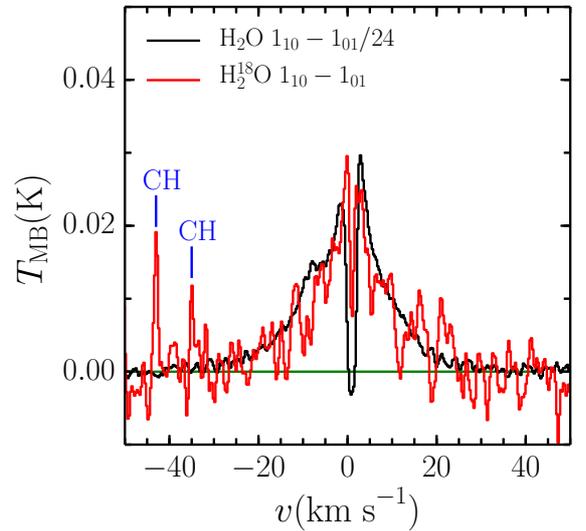}
\caption{Comparison of the H$_{2}$O (black) and H$_{2}^{18}$O (red) 1$_{10}-$1$_{01}$ spectra for IRAS4B where the H$_{2}$O spectrum has been scaled down such that the peak intensities are the same. The green line indicates the baseline. The blue lines indicate the approximate velocities of the CH transitions from the other sideband in the H$_{2}^{18}$O observations. The third component of the CH triplet is just beyond the plotted range but is also detected.}
\label{F:results_profiles_comparison_h2_18o}
\end{center}
\end{figure}

The FWZI for all H$_{2}^{18}$O detections except the 3$_{12}-$3$_{03}$ line towards Elias\,29 are smaller than those for the corresponding H$_{2}^{16}$O transition by a factor of 2$-$8. However, as shown in Fig.~\ref{F:results_profiles_comparison_h2_18o}, the spectra are consistent within the noise. Thus the difference is most likely a signal-to-noise issue. Comparison of the integrated intensities assuming an isotopic $^{16}$O/$^{18}$O ratio of 540 \citep[][]{Wilson1994} results in an optical depth for the H$_{2}$O transitions of order 20$-$30 assuming that H$_{2}^{18}$O is optically thin. 

Only TMC1A and Oph-IRS63, both Class I sources, were not detected in any transition at the 3$\sigma$ level (in 0.3\kms{} bins). All sources detected in the 1$_{10}-$1$_{01}$ line are also detected, where observed, in all other H$_{2}$O lines except the 3$_{12}-$2$_{21}$ transition. The non-detections in this line are likely due to the higher noise in these data as it is generally sources that are fainter in the other lines that are not detected. This is also likely the reason for the non-detections in H$_{2}^{18}$O as it is only the very brightest sources that are detected, and even then most have a peak signal-to-noise of less than 10. Of the 14 sources observed in the H$_{2}^{18}$O 1$_{10}-$1$_{01}$ transition, seven (BHR71, L1527, NGC1333-IRAS2A, NGC1333-IRAS4A, NGC1333-IRAS4B, Ser-SMM1 and Ser-SMM4) have detections of the CH triplet at 536.76$-$536.80\,GHz in emission in the other side-band. For NGC1333-IRAS4A, this is confused with the H$_{2}^{18}$O line, so the CH triplet is masked during the analysis. Analysis of the CH emission itself is beyond the scope of this paper.

The conclusion from the comparison of line profiles and FWZIs is therefore that the lower FWZI for the H$_{2}^{18}$O transitions compared to the corresponding H$_{2}^{16}$O line for a given source is just a signal-to-noise issue. However, the decrease in the average FWZI between Class 0 and I is real and not related to the sensitivity of the data.

\subsection{Line components}
\label{S:results_components}

\subsubsection{Gaussian decomposition}
\label{S:results_components_decomposition}

As can be seen in Figure~\ref{F:results_profiles_basic_comparison}, the water line profiles towards low-mass protostars are complex and generally not well reproduced by a single line shape, e.g. a single Gaussian, Lorentzian or triangular profile. However, as shown by \citet[][]{Kristensen2010,Kristensen2012} they can be decomposed into multiple components, each relating to different parts of the protostellar system. 

In reality, the detailed shape of the emission from a given region will depend on both the physics and geometry, particularly for shocks, and so a range of line shapes may indeed be present \citep[see e.g.][]{Jimenez-Serra2008a}. However, the observed H$_{2}$O line-shapes, particularly in high s/n data, appear Gaussian-like, so this is the most reasonable line-shape to assume. The reason that the emission from shocks is Gaussian-like may be due to our observations encompassing a number of shocks with a range of viewing angles. Alternatively, this may be the result of mixing and turbulence induced by Kelvin-Helmholtz instabilities along the cavity wall \citep[see e.g.][]{Bodo1994,Shadmehri2008}. 

As discussed in Sect.~\ref{S:results_profiles}, the width of the line profiles does not change significantly between the observed transitions, though the relative and absolute intensity of individual components does change. Therefore, while the physical conditions in the different regions within the protostar where water is emitting may be different, all transitions are probably emitting from the same parcels of gas in each case. 

We therefore choose to require that the line centre and width of each Gaussian component are exactly the same for all transitions observed towards a given source, though the intensity of a given component can be different for each line. In practical terms, this is achieved by creating an array which contains all H$_{2}$O and H$_{2}^{18}$O spectra for a given source and fitting a global function to this array which contains a number of Gaussians equal to the number of components multiplied by the number of transitions. For a given component, the line centre and width are common variables between the Gaussians applied to each transition. They are therefore constrained by all available data for a particular source, decreasing the uncertainties and improving the reliability of the fit, particularly in cases where the emission in some transitions is weak. For high signal-to-noise spectra, the difference between fitting each line separately and this global fitting approach is small, as shown in Figure~2 of \citet{Kristensen2013}. Those authors were able to use individual fits because they focused on the brightest Class 0 sources in the WISH sample and were interested in one relatively distinct component. Here we want to isolate and analyse all components in all sources, so a global fitting approach is preferred.

\begin{figure}
\begin{center}
\includegraphics[width=0.40\textwidth]{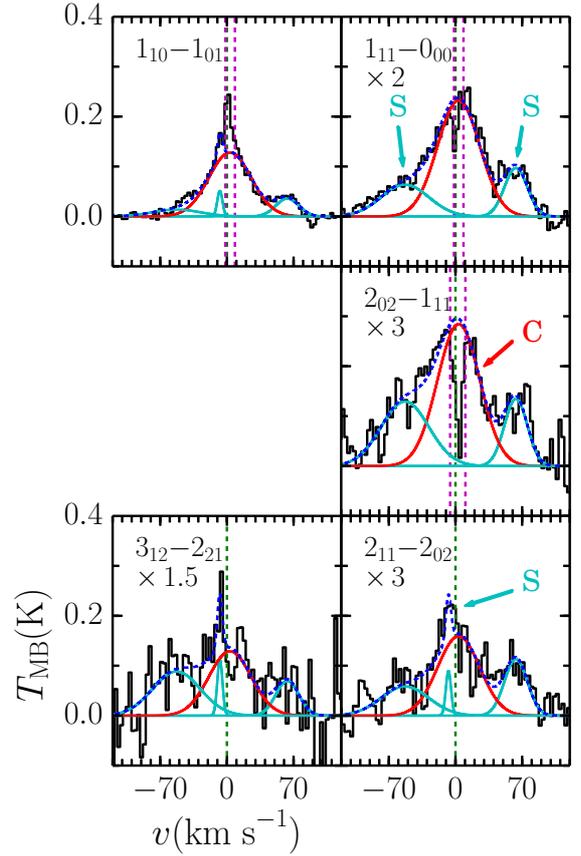}
\caption{Continuum subtracted WBS spectra for BHR71 (black) resampled to 3\kms{}. The red and cyan lines show the individual Gaussian components for the cavity shock (C) and spot shocks (S) respectively (see text and Table~\ref{T:results_components_decomposition_previous} for details) while the blue dashed line shows the combined fit for each line. All spectra have been shifted so that the source velocity is at 0\kms{}, which is indicated by the green dashed lines. At full resolution the quiescent envelope component has an inverse P-cygni profile \citep[see][]{Mottram2013} and so is masked (indicated by the magenta dashed lines) rather than being fit by multiple components during the Gaussian fitting. The broad absorption in the 2$_{02}-$1$_{11}$ transition (middle left panel) is caused by reference contamination and is also masked during the fitting process.}
\label{F:results_components_decomposition_example}
\end{center}
\end{figure}

An example best-fit result is shown in Fig.~\ref{F:results_components_decomposition_example} for BHR71, a source with a mix of low and high s/n spectra. For this source, the quiescent envelope component shows an inverse P-Cygni profile at full resolution and so is masked out from the fitting process. In other cases where only a simple emission or absorption profile from the envelope is observed, this is included in the Gaussian fit. For the 2$_{02}-$1$_{11}$ transition the absorption is due to reference contamination and so is also masked from the fitting. 

The fit results were obtained using the ordinary least-squares solver in the \textsc{python} module \textsc{scipy.odr}\footnote{http://scipy.org/} starting from an initial guess for a single Gaussian. The results and residuals of this fit were examined and the number of components increased or the initial guess modified to result in residuals below the rms. While this approach can be susceptible to finding local minima in some cases, particularly with very complex line profiles such as for BHR71, the combination of varying the initial guess and visual inspection of the residuals ensured that this returned reasonable results (e.g. combinations of large positive and negative Gaussians which mostly cancel out are excluded). In all cases the number of Gaussian components used was the minimum required for the residuals to be within the rms noise. The results of the Gaussian fitting for all sources are presented in Tables~\ref{T:Gaussians_class0_o} to \ref{T:Gaussians_h2_18o}. Where a component is not detected in a given line, a 3$\sigma$ upper limit is calculated from the noise in the spectrum. The results are consistent with those presented in previous papers \citep{Kristensen2012,Kristensen2013,Mottram2013} taking into account the latest reduction and calibration.

Having identified these components, it is then a question of attempting to relate them to the different physical components of a protostellar system. In previous work \citep{Kristensen2010,Kristensen2012,SanJoseGarcia2013,Yildiz2013} the different components have been established and named based primarily on their line-width. However, this is a rather phenomenological convention and does not always allow for clear distinction between different excitation conditions, as also noted in \citet{Kristensen2013}. We therefore prefer to use terms which indicate the most likely physical origin of the emission component \citep[c.f.][for similar terminology applied to high-mass protostars]{vanderTak2013}. Table~\ref{T:results_components_decomposition_previous} provides a summary of how these new terms are related to those used in previous papers on low-mass protostars in order to ensure continuity. 

\begin{table}
\begin{center}
\caption[]{Component terminology.}
\begin{tabular}{ccc}
\hline \noalign {\smallskip}
This paper\tablefootmark{a} & Previous papers & References \\
\hline\noalign {\smallskip}
Envelope & Narrow & 1 \\
Cavity shock & Broad or medium & 1 \\
Spot shock& Bullet or EHV, also offset & 1,2 \\
& or medium if broad also present&\\
\hline\noalign {\smallskip}
\label{T:results_components_decomposition_previous}
\end{tabular}
\tablefoot{\tablefoottext{a}{See sections~\ref{S:results_components_envelope} $-$ \ref{S:results_components_spotshock} and Fig.~\ref{F:results_components_comparison_flow} for criteria.}}
\tablebib{(1) \citet{Kristensen2012}; (2) \citet{Kristensen2013}}
\end{center}
\end{table}

\begin{figure}
\begin{center}
\includegraphics[width=0.48\textwidth]{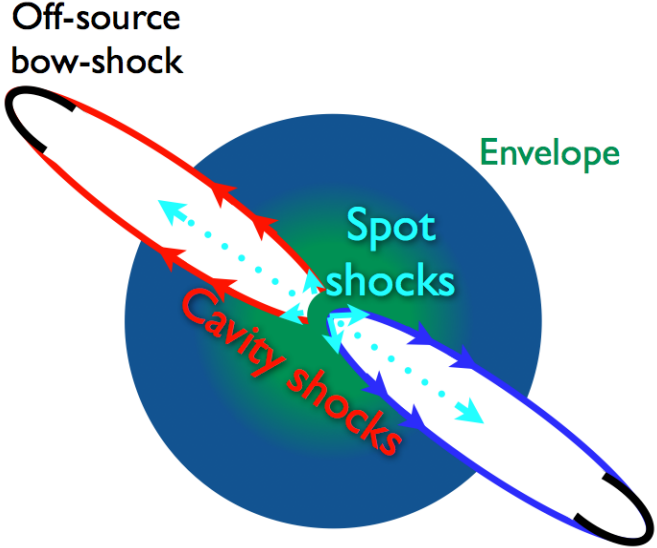}
\caption{Cartoon showing the proposed origin of the various distinct kinematic gas components observed in low-$J$ water line profiles.}
\label{F:discussion_picture_cartoon}
\end{center}
\end{figure}

The different components for each source are divided into three categories: envelope, cavity shock and spot shock, building on the work of \citet{Kristensen2012,Kristensen2013}, with the first letter of each term being used to identify them in Tables~\ref{T:Gaussians_class0_o} to \ref{T:Gaussians_h2_18o}. The following subsections (Sect.~\ref{S:results_components_envelope}$-$\ref{S:results_components_spotshock}) will discuss and motivate the definition of each of these components in turn, with Figure~\ref{F:discussion_picture_cartoon} indicating their expected physical location in a protostellar system. Following this, a summary and comparison showing how the kinematic properties of the different components relate to each other will be presented to verify that they are distinct (Sect.~\ref{S:results_components_comparison}). 

\subsubsection{Envelope}
\label{S:results_components_envelope}

Emission from the quiescent envelope is characterised by small FWHM and offset from the source velocity, thus we assign this designation to the component with the smallest FWHM for each source which has FWHM $\leq$ 5\kms{} and offset $\leq$ 2\kms{}. This can be in absorption in the ground-state lines, particularly for Class 0 sources, and even saturated where all line and continuum photons are absorbed. One confirmation that this emission and absorption comes from the envelope is that the line centres and widths are similar to those observed in C$^{18}$O towards these sources \citep{SanJoseGarcia2013}. No sources show distinct foreground absorptions offset from the source velocity, unlike HIFI spectra towards high-mass protostars \citep[e.g.][]{vanderTak2013}, primarily due to the much smaller distances to our sources. Thus most of the absorption likely comes from the protostars own envelope. Given that the sub-mm continuum and line emission from the envelope is centrally condensed \citep{Jorgensen2007,Kristensen2012,Mottram2013} we assume that the emission scales as a point-source.

While many sources also show envelope emission, it is often non-Gaussian in shape in the ground-state lines, consisting of combinations of emission and absorption in either inverse or regular P-Cygni profiles which are indicative of infall and expansion respectively. This was characterised in the 1$_{10}-$1$_{01}$ (557\,GHz) line by \citet{Kristensen2012}, and the cases showing infall profiles were analysed in more detail by \citet{Mottram2013}. In these cases, the combination of envelope emission and absorption is not a single Gaussian and so the relevant parts of the spectra are masked during the fitting process (e.g. see Fig.~\ref{F:results_components_decomposition_example}).

Absorption from the envelope is also observed in the H$_{2}^{18}$O 1$_{11}-$0$_{00}$ and 1$_{10}-$1$_{01}$ lines towards SMM1, which is consistent with the envelope of this source being particularly massive and having a relatively shallow density power-law slope \citep[c.f.][]{Kristensen2012}. The only source to show emission from the envelope in any H$_{2}^{18}$O transitions is IRAS2A, where the tentative detections in the H$_{2}^{18}$O 3$_{12}-$3$_{03}$ and 2$_{02}-$1$_{11}$ lines are the narrowest for any source and offset from the main outflow emission detected in the H$_{2}^{16}$O transitions. This emission is likely related to the hot core where $T_{\mathrm{dust}}>$100\,K \citep[see][for more details]{Visser2013} and originates on arcsecond scales based on interferometric observations \citep{Persson2012,Persson2014}. We do not study the envelope emission further in this paper. Further analysis of other sources showing absorption in the ground-state water lines, including the link with water ice, will be presented in \citet{Schmalzl2014}.

\subsubsection{Cavity shock}
\label{S:results_components_cavityshock}

Having identified any envelope contribution, we designate the remaining component which is not in absorption in any line and has the smallest ratio of offset to FWHM as the cavity shock component. This is an empirical determination based on the assumption that the average velocity offset of the currently shocked gas in the outflow cavity is lower than for more discrete and energetic shocks and that it should not be in absorption against the continuum because the emission is most likely formed on larger scales. The offset of this component is always less than 15\kms{} and decreases with smaller FWHM. That this component is Gaussian in shape, combined with the small offset compared to the FWHM, suggests that we are detecting both the red and blue-shifted lobes of the outflow cavity. 

Water emission is elongated along the direction of the outflow \citep[e.g.][]{Nisini2010, Santangelo2012} with the dominant extended component having similar velocity distributions \citep[e.g.][]{Santangelo2014} as this component. As cavity shocks also dominate the on-source line profiles, we assume that it is elongated along the outflow direction but does not fill the beam parallel to the outflow axis, as in the spectrally unresolved PACS H$_{2}$O observations.

This component should not be confused with the entrained outflow material typically probed by low-$J$ CO observations, as H$_{2}$O and low-$J$ CO emission are not spatially coincident \citep[e.g.][]{Nisini2010,Santangelo2013}. A detailed comparison between CO $J$=3$-$2 and H$_{2}$O 1$_{10}-$1$_{01}$ was presented in \citet{Kristensen2012} with the clear conclusion that these two transitions do not trace the same material. One of the main reasons is the enormous difference in critical density between these two transitions (10$^{4}$\,cm$^{-3}$ and 10$^{7}$\,cm$^{-3}$ respectively). As the gas is heated and compressed in the cavity shocks, the water abundance increases dramatically through both gas-phase synthesis and ice sputtering. During this warm and dense phase, water is one of the dominant coolants. However, as the gas cools and expands to come into pressure equilibrium with its surroundings, water excitation becomes highly inefficient due to high critical densities so little water emission originates from the cold entrained low-density outflow. Therefore, the non-coincidence of water and low-$J$ CO is consistent with the expectation that water is significantly depleted under the typical conditions in the entrained outflowing gas.

Most detections in the H$_{2}^{18}$O observations are associated with the cavity shock component, with the exception of IRAS2A as discussed above and IRAS4A, which is discussed in more detail in Appendix~\ref{S:appendix2_iras4A}. 

\subsubsection{Spot shock}
\label{S:results_components_spotshock}

All remaining components which show larger offset/FWHM are designated as spot shock components. The separation of the cavity and spot shock components is necessary because the line profiles show separate and distinct kinematic components (e.g. see Fig.~\ref{F:results_components_decomposition_example}), suggesting that they come from different shocks within the protostellar system. The use of offset/FWHM is also chosen so as to separate the component most likely associated with C-type shocks (cavity shock), where emission is centred at the source velocity, with components more likely associated with J-type shocks (spot shock), where emission is shifted away from the source velocity to the shock velocity relative to the line of sight \citep[see e.g.][]{Hollenbach1997}. 

Some spot shock components are significantly offset from the source velocity, such that they are characteristic of ``bullet'' emission with large offsets ($>$20\kms{}) from the source velocity and large FWHM (also $>$20\kms{}, e.g. see Fig.~\ref{F:results_components_decomposition_example}). These are most likely associated with J-type shocks along the jet, as they have similar kinematic properties to EHV bullet emission in CO and SiO which is spatially located in knots along the jet axis \citep[e.g.][]{Bachiller1990,Bachiller1991,Hirano2006,Santiago-Garcia2009}.

The spot shock emission with lower velocity offset may originate in J-type shocks near the base of the outflow where the wind first impacts the envelope or outflow cavity, as first suggested by \citet{Kristensen2013}. Those authors based this conclusion on: (\textit{i}) some of the spot shock components detected in water line profiles are seen in absorption against the continuum but not the outflow; (\textit{ii}) when detected in OH$^+$ and CH$^+$, the components are always in absorption against the continuum with no emission component and no outflow component. These two pieces of evidence point to an origin in front of the continuum and behind the outflow. In both cases, the velocity offset strongly suggests that the components are associated with J-type shocks \citep[e.g.][]{Hollenbach1997}.

As already noted by \citet{Kristensen2013} for NGC1333-IRAS3A and Ser-SMM3, a few sources show spot shock components in absorption. These components are too offset and/or broad to be consistent with absorption due to the envelope or foreground clouds. In addition, they are present in excited transitions which makes a foreground origin highly unlikely. Off-position contamination can also be excluded due to the offset from the source velocity and that the 1$_{10}-$1$_{01}$ position-switched observations share reference positions with other sources which do not show these components. The depth of these absorption features are consistent with absorption against the continuum only, suggesting that they originate between the observer and the continuum source, but not between the observer and the outflow emission.

We do not separate the ``bullet'' and less offset spot shocks into separate categories because the inclination of the shock relative to the line of sight plays a role in how offset a component is. However, in the cases where the offset from the source velocity is small, the spot shock components are always narrower than the cavity shock.

The suggested physical location of the spot shocks, whether in the jet or at the base of the outflow, is indicated in Fig.~\ref{F:discussion_picture_cartoon}. Bullets are observed to be small (few arcseconds) and point-like knots in interferometric observations \citep[e.g.][]{Hirano2006,Santiago-Garcia2009} and the analysis of \citet{Kristensen2013} suggests that the non-bullet spot shocks originate from very small regions ($\sim$100\,AU) near the central protostar. This is also supported by the strong similarity in line shape between the spot shock component observed in water for IRAS2A and the compact ($\sim$1\arcsec{}) emission seen in SiO and SO towards MM3 in recent interferometry observations by \citet{Codella2014}. A point-like geometry is therefore the most appropriate assumption for the spot shock component.

\subsubsection{Comparison of components}
\label{S:results_components_comparison}

\begin{figure}
\begin{center}
\includegraphics[width=0.35\textwidth]{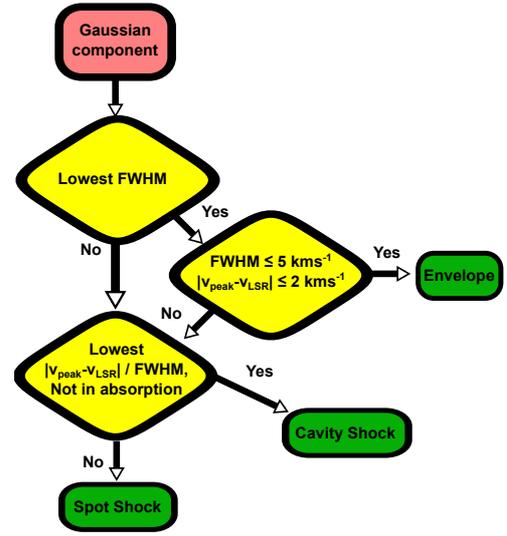}
\caption{Flow diagram for component type determination. $\mid v_{\mathrm{peak}}-v_{\mathrm{LSR}}\mid$ is the offset of the component centre from the source velocity.}
\label{F:results_components_comparison_flow}
\end{center}
\end{figure}

\begin{figure}
\begin{center}
\includegraphics[width=0.48\textwidth]{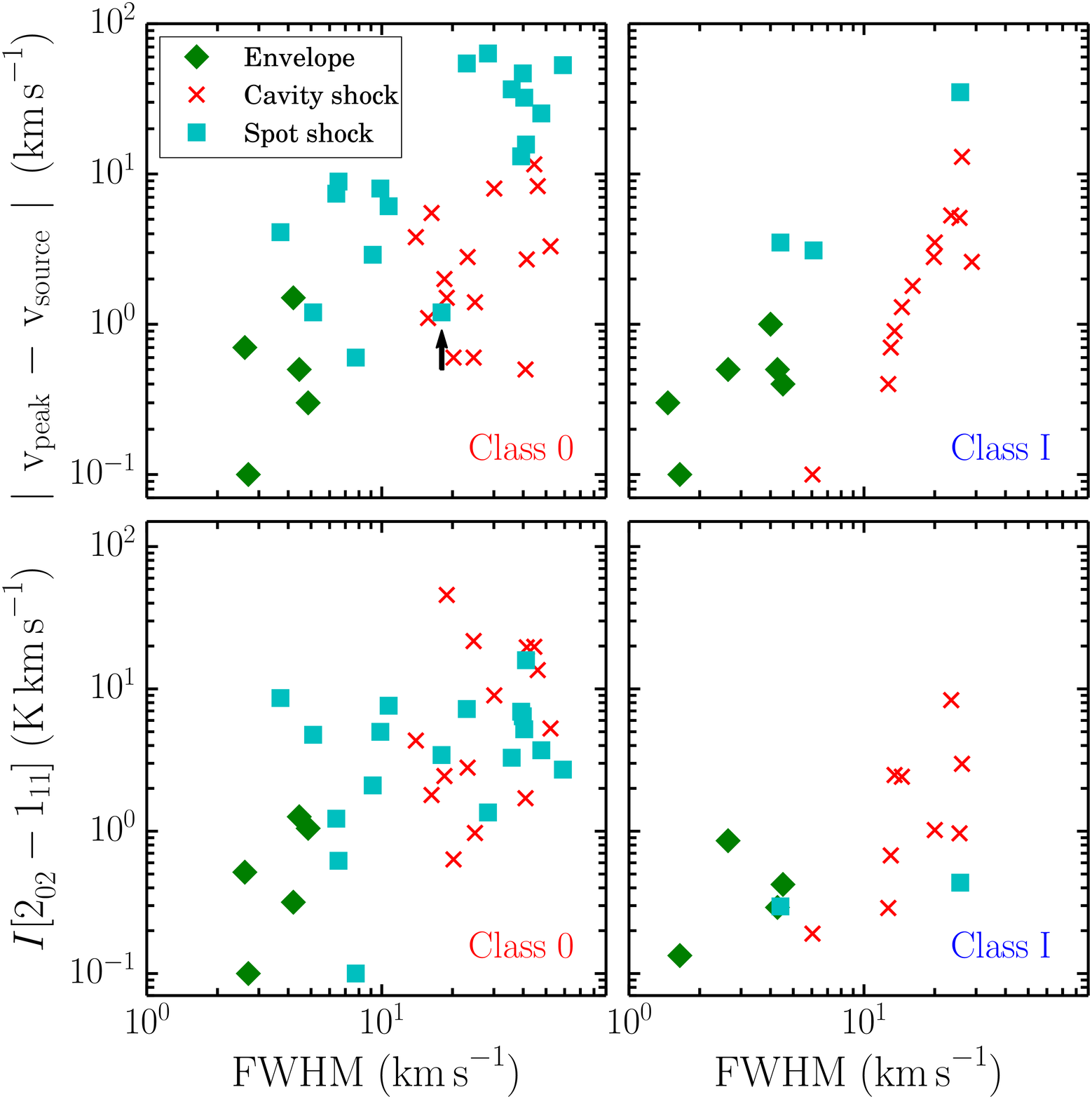}
\caption{Scatter plots of FWHM vs. offset of the peak from the source velocity (top) and intensity in the 2$_{02}-$1$_{11}$ line corrected to a common distance of 200\,pc (bottom) for the Gaussian components for Class 0 (left) and I (right) sources. When scaling the intensities, a linear scaling was used for the cavity shock components while a point-source scaling was used for the spot shock and envelope components. The black arrow indicates the broader of the two shock spots towards NGC1333-IRAS4A which is discussed further in Appendix~\ref{S:appendix2_iras4A}.}
\label{F:results_components_comparison_scatter}
\end{center}
\end{figure}

A summary of the overall classification scheme for the various components is shown in Figure~\ref{F:results_components_comparison_flow}. Fig.~\ref{F:results_components_comparison_scatter} then shows the relationship between FWHM and both velocity offset and the intensity in the 2$_{02}-$1$_{11}$ (988\,GHz) transition for the various components scaled to the typical distance for the sample of 200\,pc. As most of the outflows from these sources are larger than the \textit{Herschel} beam along the outflow axis (see Yildiz et al., subm.), the intensity of the cavity shock component was corrected using a linear scaling, i.e. assuming that the emission fills the beam in one direction and is point-like perpendicular to it ($I_{\textrm{obs}}\,(d/200)^{1}$). The spot shock and quiescent envelope components are assumed to be point-like ($I_{\textrm{obs}}\,(d/200)^{2}$). 

Though there are a few exceptions, the different components generally lie in distinct regions of the FWHM vs. offset parameter space, supporting the idea that they are formed under different conditions. In particular, the cavity shock and spot shock components are relatively well separated. The regions of FWHM vs. offset covered by the different components for the Class 0 and I sources are also similar.

The spot shock component which lies in the middle of the cluster cavity shocks in the Class 0 FWHM vs. offset plot is the broader of the two spot shocks towards NGC1333-IRAS4A, marked with a black arrow in Fig.~\ref{F:results_components_comparison_scatter}. This is likely related to bow-shocks which lie within the HIFI beam for the lower-frequency transitions (see Appendix~\ref{S:appendix2_iras4A} for more details).

In general, the intensity of the components in the Class I sources is lower than for the Class 0s. Table~\ref{T:results_components_comparison_fract} shows the number of Class 0 and I sources in which the cavity shock and spot shock components are detected for each transition, as well as the mean and standard deviation in the fractional intensity in each component with respect to the total observed intensity. For the quiescent envelope component as this can sometimes include both absorption and emission, this was calculated by subtracting the intensity of the other detected components from the total observed intensity, but may include emission and absorption which cancel each other out. Absorptions in some components can lead to other components having larger intensities than the total.

While there is significant overlap in the intensity of components in the lower panels of Fig~\ref{F:results_components_comparison_scatter}, the results in Table~\ref{T:results_components_comparison_fract} show that for a given source, the cavity shock dominates all the lines observed with HIFI, consisting of between 70 and 100$\%$ of the integrated emission. The spot shocks contribute $\sim$20$\%$ for Class 0 sources and are on average negligible for Class I sources. The detection fraction of spot shocks is also much lower for Class I sources. The quiescent  envelope does not have a strong contribution in the excited lines for Class 0 sources, though it can reduce the integrated intensity in the ground-state lines by up to 20$\%$ depending on the balance of emission and absorption. It plays a more significant role in Class I sources, contributing up to 30$\%$ of the total intensity.

\begin{table*}
\begin{center}
\caption[]{Detection statistics and average fraction of the total integrated intensity that each component contributes for each transition.}
\begin{tabular}{lccccccccccccc}
\hline \noalign {\smallskip}
Line & \multicolumn{6}{c}{Class 0} && \multicolumn{6}{c}{Class I}  \\ \cline{2-7} \cline{9-14}
\noalign {\smallskip}
 & \multicolumn{2}{c}{Envelope\tablefootmark{a}} & \multicolumn{2}{c}{Cavity Shock} & \multicolumn{2}{c}{Spot Shock} && \multicolumn{2}{c}{Envelope\tablefootmark{a}} & \multicolumn{2}{c}{Cavity Shock} & \multicolumn{2}{c}{Spot Shock}\\
\noalign {\smallskip}
 & D\tablefootmark{b} & $I_{\mathrm{comp}}/I_{\mathrm{tot}}$& D\tablefootmark{b} & $I_{\mathrm{comp}}/I_{\mathrm{tot}}$& D\tablefootmark{b} & $I_{\mathrm{comp}}/I_{\mathrm{tot}}$&& D\tablefootmark{b} & $I_{\mathrm{comp}}/I_{\mathrm{tot}}$& D\tablefootmark{b} & $I_{\mathrm{comp}}/I_{\mathrm{tot}}$& D\tablefootmark{b} & $I_{\mathrm{comp}}/I_{\mathrm{tot}}$\\
\hline\noalign {\smallskip}
H$_{2}$O 1$_{11}$-0$_{00}$&14&$-$0.1$\pm$0.1\phantom{$-$}&14&0.9$\pm$0.1&\phantom{0}9&0.2$\pm$0.1&&\phantom{0}4&0.0$\pm$0.1&\phantom{0}4&1.0$\pm$0.1&0&$-$\\
H$_{2}$O 1$_{10}$-1$_{01}$&15&$-$0.1$\pm$0.1\phantom{$-$}&15&0.9$\pm$0.1&\phantom{0}8&0.2$\pm$0.1&&12&0.1$\pm$0.1&11&1.1$\pm$0.3&3&$-$0.2$\pm$0.2\phantom{$-$}\\
H$_{2}$O 2$_{12}$-1$_{01}$&\phantom{0}5&$-$0.2$\pm$0.1\phantom{$-$}&\phantom{0}5&1.0$\pm$0.1&\phantom{0}5&0.2$\pm$0.2&&$-$&$-$&$-$&$-$&$-$&$-$\\
H$_{2}$O 2$_{02}$-1$_{11}$&14&0.0$\pm$0.1&14&0.7$\pm$0.1&10&0.3$\pm$0.1&&\phantom{0}8&0.2$\pm$0.1&\phantom{0}7&0.8$\pm$0.1&1&0.0$\pm$0.0\\
H$_{2}$O 2$_{11}$-2$_{02}$&11&0.0$\pm$0.1&11&0.7$\pm$0.1&\phantom{0}8&0.3$\pm$0.1&&\phantom{0}7&0.3$\pm$0.2&\phantom{0}5&0.7$\pm$0.2&0&$-$\\
H$_{2}$O 3$_{12}$-2$_{21}$&\phantom{0}7&0.0$\pm$0.1&\phantom{0}7&0.8$\pm$0.1&\phantom{0}5&0.2$\pm$0.1&&\phantom{0}3&0.0$\pm$0.1&\phantom{0}3&1.0$\pm$0.1&0&$-$\\
H$_{2}$O 3$_{12}$-3$_{03}$&\phantom{0}8&0.0$\pm$0.1&\phantom{0}8&0.8$\pm$0.1&\phantom{0}7&0.2$\pm$0.1&&\phantom{0}2&0.0$\pm$0.1&\phantom{0}2&1.0$\pm$0.1&0&$-$\\
\hline\noalign {\smallskip}
\label{T:results_components_comparison_fract}
\end{tabular}
\tablefoot{\tablefoottext{a}{Calculated for all sources with detected emission as $I_{\mathrm{tot}}-I_{\mathrm{cavity\,shock}}-I_{\mathrm{spot\,shock}}$. May include emission and absorption.} \tablefoottext{b}{No. of sources with detections in this component.}}
\end{center}
\end{table*}

\section{Analysis}
\label{S:analysis}

In this section we present analysis building on the results from the previous section. Discussion of the wider implication of the results and analysis, including comparison with other results in the literature, will be presented in Sect.~\ref{S:discussion}.

\subsection{Integrated intensity ratios}
\label{S:analysis_ratios}

\begin{figure}
\begin{center}
\includegraphics[width=0.40\textwidth]{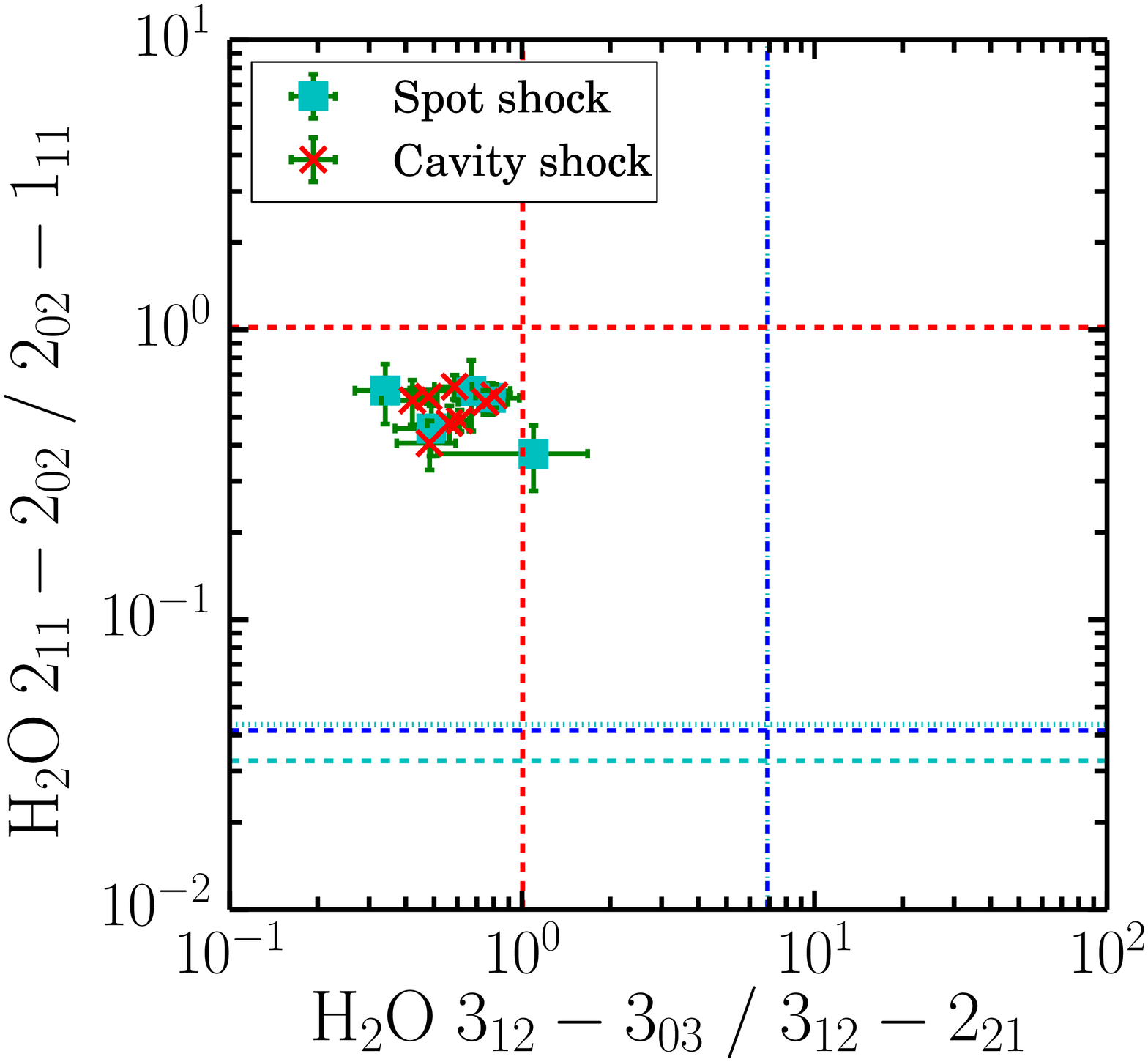}
\caption{Line ratio vs. line ratio plot. The red dashed lines indicate the limits in which both lines are in LTE and optically thick. The blue dashed, cyan dashed and cyan dotted lines indicate the limits in which both lines are in LTE and optically thin for excitation temperatures of 300, 100 and 500\,K respectively. For the x-axis, since the lines share their upper energy level then the ratio is not sensitive to temperature and the optically thin lines lie on top of each other. Observed ratios not on either line indicate either subthermal excitation and/or that one transition is optically thick while the other is optically thin. In both cases, the ratio is then dependant on the excitation conditions of the gas.}
\label{F:analysis_ratios_lratio}
\end{center}
\end{figure}

\begin{table*}
\begin{center}
\caption[]{Average H$_{2}$O line ratios.}
\begin{tabular}{lccccccc}
\hline \noalign {\smallskip}
Transitions & \multicolumn{2}{c}{\phantom{\tablefootmark{a}}D\tablefootmark{a}} & \multicolumn{2}{c}{\phantom{\tablefootmark{b}}Observed ratio\tablefootmark{b}} & \phantom{\tablefootmark{c}}Thin LTE\tablefootmark{c} & \phantom{\tablefootmark{c}}Thick LTE\tablefootmark{c} & \phantom{\tablefootmark{d}}$\theta_{1}/\theta_{2}$\tablefootmark{d}  \\ 
\noalign {\smallskip}
& \phantom{\tablefootmark{e}}C\tablefootmark{e} & \phantom{\tablefootmark{f}}S\tablefootmark{f} & \phantom{\tablefootmark{e}}C\tablefootmark{e} & \phantom{\tablefootmark{f}}S\tablefootmark{f} & & & \\
\hline\noalign {\smallskip}
1$_{10}$-1$_{01}$/2$_{12}$-1$_{01}$&\phantom{0}5&\phantom{0}4&0.43$\pm$0.06&0.58$\pm$0.17&0.40&1.10&3.00\\
3$_{12}$-3$_{03}$/3$_{12}$-2$_{21}$&\phantom{0}8&\phantom{0}5&0.61$\pm$0.04&0.62$\pm$0.19&6.91&1.00&1.05\\
1$_{11}$-0$_{00}$/2$_{02}$-1$_{11}$&19&13&1.02$\pm$0.05&1.02$\pm$0.07&0.04&0.99&0.89\\
1$_{10}$-1$_{01}$/2$_{02}$-1$_{11}$&20&12&0.79$\pm$0.07&0.73$\pm$0.11&0.26&3.11&1.77\\
2$_{12}$-1$_{01}$/2$_{02}$-1$_{11}$&\phantom{0}5&\phantom{0}4&1.50$\pm$0.12&1.15$\pm$0.28&0.65&2.84&0.59\\
2$_{11}$-2$_{02}$/2$_{02}$-1$_{11}$&15&11&0.57$\pm$0.04&0.58$\pm$0.05&0.04&1.02&1.31\\
3$_{12}$-2$_{21}$/2$_{02}$-1$_{11}$&10&\phantom{0}7&1.04$\pm$0.09&0.99$\pm$0.17&0.06&2.96&0.86\\
3$_{12}$-3$_{03}$/2$_{02}$-1$_{11}$&10&\phantom{0}8&0.52$\pm$0.04&0.50$\pm$0.05&0.39&2.97&0.90\\
\hline\noalign {\smallskip}
\label{T:analysis_ratios}
\end{tabular}
\tablefoot{\tablefoottext{a}{Number of components with detections.} \tablefoottext{b}{Mean and standard error on the mean. Not corrected for beam size.} \tablefoottext{c}{Calculated for $T_{\mathrm{ex}}$=300\,K and an ortho-to-para ratio of 3.} \tablefoottext{d}{Beam size ratio.} \tablefoottext{e}{Cavity shock component.} \tablefoottext{f}{Spot shock component.}}
\end{center}
\end{table*}

A first step in studying the excitation and physical conditions of the water-emitting gas in young protostars is to understand the opacity of the observed transitions, for which there are four regimes. Below a certain $N_{\mathrm{H}_{2}\mathrm{O}}$, a given transition will be optically thin while at high column density it will be optically thick. Both of these cases can be either in local thermodynamical equilibrium (LTE), when $n_{\mathrm{H}_{2}}$ is above the critical density for that transition, or sub-thermally excited if $n_{\mathrm{H}_{2}}<<n_{\mathrm{crit}}$. As water has large Einstein A coefficients and high critical densities, there is a significant part of realistic parameter space that is optically thick but sub-thermally excited. In this regime, the lines are said to be effectively thin because the chance of collisional de-excitation is low, so photons effectively scatter within the region and will all eventually escape the $\tau$=1 surface. As such, the intensity still scales as $N_{\mathrm{H}_{2}\mathrm{O}}\times$$n_{\mathrm{H}_{2}}$ as in the optically thin sub-critical case \citep[see e.g.][]{Linke1977} even though $\tau>$1.

As discussed in Section~\ref{S:results_profiles}, for those few sources and transitions where we can obtain H$_{2}$O/H$_{2}^{18}$O ratios, these suggest that those components detected in H$_{2}^{18}$O are optically thick in those transitions. However, the number of lines, components and sources where this is the case is small. For sources or components for which H$_{2}^{18}$O data are not available or detected, we can also use the ratios of the integrated intensity of the different components in pairs of H$_{2}^{16}$O lines which share a common level. In the limit where both lines are optically thin, in LTE and have the same beam size, following \citet{Goldsmith1999}, the line ratio becomes:

\begin{equation}
\frac{I_{1}}{I_{2}}~=~\frac{g_{\mathrm{u1}}A_{\mathrm{ul1}}}{g_{\mathrm{u2}}A_{\mathrm{ul2}}}\frac{\nu_{2}^{2}}{\nu_{1}^{2}}\,e^{(E_{\mathrm{u2}}-E_{\mathrm{u1}})/k_{\mathrm{b}}T_{\mathrm{ex}}},
\end{equation}

\noindent where, for each transition, $g_{\mathrm{u1}}$ is the statistical weight of the upper level, $A_{\mathrm{ul}}$ is the Einstein A coefficient between the two levels, $\nu$ is the frequency, $E_{\mathrm{u}}$ is the upper level energy and $T_{\mathrm{ex}}$ is the excitation temperature. Alternatively, if both lines are optically thick, in LTE and have the same beam size the line ratio is given by:

\begin{equation}
\frac{I_{1}}{I_{2}}~=~\frac{\nu_{1}}{\nu_{2}}\frac{(e^{h\nu_{2}/k_{\mathrm{b}}T_\mathrm{ex}}-1)}{(e^{h\nu_{1}/k_{\mathrm{b}}T_\mathrm{ex}}-1)}.
\end{equation}

\noindent If one line is optically thick but the other is optically thin, and/or if the transitions are sub-thermally excited, then the line ratio can take a range of values depending on the excitation conditions of the gas.

Figure~\ref{F:analysis_ratios_lratio} shows such a comparison, covering the middle and upper excitation range probed by the water transitions accessible to HIFI. The intensity ratios for all components detected in both lines are consistent with or close to the limit where all lines are optically thick. The optically thin limits have been calculated for each ratio assuming excitation temperatures of 100, 300 and 500\,K, to show that the temperature variation of this limit does not impact the result of this simple analysis. For the 3$_{12}-$3$_{03}$/3$_{12}-$2$_{21}$ ratio, the lines come from the same upper energy level, so the optically thin LTE ratio is not sensitive to temperature. What is more, a search of a wide parameter space using the non-LTE molecular line radiative transfer code \textsc{radex} \citep[][discussed in more detail in Sect.~\ref{S:analysis_excitation}]{vanderTak2007} found no non-LTE optically thin solutions where the 3$_{12}-$3$_{03}$/3$_{12}-$2$_{21}$ is below 1. We can therefore exclude both the LTE and sub-thermal optically thin regimes for these transitions.

Table~\ref{T:analysis_ratios} shows the line ratios and the standard error on the mean averaged separately for cavity and spot shock components for the transitions which share a common energy level, as well as all lines relative to the 2$_{02}-$1$_{11}$ line. The optically thick and optically thin limits are also provided, assuming an excitation temperature of 300\,K and an ortho-to-para ratio of 3. We do not present average ratios including the H$_{2}^{18}$O transitions because there are so few detections that these may not be a fair comparison. The line ratios have not been corrected for the different beam-sizes of each transition, but for many ratios the difference in beam-size is small. Correction for a point source emitting region is ($\theta_{1}/\theta_{2}$)$^{2}$, for a cylindrical emitting region which fills the beam in one axis is ($\theta_{1}/\theta_{2}$)$^{1}$ and is 1 for an emitting region which fills both beams. 

For all ratios except the 1$_{10}$-1$_{01}$/2$_{12}$-1$_{01}$ we can rule out the optically thin LTE solution. Many of the ratios are close to the optically thick LTE limit, but there are a few notable exceptions (e.g. 3$_{12}$-3$_{03}$/2$_{02}$-1$_{11}$). The average 1$_{10}$-1$_{01}$/2$_{12}$-1$_{01}$ ratio lies close to the optically thin limit, but this has the largest difference in beam size and the emitting regions are unlikely to fill the beam (discussed further in Sec.~\ref{S:analysis_excitation}). Given that the ratios of each of these transitions with the 2$_{02}-$1$_{11}$ line are not in the optically thin LTE limit, we can therefore exclude this solution for all observed lines.

Comparing the two component types, most line ratios are the same. However, those including the 1$_{10}$-1$_{01}$ and 2$_{12}$-1$_{01}$ transitions, which have the largest difference in beam size from the other lines, are slightly different. Given the similarity of the other line ratios, this probably indicates a difference in emitting area shape between the two component types rather than a large difference in excitation conditions.

\subsection{Line ratios as a function of velocity}
\label{S:analysis_vratios}

\begin{figure}[]
\begin{center}
\includegraphics[width=0.33\textwidth]{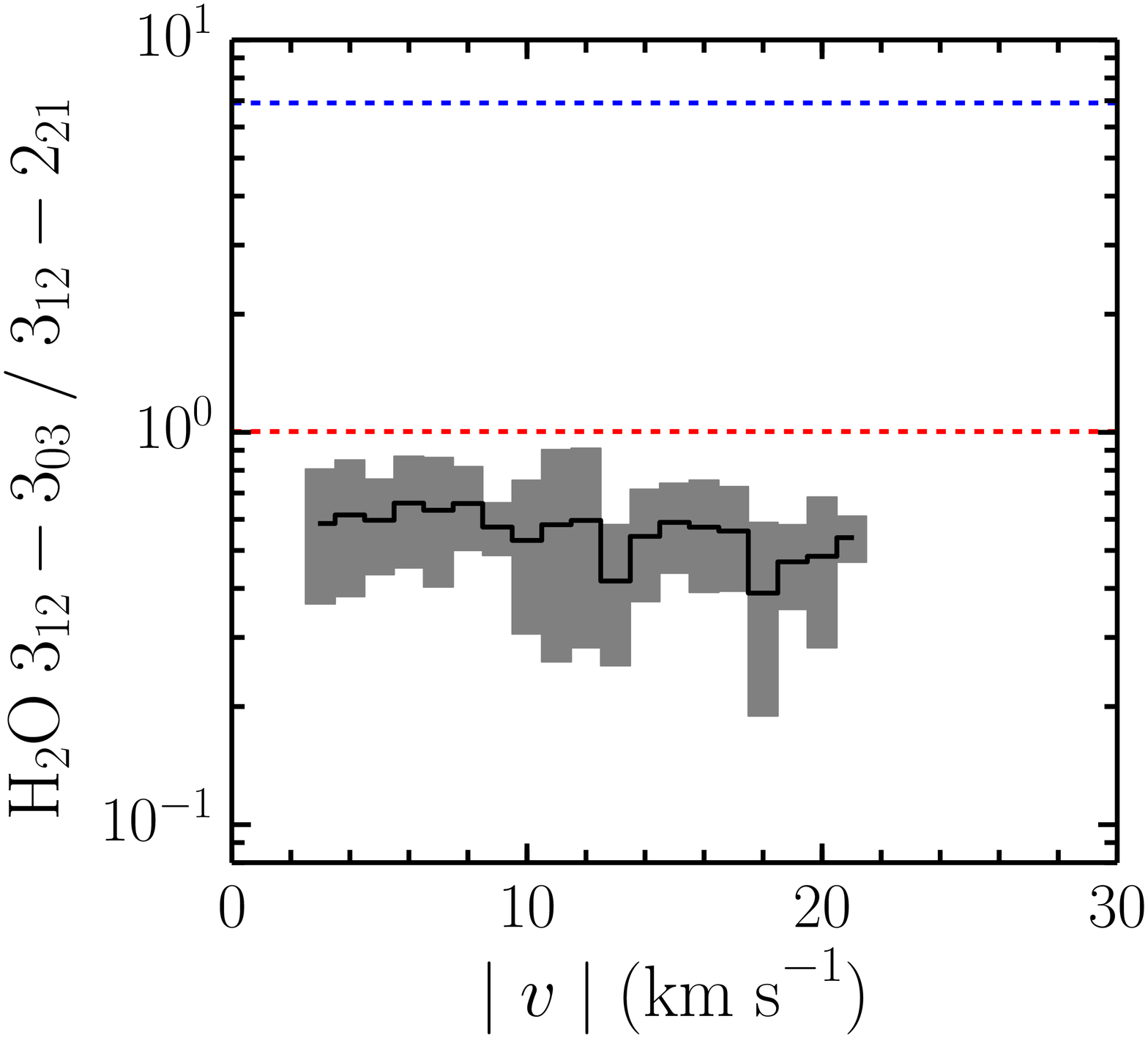}
\includegraphics[width=0.33\textwidth]{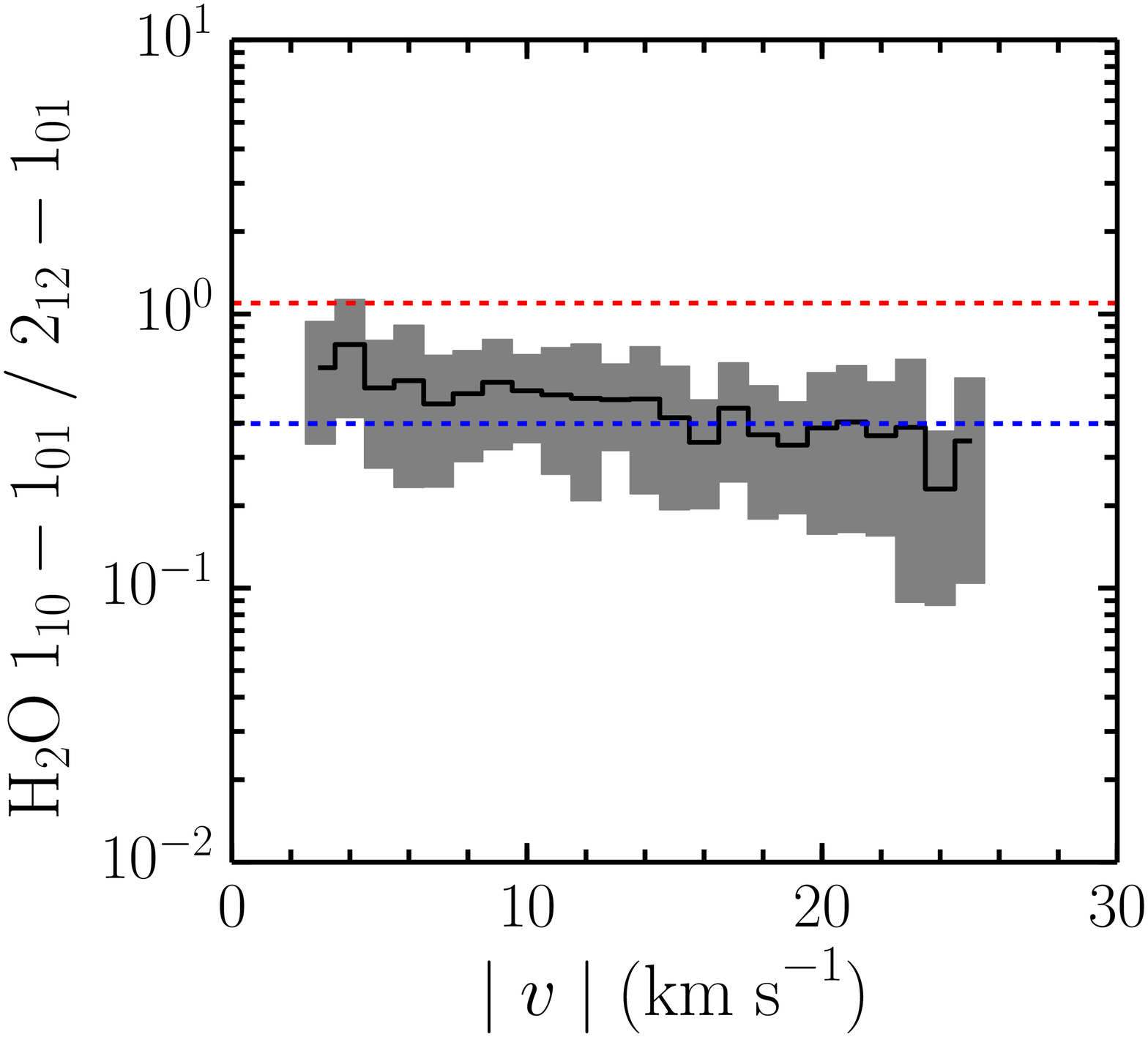}
\includegraphics[width=0.33\textwidth]{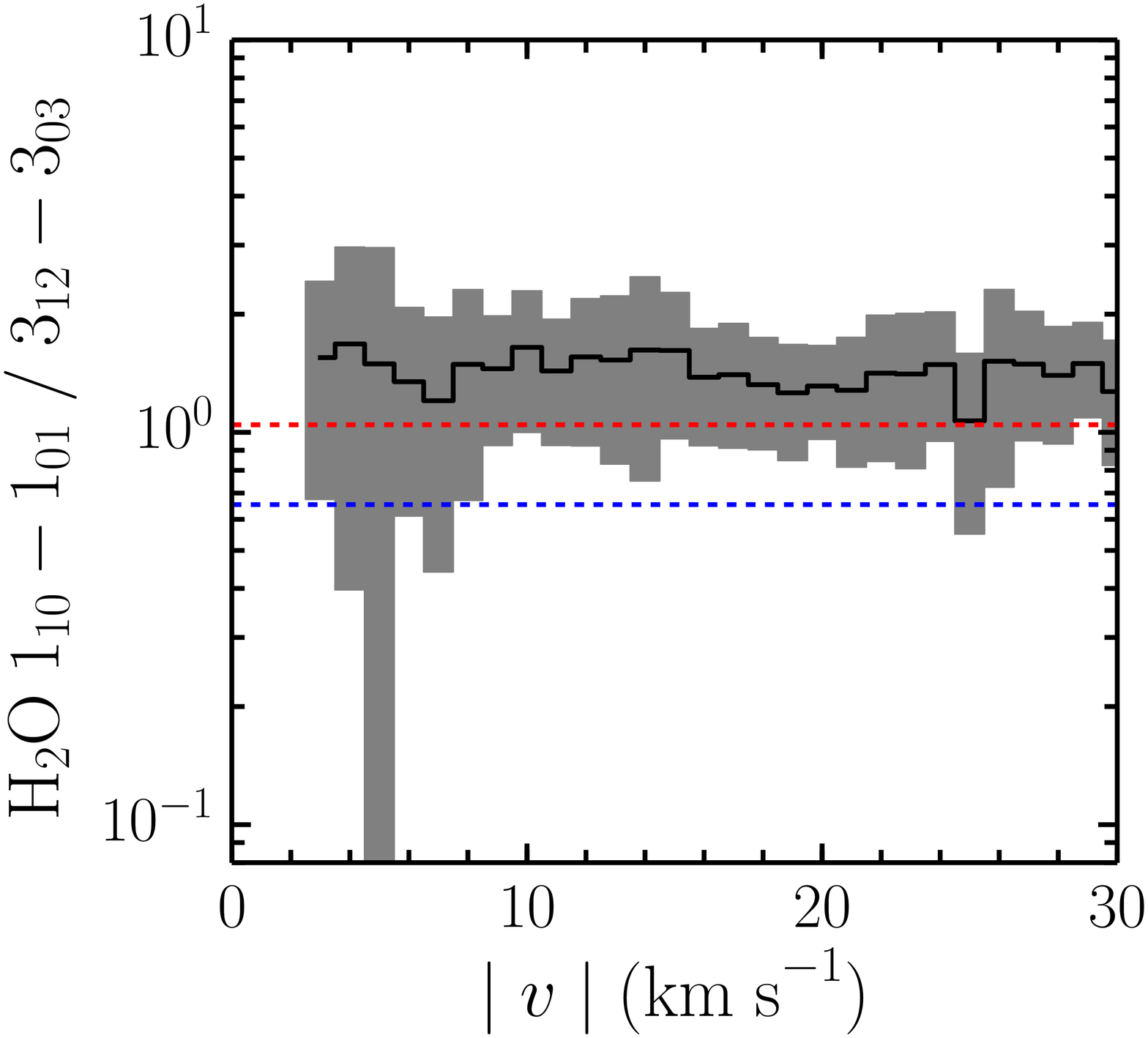}
\includegraphics[width=0.33\textwidth]{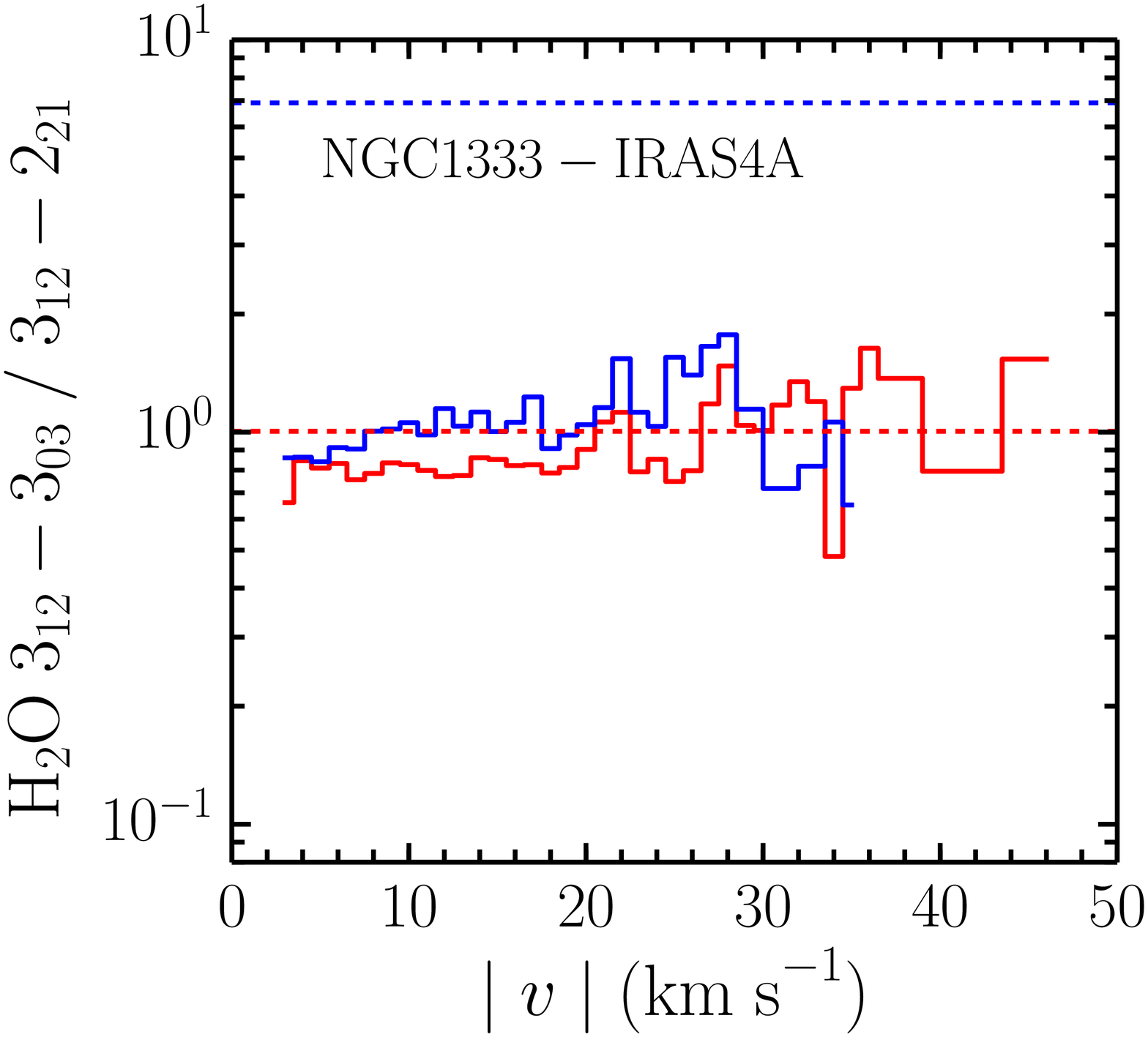}
\caption{Top and middle: Line intensity ratio as a function of velocity averaged over all sources with intensities in both lines above 3$\sigma$ after resampling to 1\kms{} bins (black). The grey region indicates the standard deviation of the sources, which is of similar magnitude to the uncertainty in the ratio for a given source. Bottom: line ratio for the red and blue wings of the Class 0 source NGC1333-IRAS4A. In all panels, the red and blue dashed lines indicate the limits in which both lines are in LTE and optically thick or optically thin respectively for an excitation temperatures of 300\,K.}
\label{F:analysis_vratios_vratio}
\vspace{-6mm}
\end{center}
\end{figure}

The intensity ratios suggest that at least some of the observed transitions are optically thick. Therefore, the next thing to consider is whether this holds for the whole line or just near the peak of the emission and whether we can distinguish between the LTE and sub-thermally excited regimes. This can be explored using the ratio of the observed water lines as a function of velocity. The top panel of Fig.~\ref{F:analysis_vratios_vratio} shows 3$_{12}-$3$_{03}$/3$_{12}-$2$_{21}$ where both lines have intensities above 3$\sigma$ after being resampled to 1\kms{} bins, averaged over all sources and both red and blue line wings. The standard deviation between the sources, shown by the grey region, is similar to the uncertainties in a single source, and the ratios are consistent with being constant as a function of velocity. The bottom panel of Fig.~\ref{F:analysis_vratios_vratio} shows the same line ratio as the top panel, but separately for the red and blue wings of NGC1333-IRAS4A. The line ratio does not change significantly between different line components, a result which is not unique to this source except in a few cases where quiescent envelope emission causes a change in the ratio near the source velocity. 

The middle panels of Fig.~\ref{F:analysis_vratios_vratio} show the 1$_{10}$-1$_{01}$/2$_{12}$-1$_{01}$ and 1$_{10}$-1$_{01}$/3$_{12}-$3$_{03}$ ratios, with the former having the largest difference in beam size and is the only ratio to show significant variation as a function of velocity, of order a factor of two. That the 1$_{10}$-1$_{01}$/3$_{12}-$3$_{03}$ ratio is constant with velocity suggests that this variation may be due to a variation in emitting region shape or position as a function of velocity. Indeed, it may be that some of the emission encompassed by all the other beams is on the edge of or outside the 2$_{12}$-1$_{01}$ beam, which is the smallest of all the observations. This certainly seems to be the case for one of the spot-shock components of NGC1333-IRAS4A which is not detected in this line and whose intensity increases with the beam-size of the transition (see Appendix~\ref{S:appendix2_iras4A}).

The 2$_{12}$-1$_{01}$ line aside, the constant line-ratios as a function of velocity suggest that the excitation conditions present hold for all velocities. This is also consistent with the H$_{2}$O and H$_{2}^{18}$O lines having the same shape (c.f. Fig~\ref{F:results_profiles_comparison_h2_18o}). The ratios do not vary from low to high velocity in contrast to low and high-$J$ CO line ratios \citep{SanJoseGarcia2013,Yildiz2013}, where the line-shape varies with $J$. This is likely caused in part by the low-$J$ CO lines being optically thick in LTE at low velocities with $\tau$ decreasing with increasing offset. Thus, CO emission from inside the $\tau=1$ surface is suppressed, with that surface varying with velocity and $J$. That this does not seem to be the case for H$_{2}$O, due to the invariant line ratio with velocity, suggests that the lines are not in the optically thick LTE solution even at low velocity. Combined with the previous analysis on the integrated intensity ratios, this suggests that the observed transitions are most likely optically thick but effectively thin, i.e. sub-thermally excited.

\subsection{Correlations}
\label{S:analysis_correlations}

\begin{figure*}
\begin{center}
\includegraphics[width=0.99\textwidth]{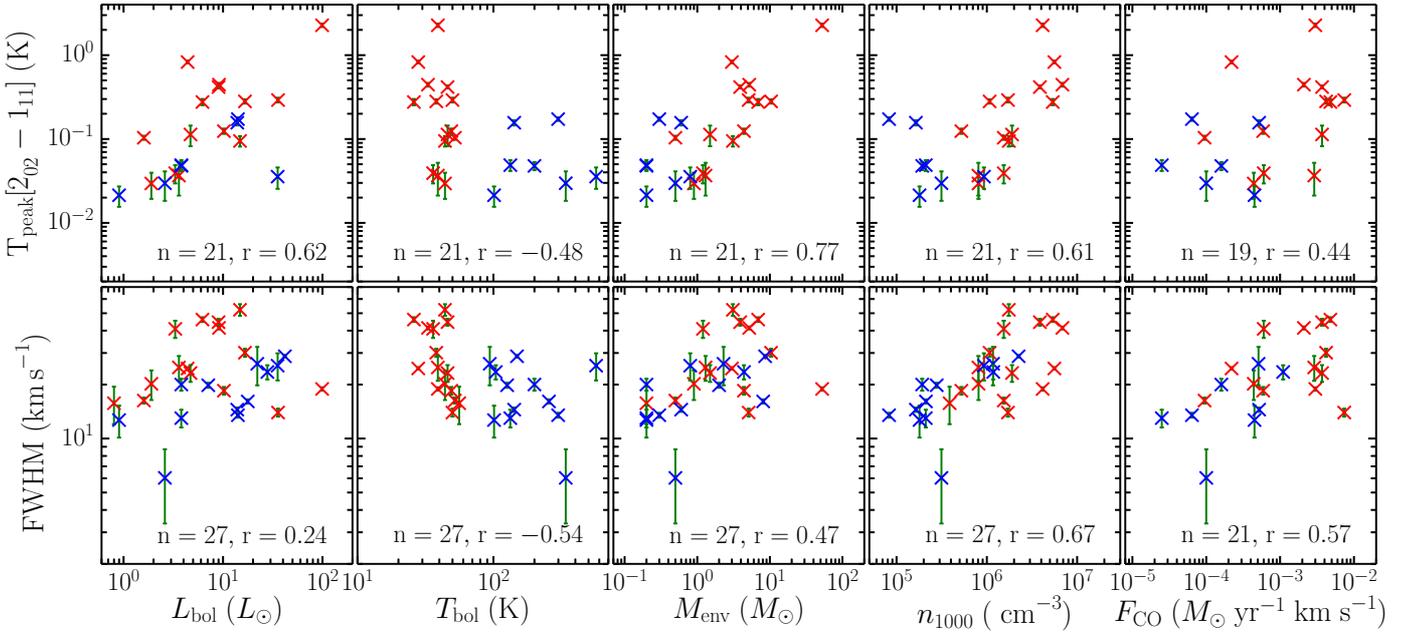}
\caption{Correlation plots of $T_{\mathrm{peak}}$ for the 2$_{02}-$1$_{11}$ line corrected to a common distance of 200\,pc assuming a linear scaling (top) and FWHM (bottom) of the cavity shock component vs., from left to right, $L_{\mathrm{bol}}$, $T_{\mathrm{bol}}$, $M_{\mathrm{env}}$, $n_{\mathrm{1000}}$ and $F_{\mathrm{CO}}$ as presented in Table~\ref{T:observations_sources}. Class 0 and I sources are coloured red and blue respectively. The number of sources and Pearson correlation coefficient are given at the bottom of each panel.}
\label{F:analysis_correlations_corr}
\end{center}
\end{figure*}

Correlation plots comparing source properties (see Table~\ref{T:observations_sources}) for all cavity shock components with $T_{\mathrm{peak}}$ corrected to a common distance of 200\,pc for the 2$_{02}-$1$_{11}$ transition assuming a linear correction (top), and with the FWHM (bottom) are shown in Fig.~\ref{F:analysis_correlations_corr}. There is a correlation of $M_{\mathrm{envl}}$ with the peak brightness temperature of the cavity shock component (3.5$\sigma$ \footnote{The significance of a Pearson correlation coefficient $p$ for sample size $n$ is given in terms of $\sigma$ as $\mid p\mid\sqrt{n-1}$ \citep{Marseille2010}}), but not with FWHM. There is a correlation between the H$_{2}$ density at 1000\,AU ($n_{\mathrm{1000}}$) as obtained from the \textsc{dusty} continuum models of \citet{Kristensen2012} and FWHM, and a weaker trend with $T_{\mathrm{peak}}$ (3.4 and 2.7$\sigma$ respectively). There is also a weak trend between $T_{\mathrm{peak}}$ and $L_{\mathrm{bol}}$ (2.8$\sigma$) and a weak negative trend between $T_{\mathrm{bol}}$ and FWHM (2.7$\sigma$). Finally, there is no correlation or trend between $F_{\mathrm{CO}}$ and $T_{\mathrm{peak}}$, but there is a weak trend with FWHM (2.5$\sigma$). 

The different behaviour of $T_{\mathrm{peak}}$ and FWHM explains why \citet{Kristensen2012} did not see correlations or trends between some of these properties and the integrated intensity of the 1$_{10}-$1$_{01}$ line. The integrated intensity is effectively a multiplication of these two separate quantities which, as shown in Fig.~\ref{F:analysis_correlations_corr}, have different behaviours, particularly with $F_{\mathrm{CO}}$. More sources are needed to confirm some of the weaker trends. The implications of these results will be discussed in Sect~\ref{S:discussion}.

\subsection{Excitation conditions}
\label{S:analysis_excitation}

\subsubsection{Method}
\label{S:analysis_excitation_method}

In order to constrain the excitation conditions (e.g. $n_{\mathrm{H}_{2}}$, $N_{\mathrm{H}_{2}\mathrm{O}}$, $T$) under which water is excited in the cavity and spot shock components, a series of calculations were run using \textsc{radex}. This assumes that the various transitions of a given species have the same line width, as imposed during the Gaussian fitting, and returns the integrated intensity and optical depth for each transition for a given H$_{2}$ volume density, molecule column density and temperature. We assume plane-parallel geometry, that the ortho-to-para ratios of H$_{2}$O and H$_{2}$ are both in the high-temperature limit of 3, a H$_{2}$O/H$_{2}^{18}$O ratio of 540 \citep{Wilson1994}, and use the latest collisional rate coefficients from \citet{Daniel2011} and \citet{Dubernet2009} and molecular spectroscopy from the Cologne Database for Molecular Spectroscopy \citep[CDMS][]{Muller2005} as collected from the Leiden Atomic and Molecular Database \citep[LAMDA\footnote{http://home.strw.leidenuniv.nl/$\sim$moldata/}][]{Schoier2005}. Even if the pre-shock ortho-to-para ratio for H$_{2}$ is as low as 10$^{-3}$, as can be the case in the cold envelope \citep[e.g.][]{Pagani2009}, shocks which are fast enough to sputter water from the grains are also efficient at ortho-to-para conversion of H$_{2}$ \citep{Kristensen2007}. A value of 3 is therefore not unreasonable even for shocks in pristine envelope material.

Grids of \textsc{radex} models were run both for the average line ratios presented in Table~\ref{T:analysis_ratios} and for each component with $n_{\mathrm{H}_{2}}$ varying from 10$^{2}$ to 10$^{10}$\,cm$^{-3}$ and $N_{\mathrm{H}_{2}\mathrm{O}}$ varying from 10$^{12}$ to 10$^{20}$\,cm$^{-2}$ for six representative temperatures (100, 300, 500, 750, 1000 and 1500\,K). For the individual components the line-width in the models was set to the value derived from the Gaussian fitting while a typical line-width of 20\kms{} was used for the average ratios. The density in these calculations is that of the material that has already passed through the shock, i.e. the post-shock gas. This is therefore different from the (pre-shock) density in the envelope as given by $n_{\mathrm{1000}}$ in the Sect.~\ref{S:analysis_correlations}, and for shocks in the jet can be entirely unrelated.

In order to compare the model ratios for each grid to the observed ratios, the observations must be corrected for differences in beam-size. All observed intensities and upper limits are corrected to the $2_{02}-1_{11}$ (988\,GHz) beam and ratioed to that line. For the cavity shock component this is done assuming that the emission comes from a 1-D structure (i.e. $I\propto\theta_{\mathrm{beam}}$) due to the extended nature of the outflows as discussed in Section~\ref{S:results_components_cavityshock}. The spot shock components are assumed to be point-like (i.e. $I\propto\theta^{2}_{\mathrm{beam}}$) as discussed in Section~\ref{S:results_components_spotshock}. 

The best-fit and significance of the models are found using $\chi^{2}$ minimisation of the observed and model line ratios with respect to the $2_{02}-1_{11}$ line. For most lines, the beam correction is relatively small, particularly for the cavity shock (see Table~\ref{T:analysis_ratios}). However, as already discussed in Sect.~\ref{S:analysis_ratios} and \ref{S:analysis_vratios}, there may be some emission which is not included in the 2$_{12}-$1$_{01}$ line but is inside the beam for the other lines, or equally is included in the 1$_{10}-$1$_{01}$ line but none of the others. In theory this should be accounted for by the beam correction, but since we do not know the true spatial distribution of the emission then we do not know how good our assumption of point-like and linear emission is for the spot and cavity shocks respectively. 

We therefore include an uncertainty in the beam correction factor in our calculation of the $\chi^{2}$ when this correction is more than a factor of 1.5, added in quadrature with the uncertainties on the intensities. The exponent of the beam correction factor can only be between 0 (for uniform emission) and 2 (for a point source), so for the shock components where we assume a point-source emitting region this uncertainty is only applied in the direction of a smaller correction exponent. The effect of this additional uncertainty is to give less weight to those lines which have large beam correction factors with respect to the 2$_{02}-$1$_{11}$ line.

The area of the emitting region in the plane of the sky is then calculated from the ratio of the model and observed $2_{02}-1_{11}$ integrated intensity, i.e. the fraction of the beam that can be at the model intensity in order to match the observed value. This is converted to a radius assuming a circular emitting region at the distance of the object for ease of comparison.

In certain parts of the parameter space searched with \textsc{radex}, certain water transitions show strong maser activity \citep[see e.g.][]{Kaufman1996}. These are mostly models with low post-shock density and high water column density. As described in \citet{vanderTak2007}, \textsc{radex} is not well suited to modelling maser activity. While none of the fitted lines are affected, masing in other lines can hamper convergence of the calculation. We have therefore limited the opacity to a certain negative value (-10) as also implemented in \textsc{ratran} \citep{Hogerheijde2000}. Changing this value does not affect the results of our fitting.

In addition, the standard version of \textsc{radex} calculates the line excitation and $\tau$ using a Gaussian profile, but the integrated intensity assuming a box-line-profile and thus $I \propto (1-\exp(-\tau))$ which is almost independent of $\tau$ for $\tau >$ a few. However, for a Gaussian line profile, $I \propto \int (1-\exp^{-\tau\exp(-k(\varv/\sigma)^{2})})d\varv$ \citep[see e.g ][]{Avrett1965}. This leads to \textsc{radex} underestimating the line intensity for high opacity, which is relevant for water. We therefore correct the line fluxes in the vein of a curve-of-growth analysis by multiplying those output by \textsc{radex} by a factor $\alpha$ given by:

\begin{equation}
\alpha = \frac{\int_{-\infty}^{\infty} 1 - \exp\left(- \tau e^{-x^2}\right) dx}{\sqrt{\pi}\left( 1 - e^{-\tau} \right)},
\label{E:correction}
\end{equation}

\noindent where $x=\sqrt{k}\varv/\sigma$. For $\tau \lesssim 0.2$, $\alpha \simeq 1$ because no correction is required. The largest correction at high $\tau$ (e.g. 10$^{4}$) is approximately a factor of three.

In some extra-galactic sources, pumping of the higher-excited water lines by the far-IR dust continuum is required to reproduce the line ratios \citep[e.g.][]{Gonzalez-Alfonso2010}. We therefore ran grids of models including the far-IR continuum from the SEDs reported by \citet{Kristensen2012} scaled by a range of scaling factors to test if this is important in low-mass protostars. The radiation field must be smaller than 2$\times$10$^{-5}$ times the bolometric luminosity before any reasonable fits to the observations could be found. Even for these low radiation levels, the best fits were not significantly different or better than those without the continuum radiation field included. In particular, there are no moderate-density ($n_{\mathrm{H}_{2}}$ of order 10$^{5-6}$), high radiation field solutions which fit the data, as is the case for external galaxies. In addition, our observed line ratios, particularly for 1$_{11}-$0$_{00}$/2$_{02}-$1$_{11}$ and 2$_{11}-$2$_{02}$/2$_{02}-$1$_{11}$, differ from the extragalactic case, where the 2$_{02}-$1$_{11}$ transition is significantly enhanced with respect to all other lines. We conclude that the far-IR radiation field does not play an important role in the excitation of water in low-mass sources, and is therefore not considered further.

\subsubsection{Results}
\label{S:analysis_excitation_results}

\begin{figure}
\begin{center}
\includegraphics[width=0.38\textwidth]{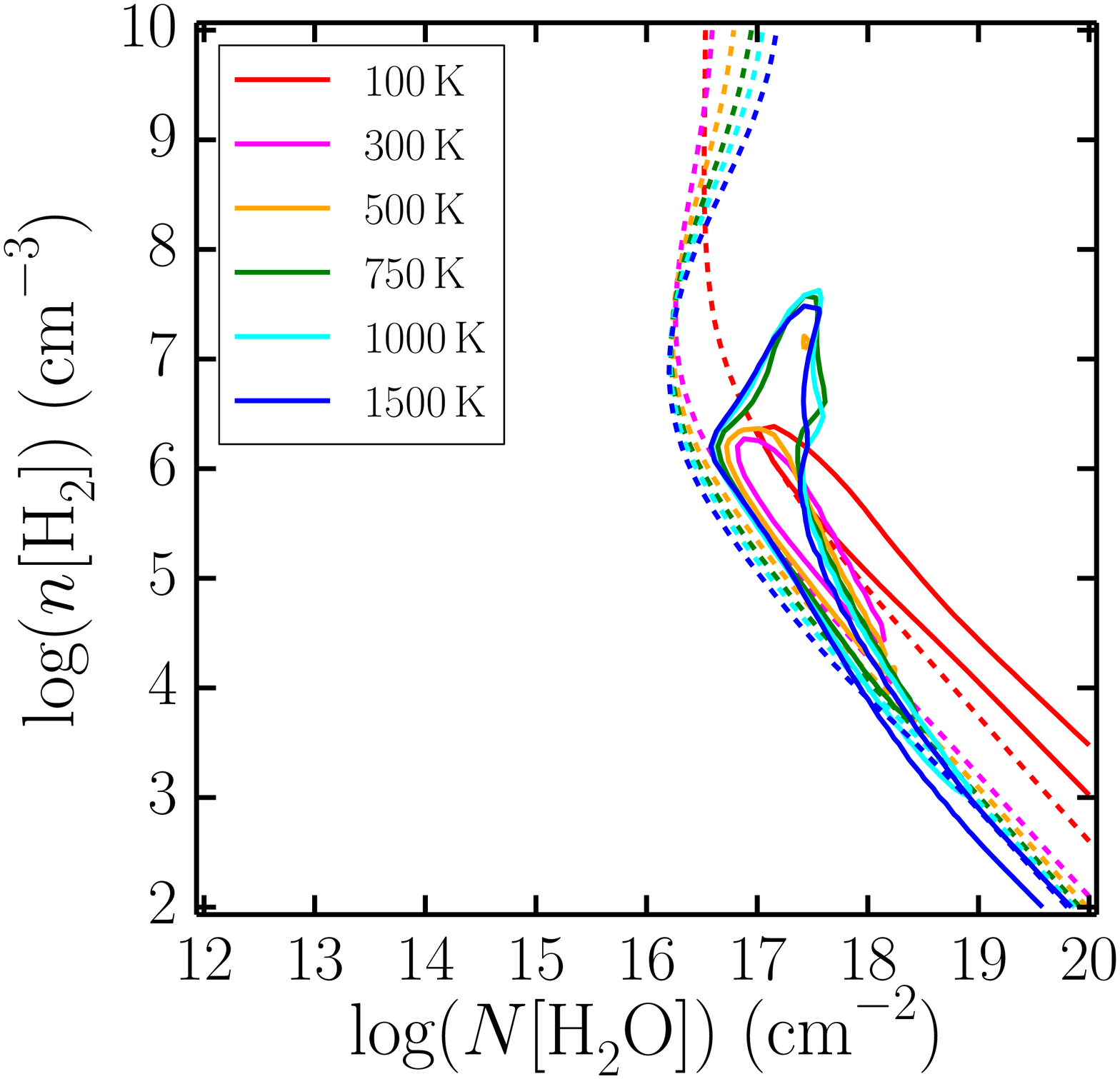}
\includegraphics[width=0.38\textwidth]{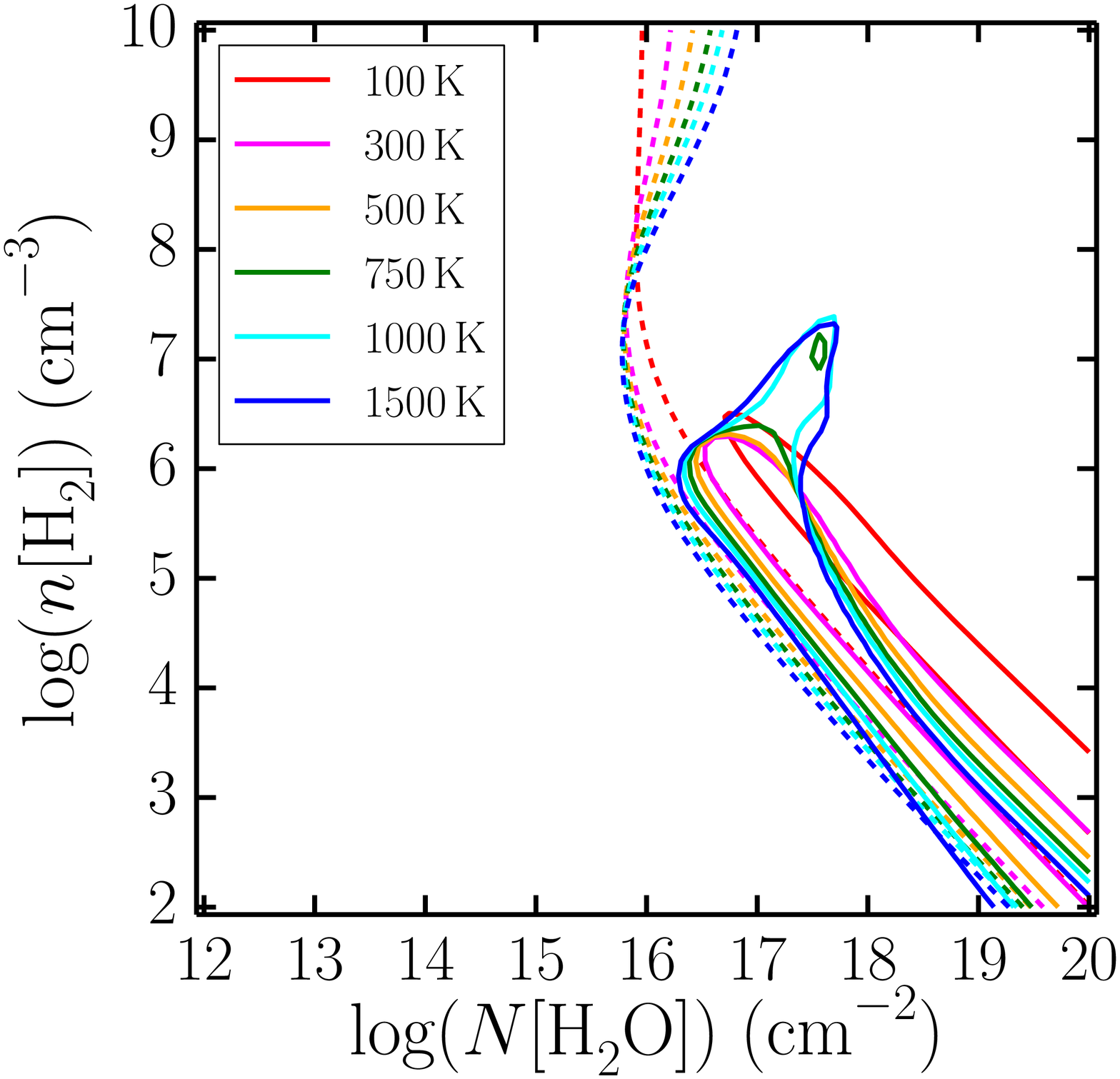}
\caption{Results of comparison between the average line ratios presented in Table~\ref{T:analysis_ratios} with \textsc{radex} models for a range of $n_{\mathrm{H}_{2}}$ and $N_{\mathrm{H}_{2}\mathrm{O}}$ for six different temperatures ranging from 100 to 1500\,K. The solid lines indicate the region within 1$\sigma$ while the dashed lines show where the plane-of-the-sky emitting area, obtained from the ratio of the observed and model intensity in the 2$_{02}-$1$_{11}$ line, corresponds to a circle with radius 100\,AU.}
\label{F:analysis_excitation_results_temperature}
\end{center}
\end{figure}

Fig.~\ref{F:analysis_excitation_results_temperature} shows, in solid lines, a comparison of the 1$\sigma$ $\chi^{2}$ contours for fits between grids of \textsc{radex} models over $N_{\mathrm{H}_{2}\mathrm{O}}$ and $n_{\mathrm{H}_{2}}$ at a range of temperatures from 100$-$1500\,K to the average line ratios for cavity and spot shocks given in Table~\ref{T:analysis_ratios}. A typical FWHM of 20\kms{}, distance of 200\,pc and intensities of the 988\,GHz line of 5 and 2.4\,K\kms{} were used based on the average of the sample. For the H$_{2}^{18}$O lines, since most sources show non-detections in all lines, the $\sigma_{\mathrm{rms}}$ values for Class 0 sources from Table~\ref{T:results_profiles_basic_fwzi} were converted to upper-limit intensities. The dashed lines show where the ratio of the model and observed fluxes corresponds to a radius of 100\,AU in the plane of the sky for a circular emitting region, with smaller regions to the upper right of this line and larger to the lower left. 

Aside from the 100\,K model, there is a slight trend towards higher $n_{\mathrm{H}_{2}}$ and lower $N_{\mathrm{H}_{2}\mathrm{O}}$ with increasing temperature but the contours and emitting region sizes derived from the different models are effectively the same within the uncertainties. This insensitivity of low-$J$ water emission lines to temperature within the post-shock density and column density regime present towards our sources has also been seen in other studies \citep[see also][]{Kristensen2013,Santangelo2014}. The temperature and density cannot both be the same as traced by low-$J$ CO, which traces temperatures of $\sim$100\,K but is insensitive to density \citep{Yildiz2013}, because otherwise water would trace similar gas (and thus have similar line profiles), which is not the case \citep[e.g][]{Nisini2010,Kristensen2012}. Higher-$J$ line profiles, such as $^{12}$CO 16$-$15 are more similar to water \citep[][Kristensen et al., in prep.]{Kristensen2013}, and trace warmer gas \citep[e.g.][]{Karska2013}, supporting the idea that water is tracing material which is warmer than 100\,K.

We therefore follow \citet{Kristensen2013} in assuming a temperature of 750\,K for the spot shock components. This is also similar to the hot component observed in CO rotation diagrams with PACS \citep[e.g.][]{Karska2013}. These same observations also show a warm component in CO with a temperature of $\sim$300\,K, which the observations of \citet{Santangelo2013} show is spatially associated with H$_{2}$O in the outflow cavity. We therefore assume a temperature of 300\,K for the cavity shock components. 

\begin{figure*}[]
\begin{center}
\includegraphics[width=0.95\textwidth]{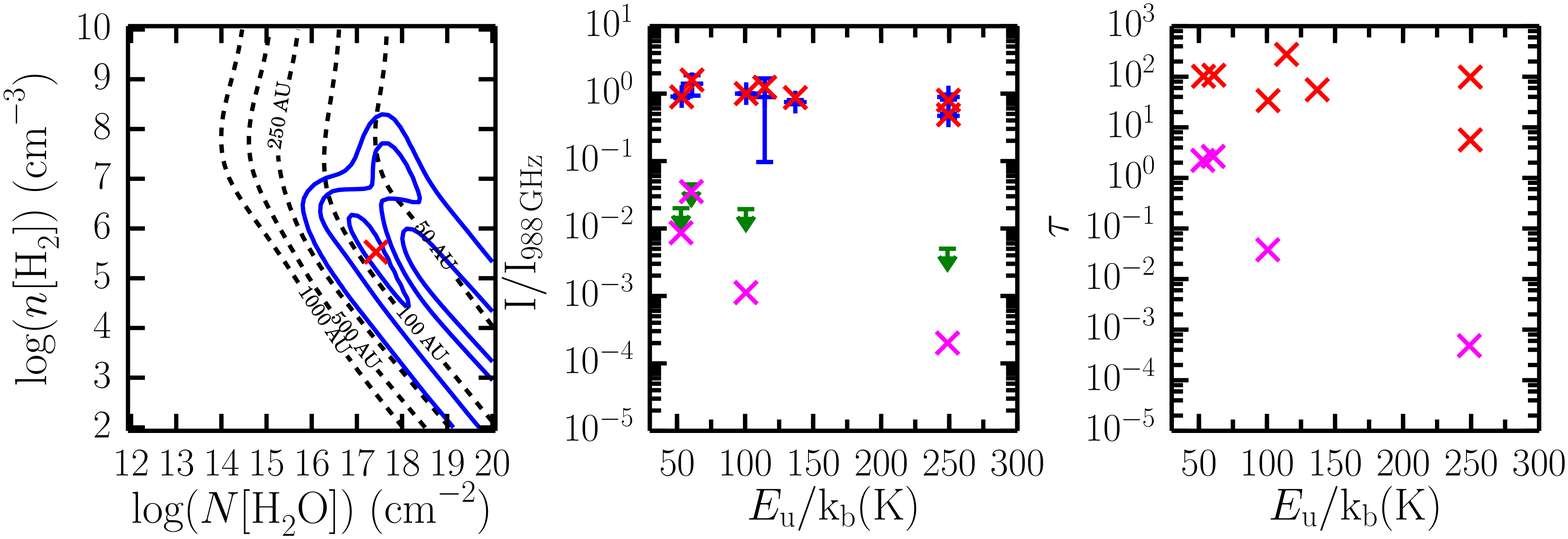}
\includegraphics[width=0.95\textwidth]{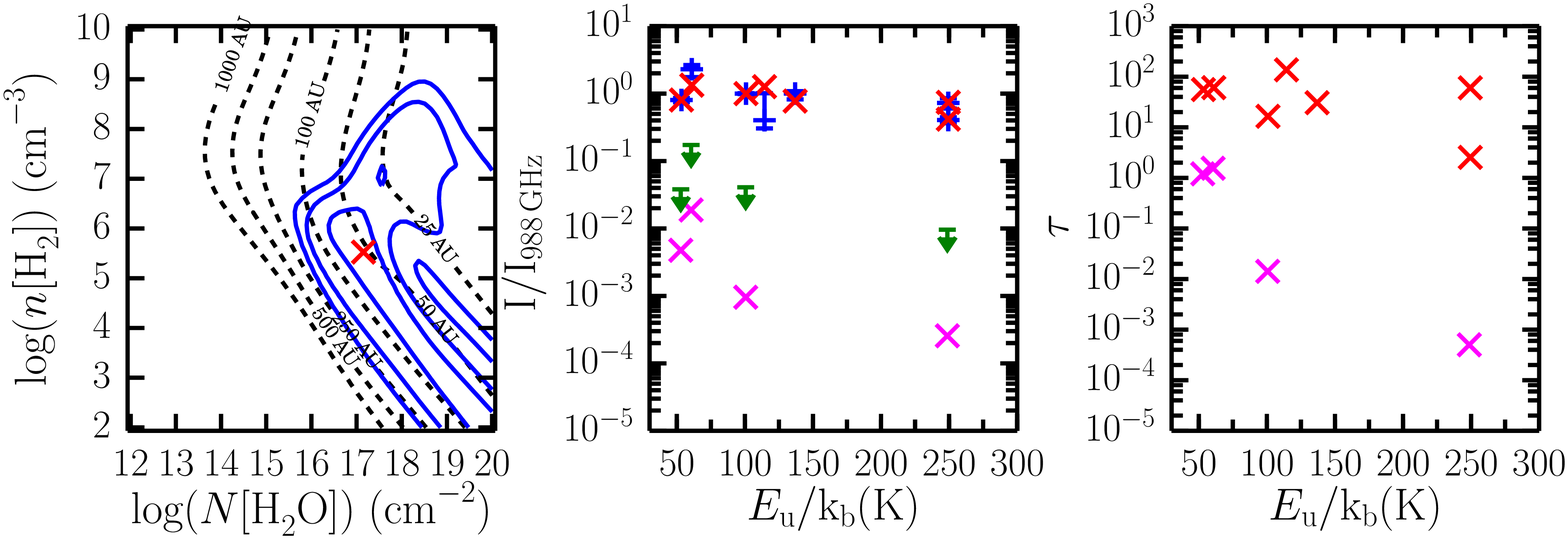}
\caption{\textsc{radex} results for the average line ratios for the cavity shock (top) and spot shock (bottom) components assuming $T$=300 and 750\,K respectively. The left-hand panels show the best-fit (red cross) and 1, 3 and 5$\sigma$ confidence limits (blue contours) for a grid in $n_{\mathrm{H}_{2}}$ and $N_{\mathrm{H}_{2}\mathrm{O}}$. The black dashed contours show the corresponding radius of the emitting region. The middle panels show a spectral line energy distribution comparing the observed (blue for H$_{2}$O, green for H$_{2}^{18}$O) and best-fit model (red for H$_{2}$O, magenta for H$_{2}^{18}$O) results. The right-hand panels show the optical depth for each line for the best-fit model.}
\label{F:analysis_excitation_results_example}
\end{center}
\end{figure*}

The fit to the average line ratios for the cavity and spot shock components are shown for 300\,K and 750\,K respectively in Fig.~\ref{F:analysis_excitation_results_example}. The left-hand panel of each row shows 1, 3 and 5$\sigma$ contours in blue with the best-fit (i.e. model with the lowest $\chi^{2}$) marked with a red cross. These typically centre around two diagonal solutions; one with $n_{\mathrm{H}_{2}}\lesssim$10$^{6}$\,cm$^{-3}$ and $N_{\mathrm{H}_{2}\mathrm{O}}\gtrsim$10$^{16}$\,cm$^{-2}$ and a parallel solution with $n_{\mathrm{H}_{2}}\lesssim$10$^{7}$\,cm$^{-3}$ and $N_{\mathrm{H}_{2}\mathrm{O}}\gtrsim$10$^{17}$\,cm$^{-2}$. In the first solution, the density is well below the critical density for all lines and the excitation is sub-thermal. For the second solution at least some of the lines are nearing thermal equilibrium. In both cases all H$_{2}$O lines, and even the lowest two non-detected H$_{2}^{18}$O transitions, are optically thick (right-hand panels). Despite this, both the higher-density thermal solution and low lower-density non-thermal solution are likely still in the effectively thin regime. 

The emitting region area for the average line ratios in the plane of the sky at the typical distance for our sources (200\,pc) is relatively small, equivalent to a circular radius of order 50-100\,AU. The solutions for the cavity and spot-shocks are very similar in terms of $n_{\mathrm{H}_{2}}$ and $N_{\mathrm{H}_{2}\mathrm{O}}$, with the spot shocks having a slightly smaller emitting region due to the smaller absolute fluxes.

Though the tail of possible solutions within 1$\sigma$ extends to lower densities and higher column densities, it is unlikely that the emission comes from a long cylinder in all sources, given the range in viewing angles within the sample. For example, assuming an abundance for water of 10$^{-4}$ with respect to H$_{2}$, a water column density of 10$^{19}$\,cm$^{-2}$ at a molecular hydrogen density of 10$^{3}$\,cm$^{-3}$ corresponds to a length along the column of 7$\times$10$^{6}$\,AU which is physically unlikely. Therefore, though there are formally solutions extending to the lower right in Fig.~\ref{F:analysis_excitation_results_temperature} and the left-hand panels of Fig.~\ref{F:analysis_excitation_results_example}, the models at the upper left part of the solution are more likely from a geometrical point of view. The best-fit therefore provides a characteristic determination, with uncertainties in $n_{\mathrm{H}_{2}}$ and $N_{\mathrm{H}_{2}\mathrm{O}}$ typically half to one order of magnitude if the other property is held constant. In comparison, a factor of two change in the temperature results in less than a factor of three change in $n_{\mathrm{H}_{2}}$ and $N_{\mathrm{H}_{2}\mathrm{O}}$, as does changing the assumed beam-correction from linear to point-like or vice-versa.

The same analysis was also performed separately for all individual source components, with the average ratios used in cases where specific lines were not observed. We also restrict the best-fit solution to have a length along the column of no more than 5000\,AU, assuming a water abundance of 10$^{-4}$, as lengths larger than this would be larger than the beam if rotated to the plane of the sky. Table~\ref{T:analysis_excitation_results} presents the best-fit results for the average line ratios and those sources where the 1$\sigma$ contour covers less than 10$\%$ of the probed parameter space. Figures of the same form as Figure~\ref{F:analysis_excitation_results_temperature} for these individual components are shown in Figures~\ref{F:analysis_excitation_results_full1} to \ref{F:analysis_excitation_results_full9} in Appendix~\ref{S:appendix1}. For the remaining sources the excitation conditions cannot be well-constrained, usually due to upper limits in multiple H$_{2}$O transitions. The $\chi^{2}$ contours are usually elongated as also seen in Fig.~\ref{F:analysis_excitation_results_example}. 

\begin{figure}
\begin{center}
\includegraphics[width=0.40\textwidth]{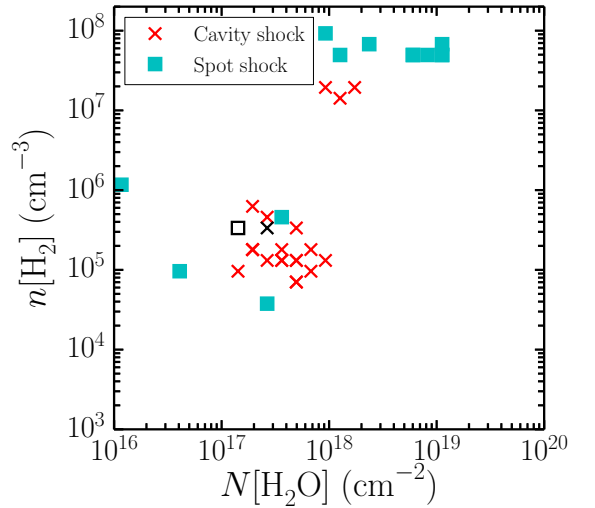}
\caption{Best-fit $N_{\mathrm{H}_{2}\mathrm{O}}$ vs. $n_{\mathrm{H}_{2}}$ from the \textsc{radex} model grids for sources with well-constrained best-fits (i.e. 1$\sigma$ contours include $<$10$\%$ of all models). The black square and cross indicate the best fit for the average spot and cavity shock line ratios.}
\label{F:analysis_excitation_results_spread}
\end{center}
\end{figure}

 The overall spread of best-fit H$_{2}$ number density vs. H$_{2}$O column density is shown in Fig.~\ref{F:analysis_excitation_results_spread}. The column densities given are over the emitting region in the plane of the sky, which have radii of order 5$-$300\,AU assuming a circular shape (see Table~\ref{T:analysis_excitation_results}). In all cases the observations exclude water column densities below $\sim$10$^{15}$\,cm$^{-2}$. For those shock components already studied by \citet{Kristensen2013} we derive slightly lower densities, lower emitting region sizes and higher column densities. This partly stems from our correction of the \textsc{radex} intensities using Eqn.~\ref{E:correction}, but is also because \citet{Kristensen2013} only considered water column densities of 4$\times$10$^{15}-$10$^{17}$\,cm$^{-2}$ and H$_{2}$ densities of 10$^{6}-$10$^{9}$\,cm$^{-3}$ and therefore only considered the thermal solution.

For most cavity shock components, the best fit favours the sub-thermal solution but there is usually a thermal solution within the 3 or 5$\sigma$ contours as well. The analysis for the majority of spot shock components favours the thermal solution, resulting also in smaller emitting regions sizes on the sky, but in all components where this is the case there are also solutions within the 1$\sigma$ contours in the sub-thermal solution. Therefore, while there is some spread in best-fit results, we cannot conclude that there is a significant difference between cavity and spot shock results when considering the 1$\sigma$ results, as also seen for the average line ratios. The Class I sources tend to have smaller emitting region sizes compared to the Class 0 sources, but there is no significant difference in $n_{\mathrm{H}_{2}}$ and $N_{\mathrm{H}_{2}\mathrm{O}}$.

\begin{table*}[ph!]
\caption[]{\textsc{radex} results for sources where the 1$\sigma$ contours include $<$10$\%$ of all models.}
\centering
\begin{tabular}{lccccccccc}
\hline
\hline \noalign {\smallskip}
\centering
Source & Comp.\tablefootmark{a} & FWHM & $\chi^{2}_{\mathrm{best}}$ & $T$ & $N_{\mathrm{best}}$\tablefootmark{b} & $n_{\mathrm{best}}$ & $r$ & $\tau_{\mathrm{988\,GHz}}$ \\\noalign {\smallskip}
&&(\kms{})&&(K)&cm$^{-2}$&cm$^{-3}$&(AU)&\\\noalign {\smallskip}
\hline \noalign {\smallskip}
Average&C&20.0&\phantom{0}0.8&300&3$\times$10$^{17}$&3$\times$10$^{ 5}$&\phantom{0}78.7&33.7\\
&S&20.0&\phantom{0}0.8&750&1$\times$10$^{17}$&3$\times$10$^{ 5}$&\phantom{0}54.2&16.5\\
\hline \noalign {\smallskip}
L1448-MM&S&23.0&\phantom{0}7.0&750&3$\times$10$^{17}$&4$\times$10$^{ 4}$&126.0&54.0\\
&S&39.8&\phantom{0}2.8&750&1$\times$10$^{19}$&7$\times$10$^{ 7}$&\phantom{0}15.6&\phantom{0}5.5\\
&C&44.6&\phantom{0}1.8&300&7$\times$10$^{17}$&1$\times$10$^{ 5}$&136.9&58.0\\
IRAS2A&C&14.0&\phantom{0}1.3&300&2$\times$10$^{17}$&2$\times$10$^{ 5}$&103.4&43.8\\
&S&39.2&\phantom{0}1.0&750&4$\times$10$^{17}$&5$\times$10$^{ 5}$&\phantom{0}58.3&16.4\\
IRAS4A&S&\phantom{0}9.9&\phantom{0}1.0&750&4$\times$10$^{16}$&1$\times$10$^{ 5}$&175.7&20.0\\
&C&41.4&\phantom{0}0.1&300&5$\times$10$^{17}$&3$\times$10$^{ 5}$&118.6&31.5\\
IRAS4B&S&\phantom{0}5.1&\phantom{0}1.0&750&1$\times$10$^{16}$&1$\times$10$^{ 6}$&166.2&\phantom{0}5.0\\
&C&24.6&\phantom{0}7.0&300&3$\times$10$^{17}$&5$\times$10$^{ 5}$&157.6&26.2\\
L1527&C&20.2&\phantom{0}2.3&300&4$\times$10$^{17}$&1$\times$10$^{ 5}$&\phantom{0}25.4&58.1\\
BHR71&S&28.3&\phantom{0}2.0&750&6$\times$10$^{18}$&5$\times$10$^{ 7}$&\phantom{00}8.5&\phantom{0}4.3\\
&C&52.3&\phantom{0}2.5&300&9$\times$10$^{17}$&1$\times$10$^{ 5}$&\phantom{0}54.6&57.5\\
&S&59.0&\phantom{0}1.4&750&1$\times$10$^{19}$&5$\times$10$^{ 7}$&\phantom{00}8.4&\phantom{0}3.8\\
IRAS15398&C&16.3&\phantom{0}4.3&300&3$\times$10$^{17}$&1$\times$10$^{ 5}$&\phantom{0}46.9&54.4\\
L483&C&18.5&\phantom{0}2.1&300&2$\times$10$^{18}$&2$\times$10$^{ 7}$&\phantom{0}28.4&25.8\\
Ser-SMM1&S&\phantom{0}3.7&\phantom{0}1.2&750&9$\times$10$^{17}$&9$\times$10$^{ 7}$&\phantom{0}60.5&\phantom{0}4.6\\
&C&18.9&\phantom{0}4.9&300&2$\times$10$^{17}$&6$\times$10$^{ 5}$&334.8&22.5\\
Ser-SMM3&C&30.1&\phantom{0}0.8&300&5$\times$10$^{17}$&7$\times$10$^{ 4}$&156.8&68.0\\
Ser-SMM4&S&10.7&\phantom{0}1.7&750&2$\times$10$^{18}$&7$\times$10$^{ 7}$&\phantom{0}32.9&\phantom{0}4.1\\
&C&46.1&\phantom{0}2.0&300&5$\times$10$^{17}$&7$\times$10$^{ 4}$&176.3&50.9\\
L723&C&24.9&\phantom{0}2.3&300&5$\times$10$^{17}$&1$\times$10$^{ 5}$&\phantom{0}40.5&62.3\\
B335&S&\phantom{0}6.5&\phantom{0}1.1&750&1$\times$10$^{18}$&5$\times$10$^{ 7}$&\phantom{0}12.0&\phantom{0}3.9\\
&C&40.9&\phantom{0}2.4&300&7$\times$10$^{17}$&2$\times$10$^{ 5}$&\phantom{0}37.6&49.2\\
L1157&C&23.2&\phantom{0}2.3&300&4$\times$10$^{17}$&2$\times$10$^{ 5}$&\phantom{0}73.9&47.3\\
&S&35.7&\phantom{0}1.5&750&8$\times$10$^{18}$&5$\times$10$^{ 7}$&\phantom{0}11.7&\phantom{0}4.7\\
&S&47.7&\phantom{0}1.2&750&1$\times$10$^{19}$&5$\times$10$^{ 7}$&\phantom{0}10.8&\phantom{0}4.8\\
\hline \noalign {\smallskip}
L1489&C&20.0&\phantom{0}2.4&300&4$\times$10$^{17}$&1$\times$10$^{ 5}$&32.3&58.5\\
TMR1&C&13.0&\phantom{0}3.4&300&2$\times$10$^{17}$&2$\times$10$^{ 5}$&32.1&45.9\\
TMC1&C&12.7&\phantom{0}2.4&300&2$\times$10$^{17}$&2$\times$10$^{ 5}$&21.2&46.7\\
IRAS12496&C&25.5&\phantom{0}3.1&300&5$\times$10$^{17}$&1$\times$10$^{ 5}$&30.9&61.3\\
&S&25.7&\phantom{0}1.4&750&6$\times$10$^{18}$&5$\times$10$^{ 7}$&\phantom{0}5.0&\phantom{0}4.7\\
GSS30-IRS5&C&14.5&\phantom{0}2.0&300&1$\times$10$^{18}$&1$\times$10$^{ 7}$&25.6&26.3\\
Elias29&C&13.5&\phantom{0}2.6&300&9$\times$10$^{17}$&2$\times$10$^{ 7}$&27.2&20.0\\
RNO91&C&\phantom{0}6.0&\phantom{0}2.3&300&1$\times$10$^{17}$&1$\times$10$^{ 5}$&24.1&78.2\\
\hline \noalign {\smallskip}
\end{tabular}
\label{T:analysis_excitation_results}
\tablefoot{\tablefoottext{a}{Component Type: C=cavity shock and S=spot shock.}\tablefoottext{b}{Column density over the emitting region.}}
\end{table*}

\section{Discussion}
\label{S:discussion}

The following subsections discuss separately the variation in properties between the different shock components (Sect.~\ref{S:discussion_components}) and as a function of source evolutionary stage (Sect.~\ref{S:discussion_evolution}). We then attempt to synthesise this into a consistent picture in Sect.~\ref{S:discussion_picture}, before also comparing our results at the source position with others further from the central source (Sect.~\ref{S:discussion_on_vs_off}).

\subsection{Cavity and spot shocks}
\label{S:discussion_components}

The two different components seen in water spectra related to the outflow-jet system (cavity and spot shock) exhibit differences in terms of their line width and offset from the source velocity (see Fig.~\ref{F:results_components_comparison_scatter}), in part due to how the components are classified. Some spot shock components have similar widths to the cavity shock component but are significantly offset from the source velocity while other spot shock components have similar offsets but smaller FWHM than the cavity shock components. However, the cavity and spot shock components show little difference in terms of their integrated intensity ratios or their line intensity ratios as a function of velocity (see Figures~\ref{F:analysis_ratios_lratio} and \ref{F:analysis_vratios_vratio}). The relatively small spread in line ratios leads to the similarity of the physical conditions under which each component is generated, though the variation in absolute fluxes and line-widths leads to variation of $\sim$1 order of magnitude in $N_{\mathrm{H}_{2}\mathrm{O}}$ and $n_{\mathrm{H}_{2}}$ if the sub-thermal and thermal solutions are considered separately (see Fig.~\ref{F:analysis_excitation_results_spread}). Thus, while the velocities that the gas is subject to may be different between these two components, the excitation is not.

The comparison with H$_{2}^{18}$O, the line intensity ratios, the line ratios as a function of velocity and the \textsc{radex} analysis (see Fig.~\ref{F:results_profiles_comparison_h2_18o}, \ref{F:analysis_ratios_lratio}, \ref{F:analysis_vratios_vratio} and Table~\ref{T:analysis_excitation_results} respectively) all agree that all H$_{2}$O transitions are likely optically thick across the whole line. However, the \textsc{radex} determinations for the density lie below the critical density of some or all of the observed transitions, so the lines are optically thick but effectively thin. The same determinations also suggest that the ground-state H$_{2}^{18}$O may be marginally optically thick ($\tau\sim$1), explaining why the determinations of optical depth from H$_{2}$O/H$_{2}^{18}$O ratios are lower than those suggested by \textsc{radex}.

The lack of variation in line ratio as a function of velocity is in marked contrast to what is found for CO, for which line ratios between lower and higher energy transitions, for example between $J$=3$-$2 and 10$-$9, often vary by a factor of 2 or more, at least for Class 0 sources, between offsets of 5 and 20\kms{} from the source velocity \citep[e.g. see Fig.~13 of][]{Yildiz2013}. This is likely because the low and higher $J$ CO transitions trace different parts of the outflow as a function of both excitation and velocity, while the water emission studied here is coming from gas under similar conditions for all velocities and transitions.

The spot shock components contribute 20$-$30$\%$ of the total integrated emission in H$_{2}$O lines, while the cavity shock component provides $\sim$70$-$80$\%$ (see Table~\ref{T:results_components_comparison_fract}). This is broadly consistent with the ratio of the number of molecules in the warm ($\sim$300\,K) and hot ($\sim$750\,K) CO components identified by \citet[][see their Table~3]{Karska2013} in high-$J$ PACS observations of the same sample. We therefore suggest that the warm and hot components in the PACS CO observations are related to the cavity and spot shock components respectively in our observations, a conclusion also reached for the spot shock components by \citet{Kristensen2013}. 

The low line offset with respect to the width of the line is consistent with the cavity shock emission originating in a C-type shock. In contrast, the larger offset with respect to the width for spot shock components, in some cases such that there is little or no contribution at the source velocity, suggests that they originate in J-type shocks. The small beam filling factors, of similar order for all components, point towards compact emitting regions. This, combined with light hydrides seen in absorption against the far-IR continuum led \citet{Kristensen2013} to argue that some of the spot shock components are located close to the central source where the impact from any wind will have the highest energy. Some spot shocks, particularly those with large velocity offsets, likely originate in shocks within the protostellar jet, as observed in other species at higher angular resolution \citep[e.g.][]{Santiago-Garcia2009}. Spot shock components originating in either location can have low offsets from the source velocity if there is a large angle between the line of sight and the direction of motion caused by the shock. In both regions, water is being (re-)formed via high-temperature ($\gtrsim$300\,K) gas phase chemistry. 

For the cavity shock component, we assumed during the \textsc{radex} calculations that the emission is extended along one axis due to the extended H$_{2}$O 2$_{12}-$1$_{01}$ (1670\,GHz) emission observed towards several of the sources in our sample \citep[see e.g.][]{Vasta2012,Santangelo2012,Santangelo2013,Nisini2013}. For this to be consistent with the small emitting regions derived, this suggests a very small extent (1-30\,AU) perpendicular to the outflow axis. The width of a C-type shock depends on the ion-neutral coupling length, which is proportional to ($n_{\mathrm{H}_{2}}\,\times\,x_{\mathrm{i}}$)$^{-1}$, where $x_{\mathrm{i}}$ the degree of ionisation \citep{Draine1980}. Thus shocks are narrower for higher densities. For the densities inferred here ($\sim$ 10$^{5}$ - 10$^{8}$ cm$^{-3}$; Table~\ref{T:analysis_excitation_results}), typical widths of the H$_2$O emitting regions range from a few hundred AU to less than ten AU \citep{Visser2012}. If the C-type shocks are irradiated as suggested by \citet{Karska2014}, the degree of ionisation increases which leads to narrower shocks. Thus, the narrow width of the emitting region inferred here ($<$ 30 AU) are consistent with either high densities or irradiated shocks; since the excitation analysis generally points to a lower density, irradiated shocks are the preferred solution.

\subsection{Class 0 vs. Class I}
\label{S:discussion_evolution}

\begin{figure}
\begin{center}
\includegraphics[width=0.46\textwidth]{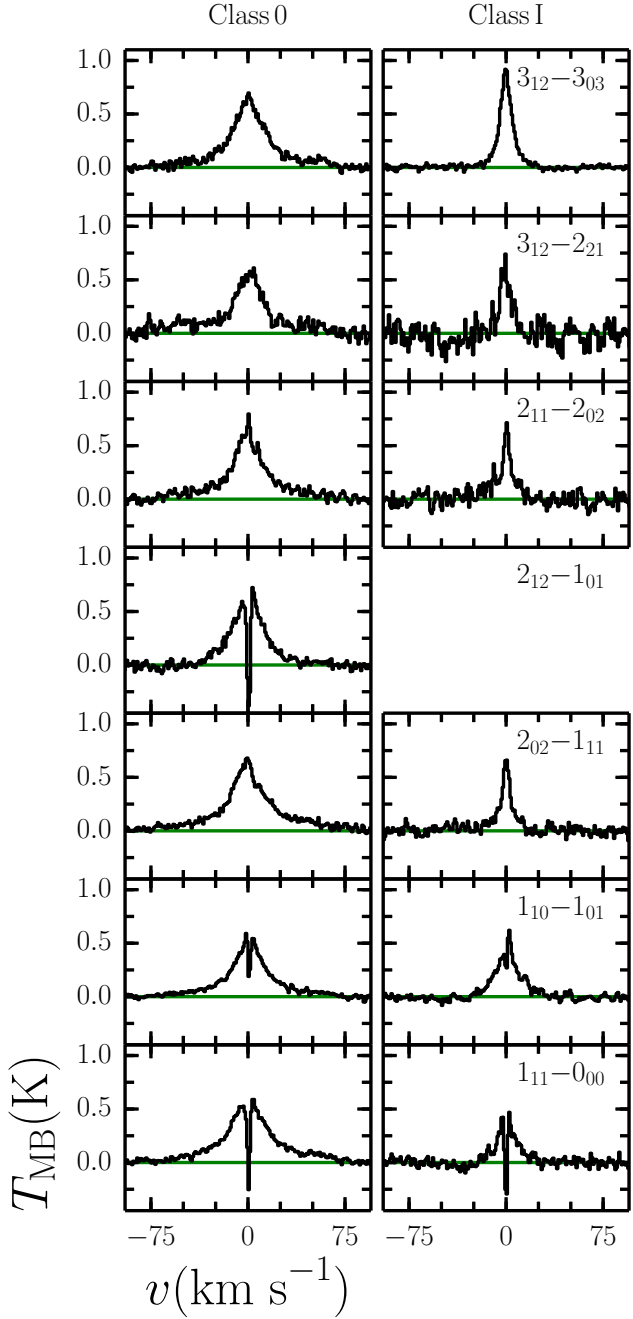}
\caption{Normalised average spectra for Class 0 (left) and I (right) sources. The green line indicates the baseline.}
\label{F:discussion_evolution_average}
\end{center}
\end{figure}

The median FWZI of the observed water lines drops significantly from Class 0 to Class I sources (see Table~\ref{T:results_profiles_basic_fwzi}). This is also seen in the average spectra (Fig.~\ref{F:discussion_evolution_average}), which are produced by averaging over all spectra with a peak s/n$\geq$4 after normalising to the peak intensity, and in the FWHM and peak intensity for the Gaussian components (see Figures~\ref{F:results_components_comparison_scatter}). This is relatively independent of the source luminosity or envelope mass, but is related to the pre-shock density of the envelope as probed by $n_{\mathrm{1000}}$.

The peak brightness of the water lines also decreases, on average, from Class 0 to Class I, as does the fraction of the total intensity in J-shock related spot shock components. There is also a shift from the quiescent envelope appearing in absorption to emission. However, as shown in Fig.~\ref{F:analysis_correlations_corr}, this is not directly related to $T_{\mathrm{bol}}$ or $F_{\mathrm{CO}}$, but rather to $M_{\mathrm{env}}$ and $n_{\mathrm{1000}}$. This correlation holds even though the lines are optically thick because they are likely effectively thin and thus the intensity still scales with $N_{\mathrm{H}_{2}\mathrm{O}}\times$$n_{\mathrm{H}_{2}}$ which is related to $M_{\mathrm{env}}$. While the average $M_{\mathrm{env}}$ and $n_{\mathrm{1000}}$ values for Class I sources are lower than for Class 0 sources, there is considerable overlap without clear segregation of the two populations. That the available reservoir is only weakly related to the evolutionary stage of the source is further confirmed by the very similar $N_{\mathrm{H}_{2}\mathrm{O}}$ and $n_{\mathrm{H}_{2}}$ results obtained from the \textsc{radex} analysis (Fig.~\ref{F:analysis_excitation_results_spread}). That said, the emitting region sizes for the Class I sources are on the small end of the distribution seen for the Class 0 sources.

We suggest that the decrease in intensity of water emission ($\propto$~$T_{\mathrm{peak}}$$\times$FWHM) as the source evolves from Class 0 to Class I is primarily related to the decrease in the velocity of the wind which drives the outflow. However, Herbig-Haro objects related to the jet in more evolved Class II/III systems have similar fast velocities to the jets in Class 0/I sources. 

One scenario which seems consistent with our findings would be that the velocity of the wind perpendicular to the jet drops as it moves further from the jet \citep[see, e.g.][; Yvart et al., in prep.]{Panoglou2012,Agra-Amboage2014} and that as the source evolves the outflow cavity widens, leading to weaker impacts on the cavity wall. In many cases it may have decreased to such an extent as the sources age that it is no longer fast enough to cause a significant J-shock even at the base of the outflow, though some emission related to the jet may remain. The cavity shock layer is also likely to get thinner as the shock cannot penetrate as far into the envelope, again leading to smaller filling factors. Indeed, at some distance from the source the oblique velocity of the wind probably becomes so low that it cannot even sputter ice from the grain mantle. The source is probably still capable of powering an entraining layer in CO since this is still in the gas phase, which is why peak H$_{2}$O brightness and outflow force from low-$J$ CO show little segregation with evolution. The decrease in envelope mass is then a weaker factor, as long as there is still enough gas with high enough density and gas-phase water abundance. We speculate that a lack of reservour may become more important for source evolving from Class I to Class II. 

\subsection{A consistent picture of water emission}
\label{S:discussion_picture}

A consistent picture, summarised in Figure~\ref{F:discussion_picture_cartoon}, is one where H$_{2}$O emission from the cavity shock component traces C-type shocks in a thin layer (of thickness only a few AU) at the interface with the cavity walls. Close to the central source this becomes a J-shock as traced by the spot shock component (see also Fig.~9 of \citet{Suutarinen2014}). This is because the angle of impact between the wind and the cavity wall is large close to the base of the outflow and becomes more oblique with distance from the central source. The pre-shock conditions probably transition smoothly between these two regimes, given the constant line ratios as a function of velocity (Fig.~\ref{F:analysis_vratios_vratio}) and similar densities and column densities of the spot and cavity shock components (Fig.~\ref{F:analysis_excitation_results_spread}). However, the change from a C to a J-type shock causes a dramatic difference in the velocity distribution of the post-shock material, resulting in the distinctly separate components we observe. This active shocked layer is distinct from the cooler entraining turbulent layer traced by low-$J$ CO. It is not clear whether water is sputtered from the ice mantles of dust grains or formed via gas-phase chemistry in this actively shocked region. The answer is probably some combination of both mechanisms, the proportion of each varying smoothly with velocity \citep[][]{VanLoo2013,Suutarinen2014}.

PACS CO observations of low-mass protostars generally find two temperature components (warm$\approx$300\,K and hot$\approx$750\,K). The emitting region sizes, column and post-shock densities we have obtained for the different components are similar to those for IRAS4B by \citet{Herczeg2012}, for L1448-MM by \citet{Lee2013} and the hot component for Ser-SMM1 by \citet{Goicoechea2012} from PACS observations of H$_{2}$O. We exclude the cold component for \citet{Goicoechea2012} because their analysis tied H$_{2}$O to low-$J$ CO lines which do not trace the same gas \citep[e.g.][]{Santangelo2013}. In particular, the J-shock related spot shock components and the C-shock related cavity shock components match well to the hot and warm PACS components respectively, both in their individual properties and their relative fractions of the total intensity. That there are fewer spot shocks observed in emission towards Class I sources is consistent with the lack of hot CO in Class I sources observed by \citet{Karska2013}. \citet{Green2013} detect hot CO towards more sources in their sample, but this is still consistent with our results as emission at 20$\%$ of the cavity shock component intensity would be too weak to be detected for those sources which are in both samples.

What is clear is that a considerable amount of the energy injected by the jet and/or outflow into the envelope is not traced by the entrained outflowing gas, but rather in the various shocks traced by water. Given that the shock related H$_{2}$O components have larger line-widths than seen in low-$J$ CO \citep[c.f.][]{Yildiz2013}, there may be a significant amount of momentum and energy carried away by these components compared to that in classical CO outflows. \citet{Bjerkeli2012} compared the momentum and energy in H$_{2}$O and CO for the VLA1623 outflow using H$_{2}$ observations to calculate abundances for water. They found water abundances with respect of H$_{2}$ of (1$-$8)$\times$10$^{-7}$, that the momentum in water was $\sim$25$\%$ that of CO and that the energies were comparable. However, the water abundances they derive are quite low compared to determinations for other sources \citep[e.g.][find 0.3$-$1$\times$10$^{-5}$]{Santangelo2013}, so the mass, momentum and energy calculated from H$_{2}$O may be an overestimate. Even so, studies which only use low-$J$ CO to quantify the impact of outflows at the source position probably underestimate the true mass, momentum, angular momentum and kinetic energy injected into the envelope and surrounding molecular cloud. Addressing this in a quantitative way requires a determination of the H$_{2}$O abundance relative to H$_{2}$ as a function of velocity, something that cannot be done with the observations presented here alone. This will be the subject of a future paper using HIFI observations of the high-$J$ CO 16-15 line (Kristensen et al., in prep.). 

\subsection{On source vs. off source}
\label{S:discussion_on_vs_off}

Having compared the various on-source components and their properties, it is also important to consider how our results compare to those obtained for shock positions further away from the central source. In general, the line profiles for off-source emission have similar maximum velocities but are less symmetric than at the source position \citep{Vasta2012,Santangelo2012,Busquet2014}, as might be expected for regions with only red or blue shifted outflow emission. While Gaussian decomposition similar to that used here has not been presented for those observations, some additional slightly offset features can be seen in some line profiles which are reminiscent of the on-source spot shock component. At some locations, particularly away from the brightest parts of the outflow, the line shape becomes more like the classical triangular shape of some CO outflows \citep{Santangelo2014}. In addition, there can be significant differences in line shape between H$_{2}$O transitions, resulting in line ratios (and thus excitation conditions) which vary with velocity.  

Table~\ref{T:discussion_on_vs_off_off} presents a summary of recent determinations of the excitation conditions towards such regions for some of the sources present in our study \citep[][]{Santangelo2013,Busquet2014,Santangelo2014}. Aside from the differences in isolating which emission to integrate over, these studies used similar Large Velocity Gradient models in their analysis, sometimes simultaneously fitting emission from water and other species also expected to originate in the outflow. As such, though the methods are not precisely the same, the results of these other studies should be comparable with the analysis presented in this paper. 

\begin{figure}
\begin{center}
\includegraphics[width=0.40\textwidth]{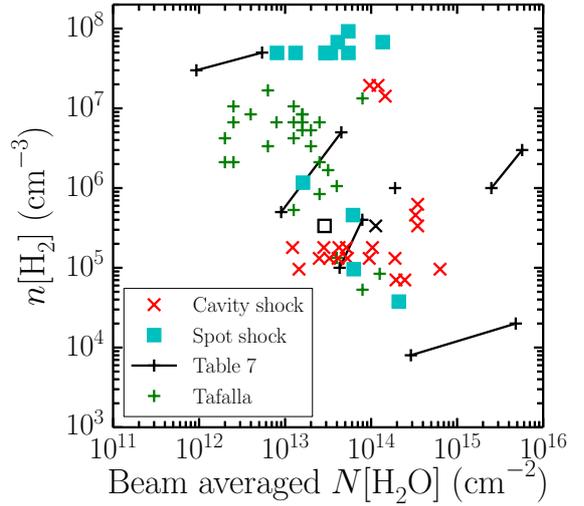}
\caption{Comparison of the conditions obtained in Sect.~\ref{S:analysis_excitation_results} with those from the literature in Table~\ref{T:discussion_on_vs_off_off} and from \citet{Tafalla2013}. The column densities have been averaged over the 1$_{10}-$1$_{01}$ beam as in \citet{Tafalla2013}. The black square and cross indicate the best fit for the average spot and cavity shock line ratios.}
\label{F:discussion_on_vs_off}
\end{center}
\end{figure}

We find much smaller emitting regions for on-source emission than derived for the off-source shock positions. Figure~\ref{F:discussion_on_vs_off} shows a comparison of the densities and beam-averaged column densities we derive with those in Table~\ref{T:discussion_on_vs_off_off} and those from \citet{Tafalla2013}. Averaged over the beam, the densities and column densities are very similar, though the absolute column densities are lower off-source due to the larger emitting regions. Lower column density at larger distances from the central source will lead to lower optical depths in the water lines. There are two possible options for this difference in column density.

\begin{table*}[]
\caption[]{Summary of off source H$_{2}$O excitation conditions.}
\centering
\begin{tabular}{lcccccc}
\hline
\hline \noalign {\smallskip}
\centering
Source & Comp.\tablefootmark{a} & $T$ & $N_{\mathrm{H}_{2}\mathrm{O}}$\tablefootmark{b} & $n_{\mathrm{H}_{2}}$ & $r$\tablefootmark{c} & Ref.  \\\noalign {\smallskip}
&&(K)&(cm$^{-2}$)&(cm$^{-3}$)&(AU) &\\\noalign {\smallskip}
\hline \noalign {\smallskip}
L1448-B2 & W & 450 & 3$\times$10$^{14}$ & 1$\times$10$^{6}$ & 2000 & 1 \\
& H & 1100 & (0.4-2)$\times$10$^{16}$ & (0.5-5)$\times$10$^{6}$ & $\sim$120 & 1 \\
L1157-B1 & W & 250-300 & (1.2-2.7)$\times$10$^{16}$ & (1-3)$\times$10$^{6}$ & 1600 & 2 \\
 & H & 900-1400 & (4.0-9.1)$\times$10$^{16}$ & (0.8-2)$\times$10$^{4}$ & 300-800 & 2 \\
IRAS4A-R2 & W & 300-500 & (1.3-2.7)$\times$10$^{13}$ & (3-5)$\times$10$^{7}$ & 1200-2000 & 3 \\
& H & 1000 & (0.7-1.3)$\times$10$^{16}$ & (1-4)$\times$10$^{5}$ & 350 & 3 \\
\hline \noalign {\smallskip}
\end{tabular}
\label{T:discussion_on_vs_off_off}
\tablefoot{\tablefoottext{a}{Component type: W = warm and H = hot.} \tablefoottext{b}{Column density over the emitting region.} \tablefoottext{c}{Radius of emitting region on the sky, calculated assuming a circular emitting area.}}
\tablebib{(1) \citet{Santangelo2013}; (2) \citet{Busquet2014}; (3) \citet{Santangelo2014}}
\end{table*}

Firstly, the water abundance could be higher at the source position than at the shock spots. Certainly lower water abundances have been found at the shock positions \citep[e.g.][]{Santangelo2013,Busquet2014} than the value of $\sim$10$^{-4}$ expected if all oxygen is forced into water by warm gas-phase chemistry. It is unlikely that H$_{2}$O could be converted into other species more efficiently at the off-source positions because the reaction rates for H$_{2}$O reacting with H or any other species are low \citep{Snell2005,McElroy2013}. The gas-phase H$_{2}$O abundance could be lower at the shock spots if sputtering is less efficient at the off-source positions because the velocity or density is lower \citep{Caselli1997}. However, the line-widths are large and post-shock densities are similar to the source position (see Fig.~\ref{F:discussion_on_vs_off}).

Alternatively, the difference may come from a change in the nature of the shocks being observed on and off source. The on-source cavity and spot shock components most likely exist as a thin layer at the boundary of the outflow cavity, as discussed above. As such, a parcel of post-shock gas cannot move azimuthally as it is in pressure equilibrium with the neighbouring parcels which are also in the post-shock. Expansion perpendicular to the outflow axis into the envelope may take place over time as the cavity opening angle increases but this will be resisted by the higher-density material in the envelope. Thus the only 'easy' expansion route for the gas will be away from the central protostar along the outflow cavity layer. This will fill a larger volume at larger distances while being subject to weaker shocks, hence the increasing beam-emitting area and decreasing column density. Indeed, this is consistent with the extended component seen in the water mapping observations of \citet{Santangelo2014}.

In contrast, the material in the post-shock of the bow-shocks is subject to the full direct impact of the jet rather than an oblique shock. It will therefore be a J-shock at or near the velocity of the jet, and may have a higher temperature and/or harsher UV field than in the on-source spot shocks. This would lead to a lower water abundance at the off-source bow-shock relative to the cavity shock. The post-shock density at the bow-shock is also higher than that of the surrounding cloud and so it can expand in most directions. Indeed, it will seek to do this as it is likely over-pressured with respect to its surroundings. The combination of a more violent and thus hotter initial shock, a pressure difference and freedom to expand in more directions, a different radiation field and different abundance would result in different excitation of the H$_{2}$O in the off-source bow-shock than in the spot-shocks on source. This in turn leads to line ratios which vary with both position and velocity, because the off-source cavity shock extends over a larger region and has different physical conditions and velocity field when compared to the bow-shocks. This is again consistent with the smaller emitting area and lower density found by \citet{Santangelo2014} towards the hot component in their maps of the NGC1333-IRAS4A shock positions compared to their warm component.

\section{Summary and Conclusions}
\label{S:conclusions}

In this paper, we have presented velocity-resolved \textit{Herschel} HIFI spectra of multiple water transitions for the whole low-mass Class 0/I subsample (29 sources) of the WISH survey. Our main findings are as follows:

\begin{itemize}

\item All water transitions for a given source studied here show very similar line profile shapes with consistent kinematic components.

\item Three distinct types of kinematic component can be identified: envelope, cavity shock and spot shock. The velocity offsets and line widths of the cavity and spot shock components are consistent with origin in C type shocks along the outflow cavity wall, and with J-shocks near the base of the outflow or in the jet respectively. The excitation conditions and relative fraction of the total intensity in each source suggest that the spot shock component is associated with the hot ($\sim$750\,K) CO component seen in PACS observations of Class 0/I protostars and the cavity shock component is associated with the warm ($\sim$300\,K) PACS CO component.

\item The line ratios are constant with velocity and similar for all sources. The emission is optically thick but effectively thin at all velocities, and traces material with post-shock densities of $n_{\mathrm{H}_{2}}$$\sim$10$^{5}-$10$^{8}$\,cm$^{-3}$ and column densities of $N_{\mathrm{H}_{2}\mathrm{O}}$$\sim$10$^{16}-$10$^{18}$\,cm$^{-2}$. All emission originates in compact emitting regions: for the spot shocks these correspond to point sources with radii of order 10-200\,AU, while for the cavity shock these come from a thin layer along the outflow cavity wall with thickness of order 1-30\,AU.

\item The excitation conditions of the different components are relatively similar at the source position. The emitting regions are also similar between the different components.

\item The major difference between the Class 0 and I sources is not in excitation conditions, but rather a decrease in line-width and intensity for more evolved sources. Coupled with the lack of J-shock components observed in older sources, this suggests that the decrease in water emission as the source evolves is primarily due to a decrease in the velocity of the wind which drives the outflow rather than the decrease in envelope density or mass. The envelope mass will likely become more important in the evolution from Class I to Class II.

\item The off-source excitation conditions reported in the literature for H$_{2}$O have similar densities but lower column densities and larger emitting regions than the on-source emission. We suggest that this difference is because the material in off-source bow-shocks has more freedom to expand and has a higher pressure difference with surrounding material than the on-source shocks in the jet or near the base of the outflow. 

\end{itemize}

Water is therefore revealing new information about the inner workings of outflows and the relationship between outflows and their driving sources. The remaining steps are to derive the water abundance, and thus obtain a quantitative comparison of the energy, momentum and mass in water compared to other tracers, and to explore these effects in a larger number of sources in order to see whether the trends that the WISH sample hint at, e.g. the separate trends between $L_{\mathrm{bol}}$ and water line intensity for Class 0 and I sources, are robust. These will be explored in upcoming papers, paving the way for a comprehensive understanding of all parts of the jet-outflow system.


\begin{acknowledgements}

The authors would like to thank the anonymous referee and Gary Melnick for helpful comments which improved the clarity and content of the paper, and U. Y{\i}ld{\i}z for assistance with \textsc{HIPE} and data reduction. JCM is funded by grant 614.001.008 from the Netherlands Organisation for Scientific Research (NWO). Astrochemistry in Leiden is supported by the Netherlands Research School for Astronomy (NOVA), by a Spinoza grant and by the European Community's Seventh Framework Programme FP7/2007-2013 under grant agreement 238258 (LASSIE). HIFI has been designed and built by a consortium of institutes and university departments from across Europe, Canada and the United States under the leadership of SRON Netherlands Institute for Space Research, Groningen, The Netherlands and with major contributions from Germany, France and the US. Consortium members are: Canada: CSA, U.Waterloo; France: CESR, LAB, LERMA, IRAM; Germany: KOSMA, MPIfR, MPS; Ireland, NUI Maynooth; Italy: ASI, IFSI-INAF, Osservatorio Astrofisico di Arcetri- INAF; Netherlands: SRON, TUD; Poland: CAMK, CBK; Spain: Observatorio Astron{\'o}mico Nacional (IGN), Centro de Astrobiolog{\'i}a (CSIC-INTA). Sweden: Chalmers University of Technology - MC2, RSS $\&$ GARD; Onsala Space Observatory; Swedish National Space Board, Stockholm University - Stockholm Observatory; Switzerland: ETH Zurich, FHNW; USA: Caltech, JPL, NHSC.

\end{acknowledgements}

\bibliography{./wish_water_excitationbib}

\begin{thebibliography}{87}
\expandafter\ifx\csname natexlab\endcsname\relax\def\natexlab#1{#1}\fi

\bibitem[{{Agra-Amboage} {et~al.}(2014){Agra-Amboage}, {Cabrit}, {Dougados},
  {Kristensen}, {Ibgui}, \& {Reunanen}}]{Agra-Amboage2014}
{Agra-Amboage}, V., {Cabrit}, S., {Dougados}, C., {et~al.} 2014, \aap, 564, A11

\bibitem[{{Andre} {et~al.}(1993){Andre}, {Ward-Thompson}, \&
  {Barsony}}]{Andre1993}
{Andre}, P., {Ward-Thompson}, D., \& {Barsony}, M. 1993, \apj, 406, 122

\bibitem[{{Avrett} \& {Hummer}(1965)}]{Avrett1965}
{Avrett}, E.~H. \& {Hummer}, D.~G. 1965, \mnras, 130, 295

\bibitem[{{Bacciotti} \& {Eisl{\"o}ffel}(1999)}]{Bacciotti1999}
{Bacciotti}, F. \& {Eisl{\"o}ffel}, J. 1999, \aap, 342, 717

\bibitem[{{Bachiller} {et~al.}(1991){Bachiller}, {Martin-Pintado}, \&
  {Fuente}}]{Bachiller1991}
{Bachiller}, R., {Martin-Pintado}, J., \& {Fuente}, A. 1991, \aap, 243, L21

\bibitem[{{Bachiller} {et~al.}(1990){Bachiller}, {Martin-Pintado}, {Tafalla},
  {Cernicharo}, \& {Lazareff}}]{Bachiller1990}
{Bachiller}, R., {Martin-Pintado}, J., {Tafalla}, M., {Cernicharo}, J., \&
  {Lazareff}, B. 1990, \aap, 231, 174

\bibitem[{{Bjerkeli} {et~al.}(2012){Bjerkeli}, {Liseau}, {Larsson}, {Rydbeck},
  {Nisini}, {Tafalla}, {Antoniucci}, {Benedettini}, {Bergman}, {Cabrit},
  {Giannini}, {Melnick}, {Neufeld}, {Santangelo}, \& {van
  Dishoeck}}]{Bjerkeli2012}
{Bjerkeli}, P., {Liseau}, R., {Larsson}, B., {et~al.} 2012, \aap, 546, A29

\bibitem[{{Bodo} {et~al.}(1994){Bodo}, {Massaglia}, {Ferrari}, \&
  {Trussoni}}]{Bodo1994}
{Bodo}, G., {Massaglia}, S., {Ferrari}, A., \& {Trussoni}, E. 1994, \aap, 283,
  655

\bibitem[{{Bontemps} {et~al.}(1996){Bontemps}, {Andre}, {Terebey}, \&
  {Cabrit}}]{Bontemps1996}
{Bontemps}, S., {Andre}, P., {Terebey}, S., \& {Cabrit}, S. 1996, \aap, 311,
  858

\bibitem[{{Busquet} {et~al.}(2014){Busquet}, {Lefloch}, {Benedettini},
  {Ceccarelli}, {Codella}, {Cabrit}, {Nisini}, {Viti}, {G{\'o}mez-Ruiz},
  {Gusdorf}, {di Giorgio}, \& {Wiesenfeld}}]{Busquet2014}
{Busquet}, G., {Lefloch}, B., {Benedettini}, M., {et~al.} 2014, \aap, 561, A120

\bibitem[{{Canto} \& {Raga}(1991)}]{Canto1991}
{Canto}, J. \& {Raga}, A.~C. 1991, \apj, 372, 646

\bibitem[{{Caselli} {et~al.}(1997){Caselli}, {Hartquist}, \&
  {Havnes}}]{Caselli1997}
{Caselli}, P., {Hartquist}, T.~W., \& {Havnes}, O. 1997, \aap, 322, 296

\bibitem[{{Chieze} {et~al.}(1998){Chieze}, {Pineau des Forets}, \&
  {Flower}}]{Chieze1998}
{Chieze}, J.-P., {Pineau des Forets}, G., \& {Flower}, D.~R. 1998, \mnras, 295,
  672

\bibitem[{{Codella} {et~al.}(2014){Codella}, {Maury}, {Gueth}, {Maret},
  {Belloche}, {Cabrit}, \& {Andr{\'e}}}]{Codella2014}
{Codella}, C., {Maury}, A.~J., {Gueth}, F., {et~al.} 2014, \aap, 563, L3

\bibitem[{{Daniel} {et~al.}(2011){Daniel}, {Dubernet}, \&
  {Grosjean}}]{Daniel2011}
{Daniel}, F., {Dubernet}, M.-L., \& {Grosjean}, A. 2011, \aap, 536, A76

\bibitem[{{de Graauw} {et~al.}(2010){de Graauw}, {Helmich}, {Phillips},
  {Stutzki}, {Caux}, {Whyborn}, {Dieleman}, {Roelfsema}, {Aarts}, {Assendorp},
  {Bachiller}, {Baechtold}, {Barcia}, {Beintema}, {Belitsky}, {Benz}, {Bieber},
  {Boogert}, {Borys}, {Bumble}, {Ca{\"i}s}, {Caris}, {Cerulli-Irelli},
  {Chattopadhyay}, {Cherednichenko}, {Ciechanowicz}, {Coeur-Joly}, {Comito},
  {Cros}, {de Jonge}, {de Lange}, {Delforges}, {Delorme}, {den Boggende},
  {Desbat}, {Diez-Gonz{\'a}lez}, {di Giorgio}, {Dubbeldam}, {Edwards},
  {Eggens}, {Erickson}, {Evers}, {Fich}, {Finn}, {Franke}, {Gaier}, {Gal},
  {Gao}, {Gallego}, {Gauffre}, {Gill}, {Glenz}, {Golstein}, {Goulooze},
  {Gunsing}, {G{\"u}sten}, {Hartogh}, {Hatch}, {Higgins}, {Honingh}, {Huisman},
  {Jackson}, {Jacobs}, {Jacobs}, {Jarchow}, {Javadi}, {Jellema}, {Justen},
  {Karpov}, {Kasemann}, {Kawamura}, {Keizer}, {Kester}, {Klapwijk}, {Klein},
  {Kollberg}, {Kooi}, {Kooiman}, {Kopf}, {Krause}, {Krieg}, {Kramer},
  {Kruizenga}, {Kuhn}, {Laauwen}, {Lai}, {Larsson}, {Leduc}, {Leinz}, {Lin},
  {Liseau}, {Liu}, {Loose}, {L{\'o}pez-Fernandez}, {Lord}, {Luinge}, {Marston},
  {Mart{\'{\i}}n-Pintado}, {Maestrini}, {Maiwald}, {McCoey}, {Mehdi}, {Megej},
  {Melchior}, {Meinsma}, {Merkel}, {Michalska}, {Monstein}, {Moratschke},
  {Morris}, {Muller}, {Murphy}, {Naber}, {Natale}, {Nowosielski}, {Nuzzolo},
  {Olberg}, {Olbrich}, {Orfei}, {Orleanski}, {Ossenkopf}, {Peacock}, {Pearson},
  {Peron}, {Phillip-May}, {Piazzo}, {Planesas}, {Rataj}, {Ravera}, {Risacher},
  {Salez}, {Samoska}, {Saraceno}, {Schieder}, {Schlecht}, {Schl{\"o}der},
  {Schm{\"u}lling}, {Schultz}, {Schuster}, {Siebertz}, {Smit}, {Szczerba},
  {Shipman}, {Steinmetz}, {Stern}, {Stokroos}, {Teipen}, {Teyssier}, {Tils},
  {Trappe}, {van Baaren}, {van Leeuwen}, {van de Stadt}, {Visser}, {Wildeman},
  {Wafelbakker}, {Ward}, {Wesselius}, {Wild}, {Wulff}, {Wunsch}, {Tielens},
  {Zaal}, {Zirath}, {Zmuidzinas}, \& {Zwart}}]{deGraauw2010}
{de Graauw}, T., {Helmich}, F.~P., {Phillips}, T.~G., {et~al.} 2010, \aap, 518,
  L6

\bibitem[{{Draine}(1980)}]{Draine1980}
{Draine}, B.~T. 1980, \apj, 241, 1021

\bibitem[{{Dubernet} {et~al.}(2009){Dubernet}, {Daniel}, {Grosjean}, \&
  {Lin}}]{Dubernet2009}
{Dubernet}, M.-L., {Daniel}, F., {Grosjean}, A., \& {Lin}, C.~Y. 2009, \aap,
  497, 911

\bibitem[{{Dzib} {et~al.}(2010){Dzib}, {Loinard}, {Mioduszewski}, {Boden},
  {Rodr{\'{\i}}guez}, \& {Torres}}]{Dzib2010}
{Dzib}, S., {Loinard}, L., {Mioduszewski}, A.~J., {et~al.} 2010, \apj, 718, 610

\bibitem[{{Frank} {et~al.}(2014){Frank}, {Ray}, {Cabrit}, {Hartigan}, {Arce},
  {Bacciotti}, {Bally}, {Benisty}, {Eisl{\"o}ffel}, {G{\"u}del}, {Lebedev},
  {Nisini}, \& {Raga}}]{Frank2014}
{Frank}, A., {Ray}, T.~P., {Cabrit}, S., {et~al.} 2014, in Protostars and
  Planets \Roman{ppnum6}, ed. H.~{Beuther}, R.~{Klessen}, C.~{Dullemond}, \&
  T.~{Henning} (Univ. of Arizona Press, Tucson), in press

\bibitem[{{Goicoechea} {et~al.}(2012){Goicoechea}, {Cernicharo}, {Karska},
  {Herczeg}, {Polehampton}, {Wampfler}, {Kristensen}, {van Dishoeck},
  {Etxaluze}, {Bern{\'e}}, \& {Visser}}]{Goicoechea2012}
{Goicoechea}, J.~R., {Cernicharo}, J., {Karska}, A., {et~al.} 2012, \aap, 548,
  A77

\bibitem[{{Goldsmith} \& {Langer}(1999)}]{Goldsmith1999}
{Goldsmith}, P.~F. \& {Langer}, W.~D. 1999, \apj, 517, 209

\bibitem[{{Gonz{\'a}lez-Alfonso} {et~al.}(2010){Gonz{\'a}lez-Alfonso},
  {Fischer}, {Isaak}, {Rykala}, {Savini}, {Spaans}, {van der Werf},
  {Meijerink}, {Israel}, {Loenen}, {Vlahakis}, {Smith}, {Charmandaris},
  {Aalto}, {Henkel}, {Wei{\ss}}, {Walter}, {Greve}, {Mart{\'{\i}}n-Pintado},
  {Naylor}, {Spinoglio}, {Veilleux}, {Harris}, {Armus}, {Lord}, {Mazzarella},
  {Xilouris}, {Sanders}, {Dasyra}, {Wiedner}, {Kramer}, {Papadopoulos},
  {Stacey}, {Evans}, \& {Gao}}]{Gonzalez-Alfonso2010}
{Gonz{\'a}lez-Alfonso}, E., {Fischer}, J., {Isaak}, K., {et~al.} 2010, \aap,
  518, L43

\bibitem[{{Green} {et~al.}(2013){Green}, {Evans}, {J{\o}rgensen}, {Herczeg},
  {Kristensen}, {Lee}, {Dionatos}, {Yildiz}, {Salyk}, {Meeus}, {Bouwman},
  {Visser}, {Bergin}, {van Dishoeck}, {Rascati}, {Karska}, {van Kempen},
  {Dunham}, {Lindberg}, {Fedele}, \& {DIGIT Team}}]{Green2013}
{Green}, J.~D., {Evans}, II, N.~J., {J{\o}rgensen}, J.~K., {et~al.} 2013, \apj,
  770, 123

\bibitem[{{Griffin} {et~al.}(2010){Griffin}, {Abergel}, {Abreu}, {Ade},
  {Andr{\'e}}, {Augueres}, {Babbedge}, {Bae}, {Baillie}, {Baluteau}, {Barlow},
  {Bendo}, {Benielli}, {Bock}, {Bonhomme}, {Brisbin}, {Brockley-Blatt},
  {Caldwell}, {Cara}, {Castro-Rodriguez}, {Cerulli}, {Chanial}, {Chen},
  {Clark}, {Clements}, {Clerc}, {Coker}, {Communal}, {Conversi}, {Cox},
  {Crumb}, {Cunningham}, {Daly}, {Davis}, {de Antoni}, {Delderfield}, {Devin},
  {di Giorgio}, {Didschuns}, {Dohlen}, {Donati}, {Dowell}, {Dowell}, {Duband},
  {Dumaye}, {Emery}, {Ferlet}, {Ferrand}, {Fontignie}, {Fox}, {Franceschini},
  {Frerking}, {Fulton}, {Garcia}, {Gastaud}, {Gear}, {Glenn}, {Goizel},
  {Griffin}, {Grundy}, {Guest}, {Guillemet}, {Hargrave}, {Harwit}, {Hastings},
  {Hatziminaoglou}, {Herman}, {Hinde}, {Hristov}, {Huang}, {Imhof}, {Isaak},
  {Israelsson}, {Ivison}, {Jennings}, {Kiernan}, {King}, {Lange}, {Latter},
  {Laurent}, {Laurent}, {Leeks}, {Lellouch}, {Levenson}, {Li}, {Li},
  {Lilienthal}, {Lim}, {Liu}, {Lu}, {Madden}, {Mainetti}, {Marliani}, {McKay},
  {Mercier}, {Molinari}, {Morris}, {Moseley}, {Mulder}, {Mur}, {Naylor},
  {Nguyen}, {O'Halloran}, {Oliver}, {Olofsson}, {Olofsson}, {Orfei}, {Page},
  {Pain}, {Panuzzo}, {Papageorgiou}, {Parks}, {Parr-Burman}, {Pearce},
  {Pearson}, {P{\'e}rez-Fournon}, {Pinsard}, {Pisano}, {Podosek}, {Pohlen},
  {Polehampton}, {Pouliquen}, {Rigopoulou}, {Rizzo}, {Roseboom}, {Roussel},
  {Rowan-Robinson}, {Rownd}, {Saraceno}, {Sauvage}, {Savage}, {Savini},
  {Sawyer}, {Scharmberg}, {Schmitt}, {Schneider}, {Schulz}, {Schwartz},
  {Shafer}, {Shupe}, {Sibthorpe}, {Sidher}, {Smith}, {Smith}, {Smith},
  {Spencer}, {Stobie}, {Sudiwala}, {Sukhatme}, {Surace}, {Stevens}, {Swinyard},
  {Trichas}, {Tourette}, {Triou}, {Tseng}, {Tucker}, {Turner}, {Vaccari},
  {Valtchanov}, {Vigroux}, {Virique}, {Voellmer}, {Walker}, {Ward}, {Waskett},
  {Weilert}, {Wesson}, {White}, {Whitehouse}, {Wilson}, {Winter}, {Woodcraft},
  {Wright}, {Xu}, {Zavagno}, {Zemcov}, {Zhang}, \& {Zonca}}]{Griffin2010}
{Griffin}, M.~J., {Abergel}, A., {Abreu}, A., {et~al.} 2010, \aap, 518, L3

\bibitem[{{Herczeg} {et~al.}(2012){Herczeg}, {Karska}, {Bruderer},
  {Kristensen}, {van Dishoeck}, {J{\o}rgensen}, {Visser}, {Wampfler}, {Bergin},
  {Y{\i}ld{\i}z}, {Pontoppidan}, \& {Gracia-Carpio}}]{Herczeg2012}
{Herczeg}, G.~J., {Karska}, A., {Bruderer}, S., {et~al.} 2012, \aap, 540, A84

\bibitem[{{Hirano} {et~al.}(2006){Hirano}, {Liu}, {Shang}, {Ho}, {Huang},
  {Kuan}, {McCaughrean}, \& {Zhang}}]{Hirano2006}
{Hirano}, N., {Liu}, S.-Y., {Shang}, H., {et~al.} 2006, \apjl, 636, L141

\bibitem[{{Hogerheijde} \& {van der Tak}(2000)}]{Hogerheijde2000}
{Hogerheijde}, M.~R. \& {van der Tak}, F.~F.~S. 2000, \aap, 362, 697

\bibitem[{{Hollenbach}(1997)}]{Hollenbach1997}
{Hollenbach}, D. 1997, in IAU Symposium, Vol. 182, Herbig-Haro Flows and the
  Birth of Stars, ed. B.~{Reipurth} \& C.~{Bertout} (Kluwer Academic
  Publishers), 181--198

\bibitem[{{Jim{\'e}nez-Serra} {et~al.}(2008{\natexlab{a}}){Jim{\'e}nez-Serra},
  {Caselli}, {Mart{\'{\i}}n-Pintado}, \& {Hartquist}}]{Jimenez-Serra2008b}
{Jim{\'e}nez-Serra}, I., {Caselli}, P., {Mart{\'{\i}}n-Pintado}, J., \&
  {Hartquist}, T.~W. 2008{\natexlab{a}}, \aap, 482, 549

\bibitem[{{Jim{\'e}nez-Serra} {et~al.}(2008{\natexlab{b}}){Jim{\'e}nez-Serra},
  {Mart{\'{\i}}n-Pintado}, {Rodr{\'{\i}}guez-Franco}, {Caselli}, {Viti}, \&
  {Hartquist}}]{Jimenez-Serra2008a}
{Jim{\'e}nez-Serra}, I., {Mart{\'{\i}}n-Pintado}, J.,
  {Rodr{\'{\i}}guez-Franco}, A., {et~al.} 2008{\natexlab{b}}, \apss, 313, 159

\bibitem[{{J{\o}rgensen} {et~al.}(2007){J{\o}rgensen}, {Bourke}, {Myers}, {Di
  Francesco}, {van Dishoeck}, {Lee}, {Ohashi}, {Sch{\"o}ier}, {Takakuwa},
  {Wilner}, \& {Zhang}}]{Jorgensen2007}
{J{\o}rgensen}, J.~K., {Bourke}, T.~L., {Myers}, P.~C., {et~al.} 2007, \apj,
  659, 479

\bibitem[{{J{\o}rgensen} {et~al.}(2009){J{\o}rgensen}, {van Dishoeck},
  {Visser}, {Bourke}, {Wilner}, {Lommen}, {Hogerheijde}, \&
  {Myers}}]{Jorgensen2009}
{J{\o}rgensen}, J.~K., {van Dishoeck}, E.~F., {Visser}, R., {et~al.} 2009,
  \aap, 507, 861

\bibitem[{{Karska} {et~al.}(2013){Karska}, {Herczeg}, {van Dishoeck},
  {Wampfler}, {Kristensen}, {Goicoechea}, {Visser}, {Nisini}, {San
  Jos{\'e}-Garc{\'{\i}}a}, {Bruderer}, {{\'S}niady}, {Doty}, {Fedele},
  {Y{\i}ld{\i}z}, {Benz}, {Bergin}, {Caselli}, {Herpin}, {Hogerheijde},
  {Johnstone}, {J{\o}rgensen}, {Liseau}, {Tafalla}, {van der Tak}, \&
  {Wyrowski}}]{Karska2013}
{Karska}, A., {Herczeg}, G.~J., {van Dishoeck}, E.~F., {et~al.} 2013, \aap,
  552, A141

\bibitem[{{Karska} {et~al.}(2014){Karska}, {Kristensen}, {van Dishoeck},
  {Drozdovskaya}, {Mottram}, {Herczeg}, {Bruderer}, {Cabrit}, {Evans},
  {Fedele}, {Gusdorf}, {Jorgensen}, {Kaufman}, {Melnick}, {Neufeld}, {Nisini},
  {Santangelo}, {Tafalla}, \& {Wampfler}}]{Karska2014}
{Karska}, A., {Kristensen}, L.~E., {van Dishoeck}, E.~F., {et~al.} 2014, \aap,
  in press.

\bibitem[{{Kaufman} \& {Neufeld}(1996)}]{Kaufman1996}
{Kaufman}, M.~J. \& {Neufeld}, D.~A. 1996, \apj, 456, 250

\bibitem[{{Kristensen} {et~al.}(2007){Kristensen}, {Ravkilde}, {Field},
  {Lemaire}, \& {Pineau Des For{\^e}ts}}]{Kristensen2007}
{Kristensen}, L.~E., {Ravkilde}, T.~L., {Field}, D., {Lemaire}, J.~L., \&
  {Pineau Des For{\^e}ts}, G. 2007, \aap, 469, 561

\bibitem[{{Kristensen} {et~al.}(2013){Kristensen}, {van Dishoeck}, {Benz},
  {Bruderer}, {Visser}, \& {Wampfler}}]{Kristensen2013}
{Kristensen}, L.~E., {van Dishoeck}, E.~F., {Benz}, A.~O., {et~al.} 2013, \aap,
  557, A23

\bibitem[{{Kristensen} {et~al.}(2012){Kristensen}, {van Dishoeck}, {Bergin},
  {Visser}, {Y{\i}ld{\i}z}, {San Jose-Garcia}, {J{\o}rgensen}, {Herczeg},
  {Johnstone}, {Wampfler}, {Benz}, {Bruderer}, {Cabrit}, {Caselli}, {Doty},
  {Harsono}, {Herpin}, {Hogerheijde}, {Karska}, {van Kempen}, {Liseau},
  {Nisini}, {Tafalla}, {van der Tak}, \& {Wyrowski}}]{Kristensen2012}
{Kristensen}, L.~E., {van Dishoeck}, E.~F., {Bergin}, E.~A., {et~al.} 2012,
  \aap, 542, A8

\bibitem[{{Kristensen} {et~al.}(2011){Kristensen}, {van Dishoeck}, {Tafalla},
  {Bachiller}, {Nisini}, {Liseau}, \& {Y{\i}ld{\i}z}}]{Kristensen2011}
{Kristensen}, L.~E., {van Dishoeck}, E.~F., {Tafalla}, M., {et~al.} 2011, \aap,
  531, L1

\bibitem[{{Kristensen} {et~al.}(2010){Kristensen}, {Visser}, {van Dishoeck},
  {Y{\i}ld{\i}z}, {Doty}, {Herczeg}, {Liu}, {Parise}, {J{\o}rgensen}, {van
  Kempen}, {Brinch}, {Wampfler}, {Bruderer}, {Benz}, {Hogerheijde}, {Deul},
  {Bachiller}, {Baudry}, {Benedettini}, {Bergin}, {Bjerkeli}, {Blake},
  {Bontemps}, {Braine}, {Caselli}, {Cernicharo}, {Codella}, {Daniel}, {de
  Graauw}, {di Giorgio}, {Dominik}, {Encrenaz}, {Fich}, {Fuente}, {Giannini},
  {Goicoechea}, {Helmich}, {Herpin}, {Jacq}, {Johnstone}, {Kaufman}, {Larsson},
  {Lis}, {Liseau}, {Marseille}, {McCoey}, {Melnick}, {Neufeld}, {Nisini},
  {Olberg}, {Pearson}, {Plume}, {Risacher}, {Santiago-Garc{\'{\i}}a},
  {Saraceno}, {Shipman}, {Tafalla}, {Tielens}, {van der Tak}, {Wyrowski},
  {Beintema}, {de Jonge}, {Dieleman}, {Ossenkopf}, {Roelfsema}, {Stutzki}, \&
  {Whyborn}}]{Kristensen2010}
{Kristensen}, L.~E., {Visser}, R., {van Dishoeck}, E.~F., {et~al.} 2010, \aap,
  521, L30

\bibitem[{{Lada} \& {Wilking}(1984)}]{Lada1984}
{Lada}, C.~J. \& {Wilking}, B.~A. 1984, \apj, 287, 610

\bibitem[{{Lee} {et~al.}(2013){Lee}, {Lee}, {Lee}, {Green}, {Evans}, {Choi},
  {Kristensen}, {Dionatos}, {J{\o}rgensen}, \& {the DIGIT Team}}]{Lee2013}
{Lee}, J., {Lee}, J.-E., {Lee}, S., {et~al.} 2013, \apjs, 209, 4

\bibitem[{{Linke} {et~al.}(1977){Linke}, {Goldsmith}, {Wannier}, {Wilson}, \&
  {Penzias}}]{Linke1977}
{Linke}, R.~A., {Goldsmith}, P.~F., {Wannier}, P.~G., {Wilson}, R.~W., \&
  {Penzias}, A.~A. 1977, \apj, 214, 50

\bibitem[{{Manoj} {et~al.}(2013){Manoj}, {Watson}, {Neufeld}, {Megeath},
  {Vavrek}, {Yu}, {Visser}, {Bergin}, {Fischer}, {Tobin}, {Stutz}, {Ali},
  {Wilson}, {Di Francesco}, {Osorio}, {Maret}, \& {Poteet}}]{Manoj2013}
{Manoj}, P., {Watson}, D.~M., {Neufeld}, D.~A., {et~al.} 2013, \apj, 763, 83

\bibitem[{{Marseille} {et~al.}(2010){Marseille}, {van der Tak}, {Herpin}, \&
  {Jacq}}]{Marseille2010}
{Marseille}, M.~G., {van der Tak}, F.~F.~S., {Herpin}, F., \& {Jacq}, T. 2010,
  \aap, 522, A40

\bibitem[{{McElroy} {et~al.}(2013){McElroy}, {Walsh}, {Markwick}, {Cordiner},
  {Smith}, \& {Millar}}]{McElroy2013}
{McElroy}, D., {Walsh}, C., {Markwick}, A.~J., {et~al.} 2013, \aap, 550, A36

\bibitem[{{Mottram} {et~al.}(2013){Mottram}, {van Dishoeck}, {Schmalzl},
  {Kristensen}, {Visser}, {Hogerheijde}, \& {Bruderer}}]{Mottram2013}
{Mottram}, J.~C., {van Dishoeck}, E.~F., {Schmalzl}, M., {et~al.} 2013, \aap,
  558, A126

\bibitem[{{M{\"u}ller} {et~al.}(2005){M{\"u}ller}, {Schl{\"o}der}, {Stutzki},
  \& {Winnewisser}}]{Muller2005}
{M{\"u}ller}, H.~S.~P., {Schl{\"o}der}, F., {Stutzki}, J., \& {Winnewisser}, G.
  2005, Journal of Molecular Structure, 742, 215

\bibitem[{{Neufeld} {et~al.}(2014){Neufeld}, {Gusdorf}, {G{\"u}sten},
  {Herczeg}, {Kristensen}, {Melnick}, {Nisini}, {Ossenkopf}, {Tafalla}, \& {van
  Dishoeck}}]{Neufeld2014}
{Neufeld}, D.~A., {Gusdorf}, A., {G{\"u}sten}, R., {et~al.} 2014, \apj, 781,
  102

\bibitem[{{Nisini} {et~al.}(2010){Nisini}, {Benedettini}, {Codella},
  {Giannini}, {Liseau}, {Neufeld}, {Tafalla}, {van Dishoeck}, {Bachiller},
  {Baudry}, {Benz}, {Bergin}, {Bjerkeli}, {Blake}, {Bontemps}, {Braine},
  {Bruderer}, {Caselli}, {Cernicharo}, {Daniel}, {Encrenaz}, {di Giorgio},
  {Dominik}, {Doty}, {Fich}, {Fuente}, {Goicoechea}, {de Graauw}, {Helmich},
  {Herczeg}, {Herpin}, {Hogerheijde}, {Jacq}, {Johnstone}, {J{\o}rgensen},
  {Kaufman}, {Kristensen}, {Larsson}, {Lis}, {Marseille}, {McCoey}, {Melnick},
  {Olberg}, {Parise}, {Pearson}, {Plume}, {Risacher}, {Santiago}, {Saraceno},
  {Shipman}, {van Kempen}, {Visser}, {Viti}, {Wampfler}, {Wyrowski}, {van der
  Tak}, {Y{\i}ld{\i}z}, {Delforge}, {Desbat}, {Hatch}, {P{\'e}ron}, {Schieder},
  {Stern}, {Teyssier}, \& {Whyborn}}]{Nisini2010}
{Nisini}, B., {Benedettini}, M., {Codella}, C., {et~al.} 2010, \aap, 518, L120

\bibitem[{{Nisini} {et~al.}(2013){Nisini}, {Santangelo}, {Antoniucci},
  {Benedettini}, {Codella}, {Giannini}, {Lorenzani}, {Liseau}, {Tafalla},
  {Bjerkeli}, {Cabrit}, {Caselli}, {Kristensen}, {Neufeld}, {Melnick}, \& {van
  Dishoeck}}]{Nisini2013}
{Nisini}, B., {Santangelo}, G., {Antoniucci}, S., {et~al.} 2013, \aap, 549, A16

\bibitem[{{Ott}(2010)}]{Ott2010}
{Ott}, S. 2010, in Astronomical Society of the Pacific Conference Series, Vol.
  434, Astronomical Data Analysis Software and Systems XIX, ed. Y.~{Mizumoto},
  K.-I. {Morita}, \& M.~{Ohishi} (Astronomical Society of the Pacific), 139

\bibitem[{{Pagani} {et~al.}(2009){Pagani}, {Vastel}, {Hugo}, {Kokoouline},
  {Greene}, {Bacmann}, {Bayet}, {Ceccarelli}, {Peng}, \&
  {Schlemmer}}]{Pagani2009}
{Pagani}, L., {Vastel}, C., {Hugo}, E., {et~al.} 2009, \aap, 494, 623

\bibitem[{{Panoglou} {et~al.}(2012){Panoglou}, {Cabrit}, {Pineau Des
  For{\^e}ts}, {Garcia}, {Ferreira}, \& {Casse}}]{Panoglou2012}
{Panoglou}, D., {Cabrit}, S., {Pineau Des For{\^e}ts}, G., {et~al.} 2012, \aap,
  538, A2

\bibitem[{{Persson} {et~al.}(2012){Persson}, {J{\o}rgensen}, \& {van
  Dishoeck}}]{Persson2012}
{Persson}, M.~V., {J{\o}rgensen}, J.~K., \& {van Dishoeck}, E.~F. 2012, \aap,
  541, A39

\bibitem[{{Persson} {et~al.}(2014){Persson}, {J{\o}rgensen}, {van Dishoeck}, \&
  {Harsono}}]{Persson2014}
{Persson}, M.~V., {J{\o}rgensen}, J.~K., {van Dishoeck}, E.~F., \& {Harsono},
  D. 2014, \aap, 563, A74

\bibitem[{{Pickett} {et~al.}(2010){Pickett}, {Poynter}, {Cohen}, {Delitsky},
  {Pearson}, \& {Muller}}]{Pickett2010}
{Pickett}, H.~M., {Poynter}, I.~R.~L., {Cohen}, E.~A., {et~al.} 2010, \jqsrt,
  111, 1617

\bibitem[{{Pilbratt} {et~al.}(2010){Pilbratt}, {Riedinger}, {Passvogel},
  {Crone}, {Doyle}, {Gageur}, {Heras}, {Jewell}, {Metcalfe}, {Ott}, \&
  {Schmidt}}]{Pilbratt2010}
{Pilbratt}, G.~L., {Riedinger}, J.~R., {Passvogel}, T., {et~al.} 2010, \aap,
  518, L1

\bibitem[{{Poglitsch} {et~al.}(2010){Poglitsch}, {Waelkens}, {Geis},
  {Feuchtgruber}, {Vandenbussche}, {Rodriguez}, {Krause}, {Renotte}, {van
  Hoof}, {Saraceno}, {Cepa}, {Kerschbaum}, {Agn{\`e}se}, {Ali}, {Altieri},
  {Andreani}, {Augueres}, {Balog}, {Barl}, {Bauer}, {Belbachir}, {Benedettini},
  {Billot}, {Boulade}, {Bischof}, {Blommaert}, {Callut}, {Cara}, {Cerulli},
  {Cesarsky}, {Contursi}, {Creten}, {De Meester}, {Doublier}, {Doumayrou},
  {Duband}, {Exter}, {Genzel}, {Gillis}, {Gr{\"o}zinger}, {Henning},
  {Herreros}, {Huygen}, {Inguscio}, {Jakob}, {Jamar}, {Jean}, {de Jong},
  {Katterloher}, {Kiss}, {Klaas}, {Lemke}, {Lutz}, {Madden}, {Marquet},
  {Martignac}, {Mazy}, {Merken}, {Montfort}, {Morbidelli}, {M{\"u}ller},
  {Nielbock}, {Okumura}, {Orfei}, {Ottensamer}, {Pezzuto}, {Popesso},
  {Putzeys}, {Regibo}, {Reveret}, {Royer}, {Sauvage}, {Schreiber}, {Stegmaier},
  {Schmitt}, {Schubert}, {Sturm}, {Thiel}, {Tofani}, {Vavrek}, {Wetzstein},
  {Wieprecht}, \& {Wiezorrek}}]{Poglitsch2010}
{Poglitsch}, A., {Waelkens}, C., {Geis}, N., {et~al.} 2010, \aap, 518, L2

\bibitem[{{Pudritz} {et~al.}(2007){Pudritz}, {Ouyed}, {Fendt}, \&
  {Brandenburg}}]{Pudritz2007}
{Pudritz}, R.~E., {Ouyed}, R., {Fendt}, C., \& {Brandenburg}, A. 2007, in
  Protostars and Planets \Roman{ppnum5}, ed. B.~{Reipurth}, D.~{Jewitt}, \&
  K.~{Keil} (Univ. of Arizona Press, Tucson), 277--294

\bibitem[{{Raga} {et~al.}(1995){Raga}, {Cabrit}, \& {Canto}}]{Raga1995}
{Raga}, A.~C., {Cabrit}, S., \& {Canto}, J. 1995, \mnras, 273, 422

\bibitem[{{Reipurth} {et~al.}(2000){Reipurth}, {Heathcote}, {Yu}, {Bally}, \&
  {Rodr{\'{\i}}guez}}]{Reipurth2000}
{Reipurth}, B., {Heathcote}, S., {Yu}, K.~C., {Bally}, J., \&
  {Rodr{\'{\i}}guez}, L.~F. 2000, \apj, 534, 317

\bibitem[{{Roelfsema} {et~al.}(2012){Roelfsema}, {Helmich}, {Teyssier},
  {Ossenkopf}, {Morris}, {Olberg}, {Shipman}, {Risacher}, {Akyilmaz},
  {Assendorp}, {Avruch}, {Beintema}, {Biver}, {Boogert}, {Borys}, {Braine},
  {Caris}, {Caux}, {Cernicharo}, {Coeur-Joly}, {Comito}, {de Lange},
  {Delforge}, {Dieleman}, {Dubbeldam}, {de Graauw}, {Edwards}, {Fich},
  {Flederus}, {Gal}, {di Giorgio}, {Herpin}, {Higgins}, {Hoac}, {Huisman},
  {Jarchow}, {Jellema}, {de Jonge}, {Kester}, {Klein}, {Kooi}, {Kramer},
  {Laauwen}, {Larsson}, {Leinz}, {Lord}, {Lorenzani}, {Luinge}, {Marston},
  {Mart{\'{\i}}n-Pintado}, {McCoey}, {Melchior}, {Michalska}, {Moreno},
  {M{\"u}ller}, {Nowosielski}, {Okada}, {Orlea{\'n}ski}, {Phillips}, {Pearson},
  {Rabois}, {Ravera}, {Rector}, {Rengel}, {Sagawa}, {Salomons},
  {S{\'a}nchez-Su{\'a}rez}, {Schieder}, {Schl{\"o}der}, {Schm{\"u}lling},
  {Soldati}, {Stutzki}, {Thomas}, {Tielens}, {Vastel}, {Wildeman}, {Xie},
  {Xilouris}, {Wafelbakker}, {Whyborn}, {Zaal}, {Bell}, {Bjerkeli}, {De Beck},
  {Cavali{\'e}}, {Crockett}, {Hily-Blant}, {Kama}, {Kaminski}, {Lefl{\'o}ch},
  {Lombaert}, {de Luca}, {Makai}, {Marseille}, {Nagy}, {Pacheco}, {van der
  Wiel}, {Wang}, \& {Y{\i}ld{\i}z}}]{Roelfsema2012}
{Roelfsema}, P.~R., {Helmich}, F.~P., {Teyssier}, D., {et~al.} 2012, \aap, 537,
  A17

\bibitem[{{San Jos{\'e}-Garc{\'{\i}}a} {et~al.}(2013){San
  Jos{\'e}-Garc{\'{\i}}a}, {Mottram}, {Kristensen}, {van Dishoeck},
  {Y{\i}ld{\i}z}, {van der Tak}, {Herpin}, {Visser}, {McCoey}, {Wyrowski},
  {Braine}, \& {Johnstone}}]{SanJoseGarcia2013}
{San Jos{\'e}-Garc{\'{\i}}a}, I., {Mottram}, J.~C., {Kristensen}, L.~E.,
  {et~al.} 2013, \aap, 553, A125

\bibitem[{{Santangelo} {et~al.}(2013){Santangelo}, {Nisini}, {Antoniucci},
  {Codella}, {Cabrit}, {Giannini}, {Herczeg}, {Liseau}, {Tafalla}, \& {van
  Dishoeck}}]{Santangelo2013}
{Santangelo}, G., {Nisini}, B., {Antoniucci}, S., {et~al.} 2013, \aap, 557, A22

\bibitem[{{Santangelo} {et~al.}(2014){Santangelo}, {Nisini}, {Codella},
  {Lorenzani}, {Y{\i}ld{\i}z}, {Antoniucci}, {Bjerkeli}, {Cabrit}, {Giannini},
  {Kristensen}, {Liseau}, {Mottram}, {Tafalla}, \& {van
  Dishoeck}}]{Santangelo2014}
{Santangelo}, G., {Nisini}, B., {Codella}, C., {et~al.} 2014, \aap, 568, A125

\bibitem[{{Santangelo} {et~al.}(2012){Santangelo}, {Nisini}, {Giannini},
  {Antoniucci}, {Vasta}, {Codella}, {Lorenzani}, {Tafalla}, {Liseau}, {van
  Dishoeck}, \& {Kristensen}}]{Santangelo2012}
{Santangelo}, G., {Nisini}, B., {Giannini}, T., {et~al.} 2012, \aap, 538, A45

\bibitem[{{Santiago-Garc{\'{\i}}a} {et~al.}(2009){Santiago-Garc{\'{\i}}a},
  {Tafalla}, {Johnstone}, \& {Bachiller}}]{Santiago-Garcia2009}
{Santiago-Garc{\'{\i}}a}, J., {Tafalla}, M., {Johnstone}, D., \& {Bachiller},
  R. 2009, \aap, 495, 169

\bibitem[{{Schmalzl} {et~al.}(2014){Schmalzl}, {Visser}, {Walsh}, {Albertsson},
  {van Dishoeck}, {Kristensen}, \& {Mottram}}]{Schmalzl2014}
{Schmalzl}, M., {Visser}, R., {Walsh}, C., {et~al.} 2014, \aap, in press

\bibitem[{{Sch{\"o}ier} {et~al.}(2005){Sch{\"o}ier}, {van der Tak}, {van
  Dishoeck}, \& {Black}}]{Schoier2005}
{Sch{\"o}ier}, F.~L., {van der Tak}, F.~F.~S., {van Dishoeck}, E.~F., \&
  {Black}, J.~H. 2005, \aap, 432, 369

\bibitem[{{Shadmehri} \& {Downes}(2008)}]{Shadmehri2008}
{Shadmehri}, M. \& {Downes}, T.~P. 2008, \mnras, 387, 1318

\bibitem[{{Shang} {et~al.}(2007){Shang}, {Li}, \& {Hirano}}]{Shang2007}
{Shang}, H., {Li}, Z.-Y., \& {Hirano}, N. 2007, in Protostars and Planets
  \Roman{ppnum5}, ed. B.~{Reipurth}, D.~{Jewitt}, \& K.~{Keil} (Univ. of
  Arizona Press, Tucson), 261--276

\bibitem[{{Snell} {et~al.}(2005){Snell}, {Hollenbach}, {Howe}, {Neufeld},
  {Kaufman}, {Melnick}, {Bergin}, \& {Wang}}]{Snell2005}
{Snell}, R.~L., {Hollenbach}, D., {Howe}, J.~E., {et~al.} 2005, \apj, 620, 758

\bibitem[{{Suutarinen} {et~al.}(2014){Suutarinen}, {Kristensen}, {Mottram},
  {Fraser}, \& {van Dishoeck}}]{Suutarinen2014}
{Suutarinen}, A.~N., {Kristensen}, L.~E., {Mottram}, J.~C., {Fraser}, H.~J., \&
  {van Dishoeck}, E.~F. 2014, \mnras, 440, 1844

\bibitem[{{Tafalla} {et~al.}(2013){Tafalla}, {Liseau}, {Nisini}, {Bachiller},
  {Santiago-Garc{\'{\i}}a}, {van Dishoeck}, {Kristensen}, {Herczeg}, \&
  {Y{\i}ld{\i}z}}]{Tafalla2013}
{Tafalla}, M., {Liseau}, R., {Nisini}, B., {et~al.} 2013, \aap, 551, A116

\bibitem[{{van der Tak} {et~al.}(2007){van der Tak}, {Black}, {Sch{\"o}ier},
  {Jansen}, \& {van Dishoeck}}]{vanderTak2007}
{van der Tak}, F.~F.~S., {Black}, J.~H., {Sch{\"o}ier}, F.~L., {Jansen}, D.~J.,
  \& {van Dishoeck}, E.~F. 2007, \aap, 468, 627

\bibitem[{{van der Tak} {et~al.}(2013){van der Tak}, {Chavarr{\'{\i}}a},
  {Herpin}, {Wyrowski}, {Walmsley}, {van Dishoeck}, {Benz}, {Bergin},
  {Caselli}, {Hogerheijde}, {Johnstone}, {Kristensen}, {Liseau}, {Nisini}, \&
  {Tafalla}}]{vanderTak2013}
{van der Tak}, F.~F.~S., {Chavarr{\'{\i}}a}, L., {Herpin}, F., {et~al.} 2013,
  \aap, 554, A83

\bibitem[{{van Dishoeck} {et~al.}(2013){van Dishoeck}, {Herbst}, \&
  {Neufeld}}]{vanDishoeck2013}
{van Dishoeck}, E.~F., {Herbst}, E., \& {Neufeld}, D.~A. 2013, Chemical
  Reviews, 113, 9043

\bibitem[{{van Dishoeck} {et~al.}(2011){van Dishoeck}, {Kristensen}, {Benz},
  {Bergin}, {Caselli}, {Cernicharo}, {Herpin}, {Hogerheijde}, {Johnstone},
  {Liseau}, {Nisini}, {Shipman}, {Tafalla}, {van der Tak}, {Wyrowski},
  {Aikawa}, {Bachiller}, {Baudry}, {Benedettini}, {Bjerkeli}, {Blake},
  {Bontemps}, {Braine}, {Brinch}, {Bruderer}, {Chavarr{\'{\i}}a}, {Codella},
  {Daniel}, {de Graauw}, {Deul}, {di Giorgio}, {Dominik}, {Doty}, {Dubernet},
  {Encrenaz}, {Feuchtgruber}, {Fich}, {Frieswijk}, {Fuente}, {Giannini},
  {Goicoechea}, {Helmich}, {Herczeg}, {Jacq}, {J{\o}rgensen}, {Karska},
  {Kaufman}, {Keto}, {Larsson}, {Lefloch}, {Lis}, {Marseille}, {McCoey},
  {Melnick}, {Neufeld}, {Olberg}, {Pagani}, {Pani{\'c}}, {Parise}, {Pearson},
  {Plume}, {Risacher}, {Salter}, {Santiago-Garc{\'{\i}}a}, {Saraceno},
  {St{\"a}uber}, {van Kempen}, {Visser}, {Viti}, {Walmsley}, {Wampfler}, \&
  {Y{\i}ld{\i}z}}]{vanDishoeck2011}
{van Dishoeck}, E.~F., {Kristensen}, L.~E., {Benz}, A.~O., {et~al.} 2011,
  \pasp, 123, 138

\bibitem[{{Van Loo} {et~al.}(2013){Van Loo}, {Ashmore}, {Caselli}, {Falle}, \&
  {Hartquist}}]{VanLoo2013}
{Van Loo}, S., {Ashmore}, I., {Caselli}, P., {Falle}, S.~A.~E.~G., \&
  {Hartquist}, T.~W. 2013, \mnras, 428, 381

\bibitem[{{Vasta} {et~al.}(2012){Vasta}, {Codella}, {Lorenzani}, {Santangelo},
  {Nisini}, {Giannini}, {Tafalla}, {Liseau}, {van Dishoeck}, \&
  {Kristensen}}]{Vasta2012}
{Vasta}, M., {Codella}, C., {Lorenzani}, A., {et~al.} 2012, \aap, 537, A98

\bibitem[{{Velusamy} {et~al.}(2007){Velusamy}, {Langer}, \&
  {Marsh}}]{Velusamy2007}
{Velusamy}, T., {Langer}, W.~D., \& {Marsh}, K.~A. 2007, \apjl, 668, L159

\bibitem[{{Visser} {et~al.}(2013){Visser}, {J{\o}rgensen}, {Kristensen}, {van
  Dishoeck}, \& {Bergin}}]{Visser2013}
{Visser}, R., {J{\o}rgensen}, J.~K., {Kristensen}, L.~E., {van Dishoeck},
  E.~F., \& {Bergin}, E.~A. 2013, \apj, 769, 19

\bibitem[{{Visser} {et~al.}(2012){Visser}, {Kristensen}, {Bruderer}, {van
  Dishoeck}, {Herczeg}, {Brinch}, {Doty}, {Harsono}, \& {Wolfire}}]{Visser2012}
{Visser}, R., {Kristensen}, L.~E., {Bruderer}, S., {et~al.} 2012, \aap, 537,
  A55

\bibitem[{{Wilson} \& {Rood}(1994)}]{Wilson1994}
{Wilson}, T.~L. \& {Rood}, R. 1994, \araa, 32, 191

\bibitem[{{Y{\i}ld{\i}z} {et~al.}(2013){Y{\i}ld{\i}z}, {Kristensen}, {van
  Dishoeck}, {San Jos{\'e}-Garc{\'{\i}}a}, {Karska}, {Harsono}, {Tafalla},
  {Fuente}, {Visser}, {J{\o}rgensen}, \& {Hogerheijde}}]{Yildiz2013}
{Y{\i}ld{\i}z}, U.~A., {Kristensen}, L.~E., {van Dishoeck}, E.~F., {et~al.}
  2013, \aap, 556, A89

\end{thebibliography}

\bibliographystyle{./aa}

\Online
\appendix

\section{Supplementary material}
\label{S:appendix1}

\begin{table*}
\begin{center}
\caption[]{Observed water lines.}
\begin{tabular}{lccccccccc}
\hline \noalign {\smallskip}
Line & Rest Frequency\tablefootmark{a}& E$_{\mathrm{u}}$/$k_{\mathrm{b}}$ & $A_{\mathrm{ul}}$ & $n_{\mathrm{cr}}$ & $\eta_{mb}$\tablefootmark{b} & $\theta_{mb}$\tablefootmark{b} & WBS resolution & HRS resolution & Obs. Time\tablefootmark{c}   \\
 & (GHz) & (K) & (s$^{-1}$) & (cm$^{-3}$) & & (\arcsec{}) & (\kms{}) & (\kms{}) & Class 0/I (min) \\
\hline\noalign {\smallskip}
H$_{2}$O 1$_{10}$-1$_{01}$ & \phantom{0}556.93607 & \phantom{0}61.0 & 3.458$\times$10$^{-3}$ & 1$\times$10$^{7}$ & 0.75 & 38.1 & 0.27 & 0.03 & 13.5 / 20.7 \\
H$_{2}$O 2$_{12}$-1$_{01}$ & 1669.90496 & 114.4 & 5.593$\times$10$^{-2}$ & 3$\times$10$^{8}$ & 0.71 & 12.7 & 0.09 & 0.02 & \phantom{$-$}23.7 / \phantom{0}$-$\phantom{23.}  \\
H$_{2}$O 3$_{12}$-2$_{21}$ & 1153.12682 & 249.4 & 2.634$\times$10$^{-3}$ & 8$\times$10$^{6}$ & 0.64 & 18.4 & 0.13 & 0.06 &  13.0 / \phantom{0}9.9 \\
H$_{2}$O 3$_{12}$-3$_{03}$ & 1097.36505 & 249.4 & 1.648$\times$10$^{-2}$ & 5$\times$10$^{7}$ & 0.74 & 19.3 & 0.14 & 0.07 & \phantom{$-$}32.5 (306\tablefootmark{d}) / \phantom{0}$-$ (306\tablefootmark{d}) \phantom{2.} \\
H$_{2}^{18}$O 1$_{10}$-1$_{01}$ & \phantom{0}547.67644 & \phantom{0}60.5 & 3.290$\times$10$^{-3}$& 1$\times$10$^{7}$ & 0.75 & 38.7 & 0.27 & 0.07 & 33.0$-$126.0 / \phantom{0}$-$\phantom{33.126.0} \\
H$_{2}^{18}$O 3$_{12}$-3$_{03}$ & 1095.62739 & 248.7 & 1.621$\times$10$^{-2}$ & 5$\times$10$^{7}$ & 0.74 & 19.4 & 0.14 & 0.07 & \phantom{$-$}32.5 (306\tablefootmark{d}) / \phantom{0}$-$ (306\tablefootmark{d}) \phantom{2.} \\
\end{tabular}
\tablefoot{\tablefoottext{a}{Taken from \citet{Daniel2011} and \citet{Dubernet2009} for H$_{2}$O and the JPL database \citep{Pickett2010} for H$_{2}^{18}$O.} \tablefoottext{b}{Calculated using equations 1 and 3 from \citet{Roelfsema2012}.} \tablefoottext{c}{Total time including on+off source and overheads.}\tablefoottext{d}{Deep observation for IRAS2A and sources observed as part of OT2 programme OT2\_rvisser\_2.}}
\end{center}
\end{table*}

\begin{sidewaystable*}[ph!]
\caption[]{Observation identification numbers.}
\centering
\begin{tabular}{lccccccccc}
\hline
\hline \noalign {\smallskip}
\centering
Source &  H$_{2}$O 1$_{10}$-1$_{01}$ & H$_{2}$O 2$_{12}$-1$_{01}$ & H$_{2}$O 3$_{12}$-2$_{21}$ & H$_{2}$O 3$_{12}$-3$_{03}$\tablefootmark{a} & H$_{2}$O 1$_{11}$-0$_{00}$\tablefootmark{b} & H$_{2}$O 2$_{02}$-1$_{11}$ & H$_{2}$O 2$_{11}$-2$_{02}$ & H$_{2}$$^{18}$O 1$_{10}$-1$_{01}$ & H$_{2}$$^{18}$O 2$_{02}$-1$_{11}$ \\
\hline\noalign {\smallskip}
L1448-MM & 1342203202 & $-$ & 1342203253 & 1342201802 & 1342201803 & 1342203182 & 1342203216 & 1342203186 & $-$\\
NGC1333-IRAS2A & 1342202067 & 1342215966 & 1342191701 & \phantom{\tablefootmark{c}}1342215968\tablefootmark{c} & 1342191657 & 1342191606 & 1342191748 & 1342192206 & 1342191607 \\
NGC1333-IRAS4A & 1342202065 & 1342203951 & 1342191721 & \phantom{\tablefootmark{e}}1342249014\tablefootmark{e} & 1342191656 & 1342191605 & 1342191749 & 1342192207 & $-$ \\
NGC1333-IRAS4B & 1342202064 & 1342203952 & 1342191722 & \phantom{\tablefootmark{e}}1342249851\tablefootmark{e} & 1342191655 & 1342191604 & 1342191750 & 1342202033 & 1342203179\\
L1527  & 1342192524 & $-$ & 1342203256 & $-$ & 1342216335 & 1342203156 & 1342203214 & 1342203188  & $-$\\
Ced110-IRS4 & 1342201525 & $-$ & 1342201734 & $-$ & 1342200765 & 1342201756 & 1342201544 & $-$ & $-$ \\
BHR71 & 1342201677 & $-$ & 1342201732 & $-$ & 1342200764 & 1342215915 & 1342201542 & 1342200755 & $-$ \\
\end{tabular}
\label{T:obsids}
\tablefoot{\tablefoottext{a}{Observation also contains H$_{2}$$^{18}$O 3$_{12}$-3$_{03}$ in same sideband.} \tablefoottext{b}{Observation also contains H$_{2}$$^{18}$O 1$_{11}$-0$_{00}$ in other sideband.} \tablefoottext{c}{Deep integration.} \tablefoottext{d}{Observed as part of OT2 programme OT2\_evandish\_4.}\tablefoottext{e}{Observed as part of OT2 programme OT2\_rvisser\_2.}}
\end{sidewaystable*}

\begin{sidewaystable*}[ph!]
\caption[]{Observed properties of ortho-water line profiles}
\centering
\begin{tabular}{l|cccc|cccc|cccc|cccc}
\hline
\hline \noalign {\smallskip}
\centering
Source & \multicolumn{4}{c|}{H$_{2}$O 1$_{10}$-1$_{01}$} & \multicolumn{4}{c|}{H$_{2}$O 2$_{12}$-1$_{01}$} & \multicolumn{4}{c|}{H$_{2}$O 3$_{12}$-2$_{21}$} & \multicolumn{4}{c}{H$_{2}$O 3$_{12}$-3$_{03}$} \\ \noalign {\smallskip}
&$\sigma_{rms}$&$T^{\mathrm{peak}}_{\mathrm{MB}}$&$\int T_{\mathrm{MB}}\,\mathrm{d}\varv$&$FWZI$&$\sigma_{rms}$&$T^{\mathrm{peak}}_{\mathrm{MB}}$&$\int T_{\mathrm{MB}}\,\mathrm{d}\varv$&$FWZI$&$\sigma_{rms}$&$T^{\mathrm{peak}}_{\mathrm{MB}}$&$\int T_{\mathrm{MB}}\,\mathrm{d}\varv$&$FWZI$&$\sigma_{rms}$&$T^{\mathrm{peak}}_{\mathrm{MB}}$&$\int T_{\mathrm{MB}}\,\mathrm{d}\varv$&$FWZI$\\\noalign {\smallskip}
&(mK)&(mK)&(K\kms{})&(\kms{})&(mK)&(mK)&(K\kms{})&(\kms{})&(mK)&(mK)&(K\kms{})&(\kms{})&(mK)&(mK)&(K\kms{})&(\kms{})\\\noalign {\smallskip}
\hline \noalign {\smallskip}
L1448-MM&\phantom{0}12&\phantom{0}330&19.31$\pm$0.08&153&\phantom{0}$-$\phantom{0}&\phantom{00}$-$\phantom{0}&\phantom{00.00}$-$\phantom{0.00}&\phantom{0}$-$\phantom{0}&103&\phantom{0}440&19.95$\pm$0.68&147&\phantom{0}22&\phantom{0}189&10.83$\pm$0.14&150\\
NGC1333-IRAS2A&\phantom{0}12&\phantom{0}325&\phantom{0}5.31$\pm$0.05&\phantom{0}69&123&\phantom{0}724&\phantom{0}8.83$\pm$0.52&\phantom{0}60&\phantom{0}99&\phantom{0}414&\phantom{0}5.40$\pm$0.40&\phantom{0}54&\phantom{0}11&\phantom{0}210&\phantom{0}4.40$\pm$0.05&\phantom{0}66\\
NGC1333-IRAS4A&\phantom{0}14&\phantom{0}918&18.17$\pm$0.08&\phantom{0}93&118&\phantom{0}850&24.12$\pm$0.60&\phantom{0}87&100&\phantom{0}656&16.52$\pm$0.46&\phantom{0}72&\phantom{0}10&\phantom{0}362&11.45$\pm$0.05&105\\
NGC1333-IRAS4B&\phantom{0}13&\phantom{0}712&10.64$\pm$0.06&\phantom{0}75&118&1804&29.39$\pm$0.58&\phantom{0}81&105&1486&23.27$\pm$0.54&\phantom{0}90&\phantom{0}10&\phantom{0}724&12.52$\pm$0.06&123\\
\end{tabular}
\label{T:int_ortho}
\end{sidewaystable*}

\begin{sidewaystable*}[ph!]
\caption[]{Observed properties of para-water line profiles}
\centering
\begin{tabular}{l|cccc|cccc|cccc}
\hline
\hline \noalign {\smallskip}
\centering
Source & \multicolumn{4}{c|}{H$_{2}$O 1$_{11}$-0$_{00}$} & \multicolumn{4}{c|}{H$_{2}$O 2$_{02}$-1$_{11}$} & \multicolumn{4}{c}{H$_{2}$O 2$_{11}$-2$_{02}$} \\
&$\sigma_{rms}$&$T^{\mathrm{peak}}_{\mathrm{MB}}$&$\int T_{\mathrm{MB}}\,\mathrm{d}\varv$&$FWZI$&$\sigma_{rms}$&$T^{\mathrm{peak}}_{\mathrm{MB}}$&$\int T_{\mathrm{MB}}\,\mathrm{d}\varv$&$FWZI$&$\sigma_{rms}$&$T^{\mathrm{peak}}_{\mathrm{MB}}$&$\int T_{\mathrm{MB}}\,\mathrm{d}\varv$&$FWZI$\\ \noalign {\smallskip}
&(mK)&(mK)&(K\kms{})&(\kms{})&(mK)&(mK)&(K\kms{})&(\kms{})&(mK)&(mK)&(K\kms{})&(\kms{})\\ \noalign {\smallskip}
\hline \noalign {\smallskip}
L1448-MM&\phantom{0}19&\phantom{0}393&22.51$\pm$0.13&156&\phantom{0}23&\phantom{0}408&25.13$\pm$0.16&159&\phantom{0}20&\phantom{0}252&14.11$\pm$0.13&144\\
NGC1333-IRAS2A&\phantom{0}19&\phantom{0}369&\phantom{0}5.44$\pm$0.09&\phantom{0}72&\phantom{0}23&\phantom{0}394&\phantom{0}8.43$\pm$0.10&\phantom{0}69&\phantom{0}21&\phantom{0}266&\phantom{0}6.69$\pm$0.10&\phantom{0}84\\
NGC1333-IRAS4A&\phantom{0}18&\phantom{0}720&19.56$\pm$0.11&108&\phantom{0}23&\phantom{0}771&22.60$\pm$0.12&\phantom{0}96&\phantom{0}20&\phantom{0}550&16.23$\pm$0.10&\phantom{0}96\\
NGC1333-IRAS4B&\phantom{0}18&1108&15.62$\pm$0.10&102&\phantom{0}22&1365&23.62$\pm$0.12&\phantom{0}96&\phantom{0}21&\phantom{0}855&13.99$\pm$0.11&\phantom{0}93\\
L1527&\phantom{0}18&\phantom{00}61&\phantom{0}0.57$\pm$0.06&\phantom{0}30&\phantom{0}22&\phantom{00}82&\phantom{0}1.05$\pm$0.08&\phantom{0}45&\phantom{0}21&$<$\phantom{0}62&$-$&$-$\\
\end{tabular}
\label{T:int_para}
\end{sidewaystable*}

\begin{sidewaystable*}[ph!]
\caption[]{Observed properties of H$_{2}$$^{18}$O line profiles}
\centering
\begin{tabular}{l|cccc|cccc|cccc|cccc}
\hline
\hline \noalign {\smallskip}
\centering
Source & \multicolumn{4}{c|}{H$_{2}$$^{18}$O 1$_{10}$-1$_{01}$} & \multicolumn{4}{c|}{H$_{2}$$^{18}$O 3$_{12}$-3$_{03}$} & \multicolumn{4}{c|}{H$_{2}$$^{18}$O 1$_{11}$-0$_{00}$} & \multicolumn{4}{c}{H$_{2}$$^{18}$O 2$_{02}$-1$_{11}$} \\
&$\sigma_{rms}$&$T^{\mathrm{peak}}_{\mathrm{MB}}$&$\int T_{\mathrm{MB}}\,\mathrm{d}\varv$&$FWZI$&$\sigma_{rms}$&$T^{\mathrm{peak}}_{\mathrm{MB}}$&$\int T_{\mathrm{MB}}\,\mathrm{d}\varv$&$FWZI$&$\sigma_{rms}$&$T^{\mathrm{peak}}_{\mathrm{MB}}$&$\int T_{\mathrm{MB}}\,\mathrm{d}\varv$&$FWZI$&$\sigma_{rms}$&$T^{\mathrm{peak}}_{\mathrm{MB}}$&$\int T_{\mathrm{MB}}\,\mathrm{d}\varv$&$FWZI$\\\noalign {\smallskip}
&(mK)&(mK)&(K\kms{})&(\kms{})&(mK)&(mK)&(K\kms{})&(\kms{})&(mK)&(mK)&(K\kms{})&(\kms{})&(mK)&(mK)&(K\kms{})&(\kms{})\\\noalign {\smallskip}
\hline \noalign {\smallskip}
L1448-MM&\phantom{00}4&$<$\phantom{0}13&\phantom{00.00}$-$\phantom{0.00}&\phantom{0}$-$\phantom{0}&\phantom{0}21&$<$\phantom{0}62&\phantom{00.00}$-$\phantom{0.00}&\phantom{0}$-$\phantom{0}&\phantom{0}18&$<$\phantom{0}54&\phantom{00.00}$-$\phantom{0.00}&\phantom{0}$-$\phantom{0}&\phantom{0}$-$\phantom{0}&\phantom{00}$-$\phantom{0}&\phantom{00.00}$-$\phantom{0.00}&\phantom{0}$-$\phantom{0}\\
NGC1333-IRAS2A&\phantom{00}4&$<$\phantom{0}10&\phantom{00.00}$-$\phantom{0.00}&\phantom{0}$-$\phantom{0}&\phantom{00}8&\phantom{00}27&\phantom{0}0.09$\pm$0.02&\phantom{0}12&\phantom{0}19&$<$\phantom{0}56&\phantom{00.00}$-$\phantom{0.00}&\phantom{0}$-$\phantom{0}&\phantom{0}16&$<$\phantom{0}49&\phantom{00.00}$-$\phantom{0.00}&\phantom{0}$-$\phantom{0}\\
\end{tabular}
\label{T:int_h218o}
\end{sidewaystable*}

\clearpage

\begin{sidewaystable*}[ph!]
\caption[]{Gaussian decomposition results}
\centering
\begin{tabular}{lcccccccccccc}
\hline
\hline \noalign {\smallskip}
\centering
Source & Comp.\tablefootmark{a} & FWHM & $\varv_{\mathrm{peak}}$ & $\varv_{\mathrm{source}}$ & \multicolumn{2}{c}{H$_{2}$O 1$_{10}$-1$_{01}$} & \multicolumn{2}{c}{H$_{2}$O 2$_{12}$-1$_{01}$} & \multicolumn{2}{c}{H$_{2}$O 3$_{12}$-2$_{21}$} & \multicolumn{2}{c}{H$_{2}$O 3$_{12}$-3$_{03}$}\\\noalign {\smallskip}
&&&&&$T^{\mathrm{peak}}_{\mathrm{MB}}$&$\int T_{\mathrm{MB}}\,\mathrm{d}\varv$&$T^{\mathrm{peak}}_{\mathrm{MB}}$&$\int T_{\mathrm{MB}}\,\mathrm{d}\varv$&$T^{\mathrm{peak}}_{\mathrm{MB}}$&$\int T_{\mathrm{MB}}\,\mathrm{d}\varv$&$T^{\mathrm{peak}}_{\mathrm{MB}}$&$\int T_{\mathrm{MB}}\,\mathrm{d}\varv$\\\noalign {\smallskip}
&&(\kms{})&(\kms{})&(\kms{})&(K)&(K\kms{})&(K)&(K\kms{})&(K)&(K\kms{})&(K)&(K\kms{})\\\noalign {\smallskip}
\hline \noalign {\smallskip}
L1448-MM&S&\phantom{0}7.8$\pm$ 0.8&\phantom{0}4.6$\pm$0.3&\phantom{0}5.2&\phantom{0}$-$0.12$\pm$0.02\phantom{$-$}&\phantom{0}$-$1.01$\pm$0.21\phantom{$-$}&$-$&$-$&$<$\phantom{0}0.26\phantom{$<$}&$<$\phantom{0}5.02\phantom{$<$}&\phantom{0}$-$0.07$\pm$0.02\phantom{$-$}&\phantom{0}$-$0.59$\pm$0.17\phantom{$-$}\\
&S&23.0$\pm$ 1.7&59.6$\pm$0.7&\phantom{0}5.2&\phantom{0}0.12$\pm$0.01&\phantom{0}3.00$\pm$0.37&$-$&$-$&\phantom{0}0.15$\pm$0.03&\phantom{0}3.71$\pm$0.73&\phantom{0}0.07$\pm$0.01&\phantom{0}1.82$\pm$0.28\\
&S&39.8$\pm$ 3.7&$-$41.5$\pm$1.6\phantom{$-$}&\phantom{0}5.2&\phantom{0}0.08$\pm$0.01&\phantom{0}3.20$\pm$0.46&$-$&$-$&\phantom{0}0.19$\pm$0.02&\phantom{0}7.88$\pm$1.18&\phantom{0}0.06$\pm$0.01&\phantom{0}2.68$\pm$0.40\\
&C&44.6$\pm$ 2.0&16.8$\pm$0.7&\phantom{0}5.2&\phantom{0}0.30$\pm$0.01&14.25$\pm$0.79&$-$&$-$&\phantom{0}0.18$\pm$0.02&\phantom{0}8.70$\pm$1.02&\phantom{0}0.15$\pm$0.01&\phantom{0}6.99$\pm$0.50\\
\end{tabular}
\label{T:Gaussians_class0_o}
\tablefoot{\tablefoottext{a}{Component type: E=envelope, C=cavity shock and S=spot shock.}\tablefoottext{b}{Non-gaussian combination of absorption and emission so excluded from fit.}}
\end{sidewaystable*}

\begin{sidewaystable*}[ph!]
\caption[]{Gaussian decomposition results}
\centering
\begin{tabular}{lcccccccccc}
\hline
\hline \noalign {\smallskip}
\centering
Source & Comp.\tablefootmark{a} & FWHM & $\varv_{\mathrm{peak}}$ & $\varv_{\mathrm{source}}$ &\multicolumn{2}{c}{H$_{2}$O 1$_{11}$-0$_{00}$} & \multicolumn{2}{c}{H$_{2}$O 2$_{02}$-1$_{11}$} & \multicolumn{2}{c}{H$_{2}$O 2$_{11}$-2$_{02}$}\\\noalign {\smallskip}
&&&&&$T^{\mathrm{peak}}_{\mathrm{MB}}$&$\int T_{\mathrm{MB}}\,\mathrm{d}\varv$&$T^{\mathrm{peak}}_{\mathrm{MB}}$&$\int T_{\mathrm{MB}}\,\mathrm{d}\varv$&$T^{\mathrm{peak}}_{\mathrm{MB}}$&$\int T_{\mathrm{MB}}\,\mathrm{d}\varv$\\\noalign {\smallskip}
&&(\kms{})&(\kms{})&(\kms{})&(K)&(K\kms{})&(K)&(K\kms{})&(K)&(K\kms{})\\\noalign {\smallskip}
\hline \noalign {\smallskip}
L1448-MM&S&\phantom{0}7.8$\pm$ 0.8&\phantom{0}4.6$\pm$0.3&\phantom{0}5.2&\phantom{0}$-$0.27$\pm$0.03\phantom{$-$}&\phantom{0}$-$2.20$\pm$0.34\phantom{$-$}&\phantom{0}$-$0.21$\pm$0.04\phantom{$-$}&\phantom{0}$-$1.75$\pm$0.37\phantom{$-$}&\phantom{0}$-$0.12$\pm$0.03\phantom{$-$}&\phantom{0}$-$1.00$\pm$0.30\phantom{$-$}\\
&S&23.0$\pm$ 1.7&59.6$\pm$0.7&\phantom{0}5.2&\phantom{0}0.18$\pm$0.02&\phantom{0}4.46$\pm$0.51&\phantom{0}0.21$\pm$0.02&\phantom{0}5.23$\pm$0.54&\phantom{0}0.10$\pm$0.01&\phantom{0}2.39$\pm$0.40\\
&S&39.8$\pm$ 3.7&$-$41.5$\pm$1.6\phantom{$-$}&\phantom{0}5.2&\phantom{0}0.09$\pm$0.01&\phantom{0}3.83$\pm$0.59&\phantom{0}0.11$\pm$0.01&\phantom{0}4.65$\pm$0.64&\phantom{0}0.07$\pm$0.01&\phantom{0}2.87$\pm$0.54\\
&C&44.6$\pm$ 2.0&16.8$\pm$0.7&\phantom{0}5.2&\phantom{0}0.36$\pm$0.01&17.02$\pm$0.99&\phantom{0}0.35$\pm$0.01&16.81$\pm$0.97&\phantom{0}0.21$\pm$0.01&\phantom{0}9.98$\pm$0.73\\
IRAS2A&E&\phantom{0}4.9$\pm$ 0.8&\phantom{0}8.0$\pm$0.3&\phantom{0}7.7&\phantom{\tablefootmark{b}}$-$\tablefootmark{b}&\phantom{\tablefootmark{b}}$-$\tablefootmark{b}&$<$\phantom{0}0.06\phantom{$<$}&$<$\phantom{0}0.76\phantom{$<$}&\phantom{0}0.07$\pm$0.03&\phantom{0}0.38$\pm$0.18\\
\end{tabular}
\label{T:Gaussians_class0_p}
\tablefoot{\tablefoottext{a}{Component type: E=envelope, C=cavity shock and S=spot shock.}\tablefoottext{b}{Non-gaussian combination of absorption and emission so excluded from fit.}}
\end{sidewaystable*}

\begin{sidewaystable*}[ph!]
\caption[]{Gaussian decomposition results}
\centering
\begin{tabular}{lcccccccccccc}
\hline
\hline \noalign {\smallskip}
\centering
Source & Comp.\tablefootmark{a} & FWHM & $\varv_{\mathrm{peak}}$ & $\varv_{\mathrm{source}}$ & \multicolumn{2}{c}{H$_{2}$O 1$_{10}$-1$_{01}$} & \multicolumn{2}{c}{H$_{2}$O 2$_{12}$-1$_{01}$} & \multicolumn{2}{c}{H$_{2}$O 3$_{12}$-2$_{21}$} & \multicolumn{2}{c}{H$_{2}$O 3$_{12}$-3$_{03}$}\\\noalign {\smallskip}
&&&&&$T^{\mathrm{peak}}_{\mathrm{MB}}$&$\int T_{\mathrm{MB}}\,\mathrm{d}\varv$&$T^{\mathrm{peak}}_{\mathrm{MB}}$&$\int T_{\mathrm{MB}}\,\mathrm{d}\varv$&$T^{\mathrm{peak}}_{\mathrm{MB}}$&$\int T_{\mathrm{MB}}\,\mathrm{d}\varv$&$T^{\mathrm{peak}}_{\mathrm{MB}}$&$\int T_{\mathrm{MB}}\,\mathrm{d}\varv$\\\noalign {\smallskip}
&&(\kms{})&(\kms{})&(\kms{})&(K)&(K\kms{})&(K)&(K\kms{})&(K)&(K\kms{})&(K)&(K\kms{})\\\noalign {\smallskip}
\hline \noalign {\smallskip}
IRAS3A&E&\phantom{0}1.5$\pm$ 0.1&\phantom{0}8.8$\pm$0.0&\phantom{0}8.5&\phantom{0}$-$0.20$\pm$0.01\phantom{$-$}&\phantom{0}$-$0.31$\pm$0.03\phantom{$-$}&$-$&$-$&$-$&$-$&$-$&$-$\\
&S&\phantom{0}6.1$\pm$ 0.3&\phantom{0}5.4$\pm$0.1&\phantom{0}8.5&\phantom{0}$-$0.19$\pm$0.00\phantom{$-$}&\phantom{0}$-$1.23$\pm$0.06\phantom{$-$}&$-$&$-$&$-$&$-$&$-$&$-$\\
&C&28.8$\pm$ 0.3&\phantom{0}5.9$\pm$0.1&\phantom{0}8.5&\phantom{0}0.26$\pm$0.00&\phantom{0}8.11$\pm$0.14&$-$&$-$&$-$&$-$&$-$&$-$\\
L1489&C&20.0$\pm$ 1.5&\phantom{0}3.7$\pm$0.6&\phantom{0}7.2&\phantom{0}0.04$\pm$0.00&\phantom{0}0.93$\pm$0.12&$-$&$-$&$<$\phantom{0}0.22\phantom{$<$}&$<$11.27\phantom{$<$}&$-$&$-$\\
L1551-IRS5&E&\phantom{0}4.3$\pm$ 0.8&\phantom{0}6.7$\pm$0.3&\phantom{0}6.2&\phantom{\tablefootmark{b}}$-$\tablefootmark{b}&\phantom{\tablefootmark{b}}$-$\tablefootmark{b}&$-$&$-$&$<$\phantom{0}0.24\phantom{$<$}&$<$\phantom{0}2.54\phantom{$<$}&$-$&$-$\\
\end{tabular}
\label{T:Gaussians_class1_o}
\tablefoot{\tablefoottext{a}{Component type: E=envelope, C=cavity shock and S=spot shock.}\tablefoottext{b}{Non-gaussian combination of absorption and emission so excluded from fit.}}
\end{sidewaystable*}

\begin{sidewaystable*}[ph!]
\caption[]{Gaussian decomposition results}
\centering
\begin{tabular}{lcccccccccc}
\hline
\hline \noalign {\smallskip}
\centering
Source & Comp.\tablefootmark{a} & FWHM & $\varv_{\mathrm{peak}}$ & $\varv_{\mathrm{source}}$ &\multicolumn{2}{c}{H$_{2}$O 1$_{11}$-0$_{00}$} & \multicolumn{2}{c}{H$_{2}$O 2$_{02}$-1$_{11}$} & \multicolumn{2}{c}{H$_{2}$O 2$_{11}$-2$_{02}$}\\\noalign {\smallskip}
&&&&&$T^{\mathrm{peak}}_{\mathrm{MB}}$&$\int T_{\mathrm{MB}}\,\mathrm{d}\varv$&$T^{\mathrm{peak}}_{\mathrm{MB}}$&$\int T_{\mathrm{MB}}\,\mathrm{d}\varv$&$T^{\mathrm{peak}}_{\mathrm{MB}}$&$\int T_{\mathrm{MB}}\,\mathrm{d}\varv$\\\noalign {\smallskip}
&&(\kms{})&(\kms{})&(\kms{})&(K)&(K\kms{})&(K)&(K\kms{})&(K)&(K\kms{})\\\noalign {\smallskip}
\hline \noalign {\smallskip}
IRAS3A&E&\phantom{0}1.5$\pm$ 0.1&\phantom{0}8.8$\pm$0.0&\phantom{0}8.5&$-$&$-$&$-$&$-$&$-$&$-$\\
&S&\phantom{0}6.1$\pm$ 0.3&\phantom{0}5.4$\pm$0.1&\phantom{0}8.5&$-$&$-$&$-$&$-$&$-$&$-$\\
&C&28.8$\pm$ 0.3&\phantom{0}5.9$\pm$0.1&\phantom{0}8.5&$-$&$-$&$-$&$-$&$-$&$-$\\
L1489&C&20.0$\pm$ 1.5&\phantom{0}3.7$\pm$0.6&\phantom{0}7.2&\phantom{0}0.07$\pm$0.01&\phantom{0}1.42$\pm$0.20&\phantom{0}0.07$\pm$0.01&\phantom{0}1.46$\pm$0.20&\phantom{0}0.05$\pm$0.01&\phantom{0}0.96$\pm$0.15\\
L1551-IRS5&E&\phantom{0}4.3$\pm$ 0.8&\phantom{0}6.7$\pm$0.3&\phantom{0}6.2&\phantom{\tablefootmark{b}}$-$\tablefootmark{b}&\phantom{\tablefootmark{b}}$-$\tablefootmark{b}&\phantom{0}0.13$\pm$0.02&\phantom{0}0.59$\pm$0.15&\phantom{0}0.07$\pm$0.02&\phantom{0}0.34$\pm$0.10\\
\end{tabular}
\label{T:Gaussians_class1_p}
\tablefoot{\tablefoottext{a}{Component type: E=envelope, C=cavity shock and S=spot shock.}\tablefoottext{b}{Non-gaussian combination of absorption and emission so excluded from fit.}}
\end{sidewaystable*}

\begin{sidewaystable*}[ph!]
\caption[]{Gaussian decomposition results}
\centering
\begin{tabular}{lcccccccccccc}
\hline
\hline \noalign {\smallskip}
\centering
Source & Comp.\tablefootmark{a} & FWHM & $\varv_{\mathrm{peak}}$ & $\varv_{\mathrm{source}}$ & \multicolumn{2}{c}{H$_{2}^{18}$O 1$_{10}$-1$_{01}$} & \multicolumn{2}{c}{H$_{2}^{18}$O 3$_{12}$-3$_{03}$} & \multicolumn{2}{c}{H$_{2}^{18}$O 1$_{11}$-0$_{00}$} & \multicolumn{2}{c}{H$_{2}^{18}$O 2$_{02}$-1$_{11}$}\\\noalign {\smallskip}
&&&&&$T^{\mathrm{peak}}_{\mathrm{MB}}$&$\int T_{\mathrm{MB}}\,\mathrm{d}\varv$&$T^{\mathrm{peak}}_{\mathrm{MB}}$&$\int T_{\mathrm{MB}}\,\mathrm{d}\varv$&$T^{\mathrm{peak}}_{\mathrm{MB}}$&$\int T_{\mathrm{MB}}\,\mathrm{d}\varv$&$T^{\mathrm{peak}}_{\mathrm{MB}}$&$\int T_{\mathrm{MB}}\,\mathrm{d}\varv$\\\noalign {\smallskip}
&&(\kms{})&(\kms{})&(\kms{})&(K)&(K\kms{})&(K)&(K\kms{})&(K)&(K\kms{})&(K)&(K\kms{})\\\noalign {\smallskip}
\hline \noalign {\smallskip}
IRAS2A&E&\phantom{0}4.9$\pm$ 0.8&\phantom{0}8.0$\pm$0.3&\phantom{0}7.7&$<$\phantom{0}0.011\phantom{$<$}&$<$\phantom{0}0.139\phantom{$<$}&\phantom{0}0.015$\pm$0.008&\phantom{0}0.079$\pm$0.045&$<$\phantom{0}0.056\phantom{$<$}&$<$\phantom{0}0.684\phantom{$<$}&$<$\phantom{0}0.048\phantom{$<$}&$<$\phantom{0}0.585\phantom{$<$}\\
&C&14.0$\pm$ 0.7&11.5$\pm$0.3&\phantom{0}7.7&$<$\phantom{0}0.011\phantom{$<$}&$<$\phantom{0}0.399\phantom{$<$}&$<$\phantom{0}0.023\phantom{$<$}&$<$\phantom{0}0.802\phantom{$<$}&$<$\phantom{0}0.056\phantom{$<$}&$<$\phantom{0}1.966\phantom{$<$}&$<$\phantom{0}0.048\phantom{$<$}&$<$\phantom{0}1.681\phantom{$<$}\\
&S&39.2$\pm$ 2.3&\phantom{0}$-$5.4$\pm$1.3\phantom{$-$}&\phantom{0}7.7&$<$\phantom{0}0.011\phantom{$<$}&$<$\phantom{0}1.121\phantom{$<$}&$<$\phantom{0}0.023\phantom{$<$}&$<$\phantom{0}2.252\phantom{$<$}&$<$\phantom{0}0.056\phantom{$<$}&$<$\phantom{0}5.517\phantom{$<$}&$<$\phantom{0}0.048\phantom{$<$}&$<$\phantom{0}4.718\phantom{$<$}\\
\end{tabular}
\label{T:Gaussians_h2_18o}
\tablefoot{\tablefoottext{a}{Component type: E=envelope, C=cavity shock and S=spot shock.}}
\end{sidewaystable*}

\begin{figure*}
\begin{center}
\includegraphics[width=0.9\textwidth]{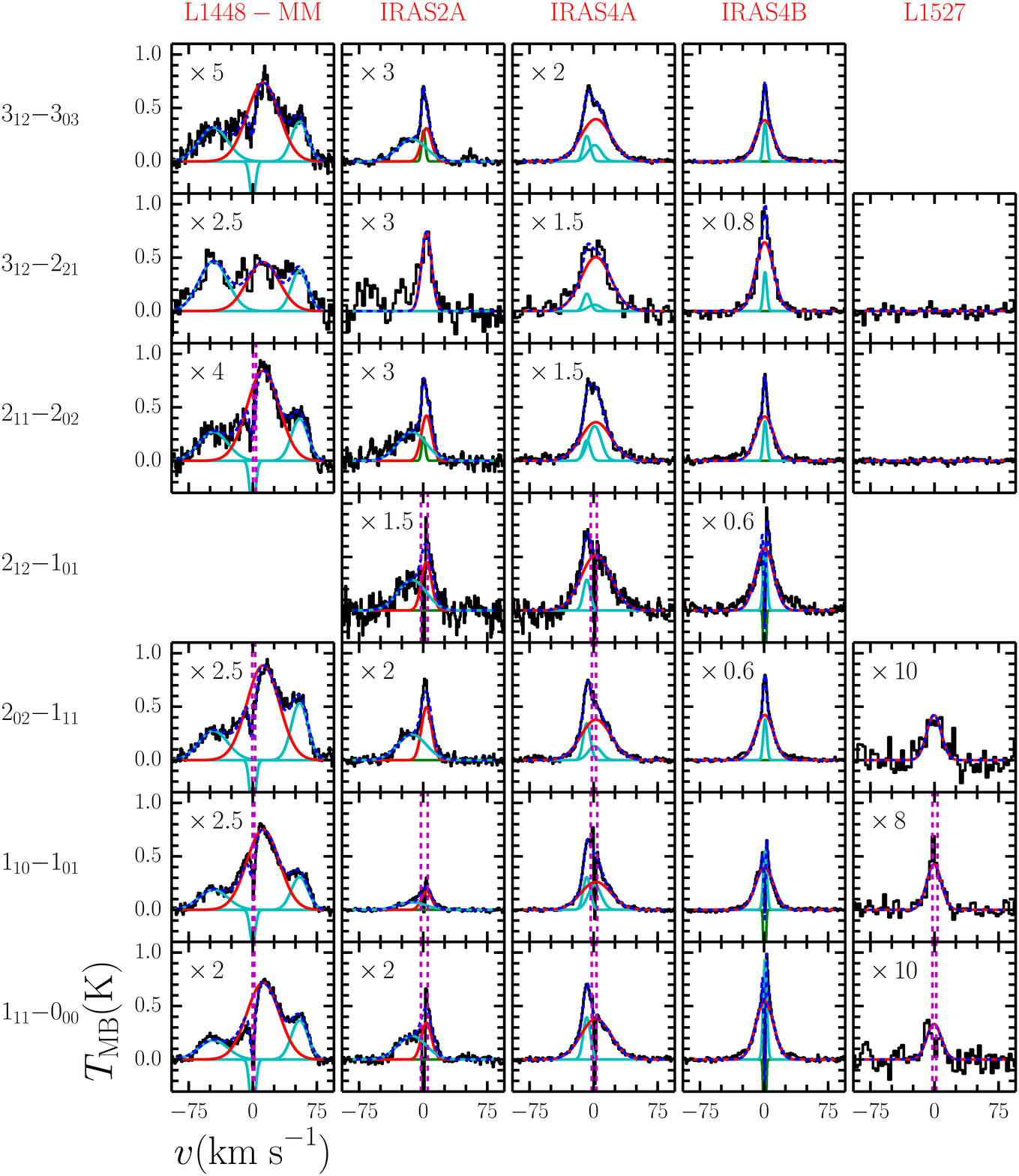}
\caption{Observed H$_{2}$O spectra for Class 0 sources in the WISH sample. All spectra are continuum-subtracted and have been recentred so that the source velocity is at 0\kms{} for ease of comparison. Some spectra have also been resampled to a lower velocity resolution for ease of comparison. The red, cyan and green lines show the individual Gaussian components for the cavity shock, spot shocks, and quiescent envelope respectively (see text and Table~\ref{T:results_components_decomposition_previous} for details) while the blue dashed line shows the combined fit for each line. The vertical dashed magenta lines indicate regions which are masked during the fitting.}
\label{F:H2O_class0_1}
\end{center}
\end{figure*}

\begin{figure*}
\begin{center}
\includegraphics[width=0.9\textwidth]{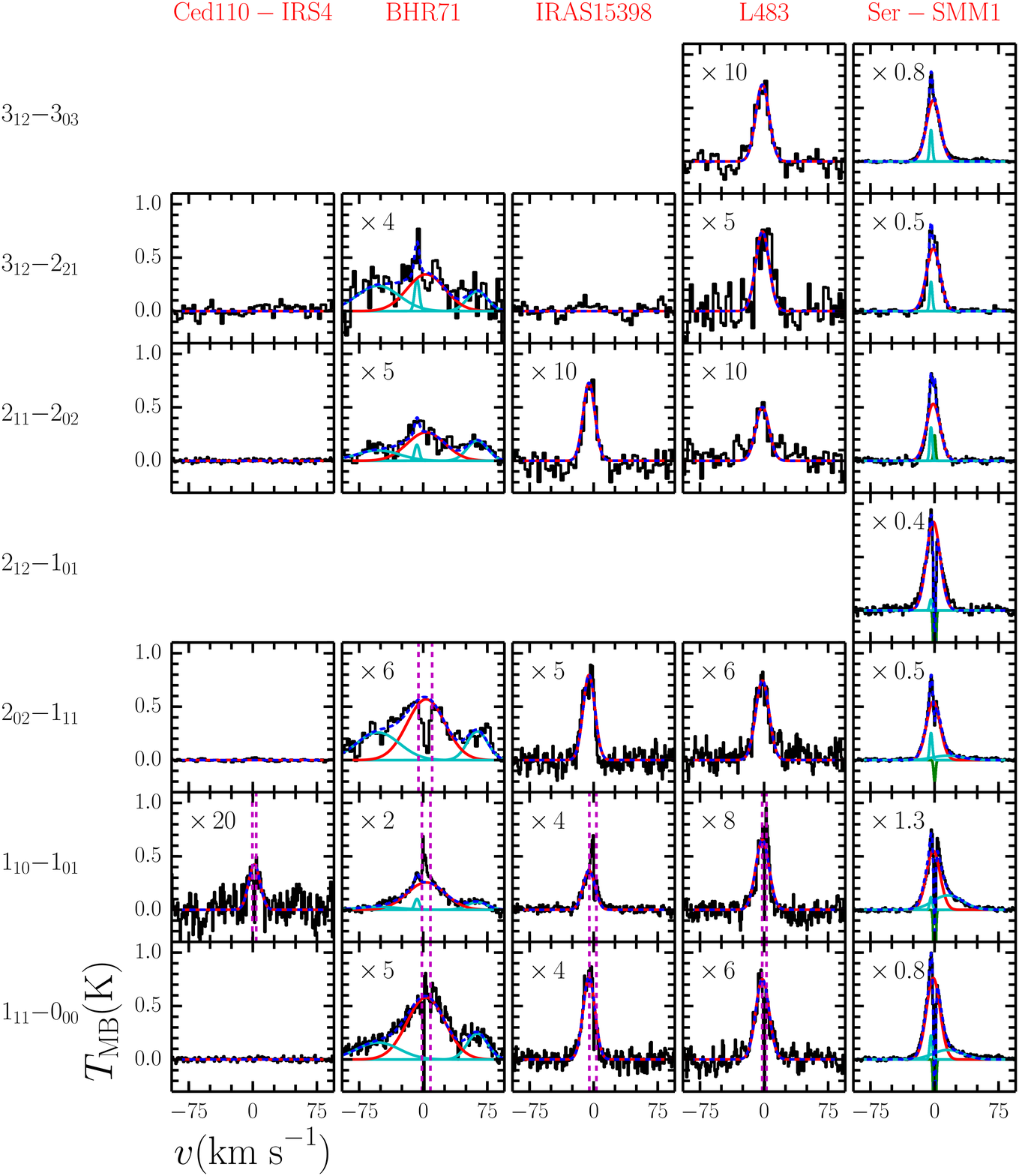}
\caption{As in Figure~\ref{F:H2O_class0_1}}
\label{F:H2O_class0_2}
\end{center}
\end{figure*}

\begin{figure*}
\begin{center}
\includegraphics[width=0.9\textwidth]{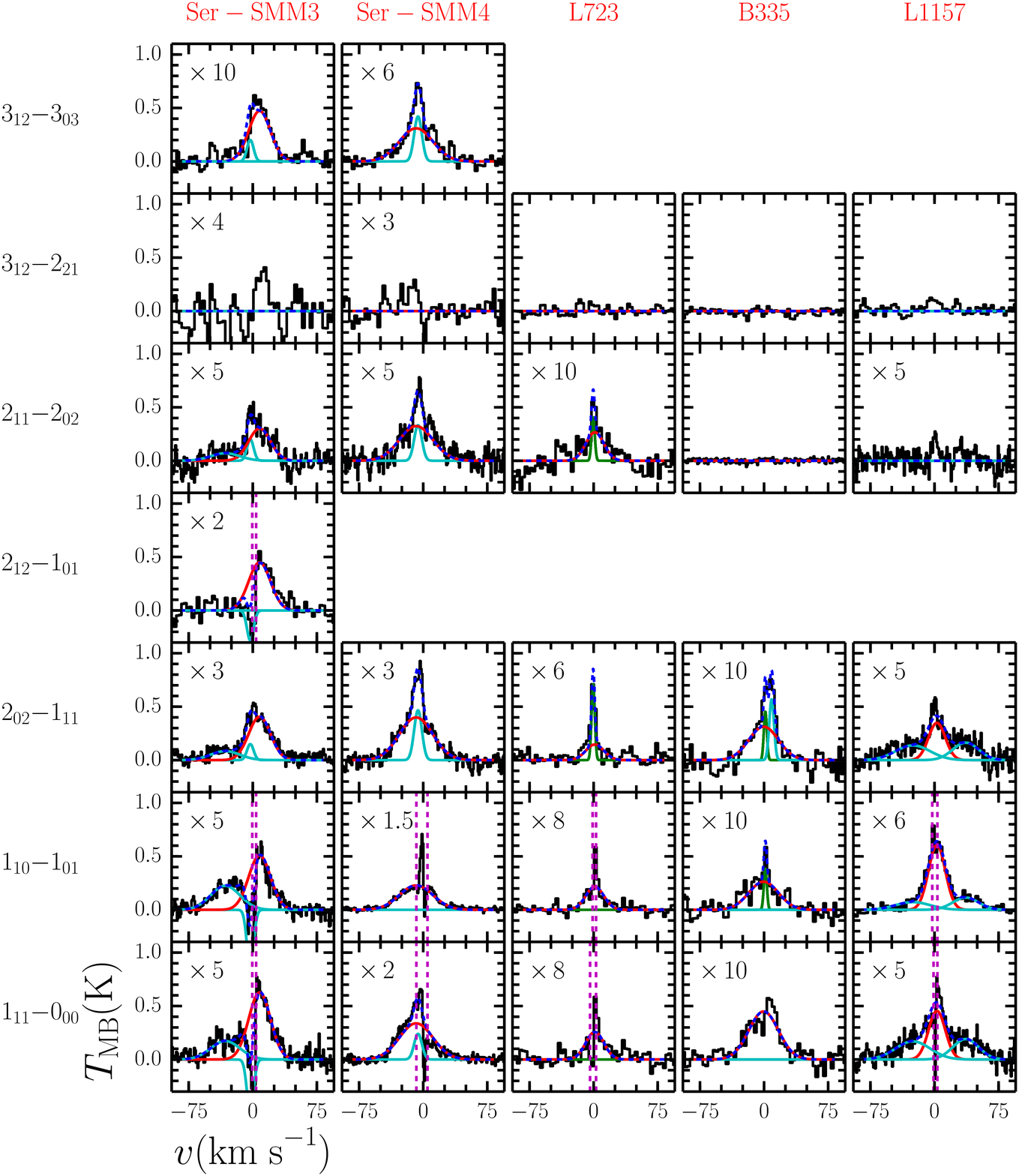}
\caption{As in Figure~\ref{F:H2O_class0_1}}
\label{F:H2O_class0_3}
\end{center}
\end{figure*}

\begin{figure*}
\begin{center}
\includegraphics[width=0.9\textwidth]{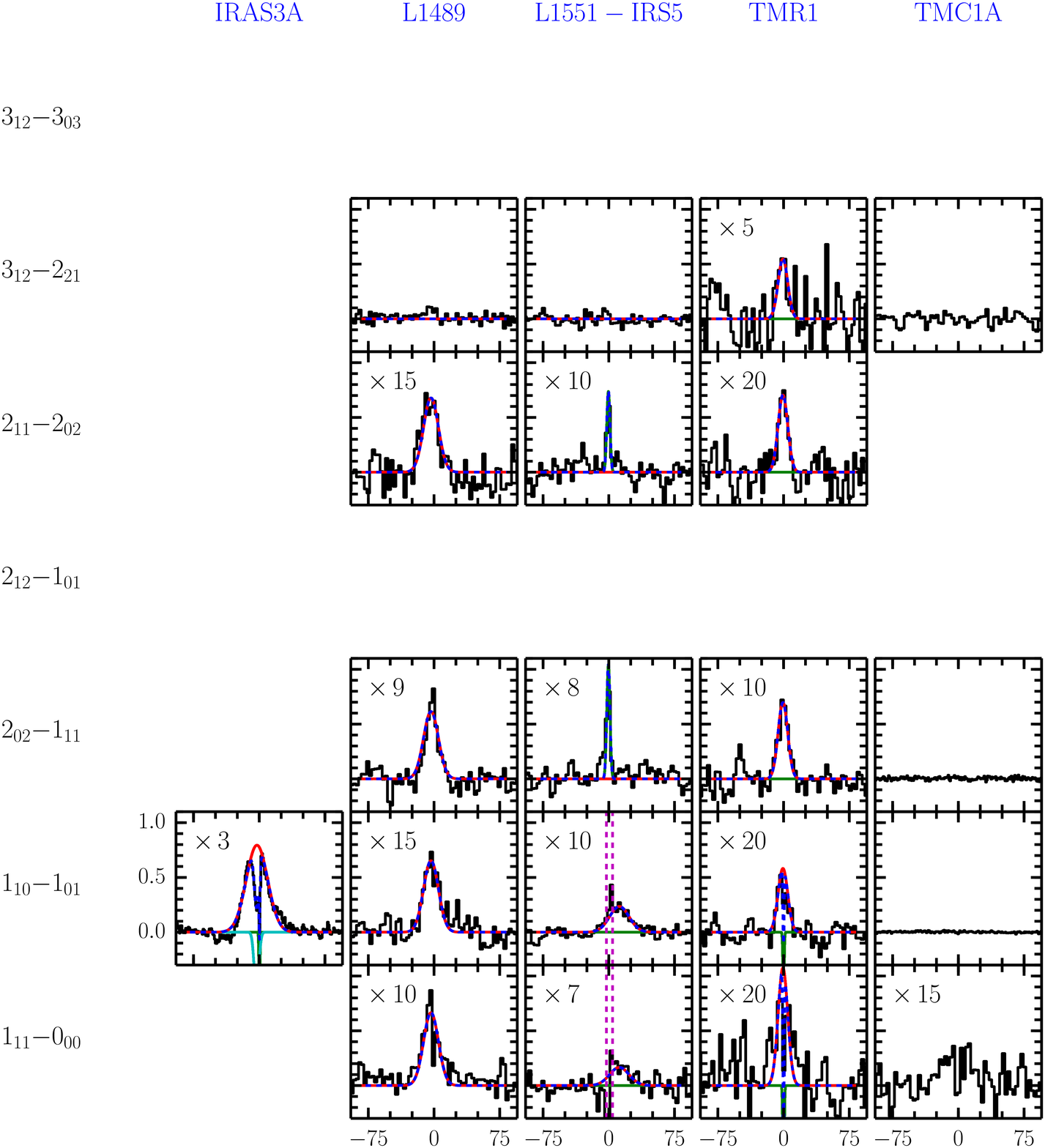}
\caption{As in Figure~\ref{F:H2O_class0_1} but for Class I sources.}
\label{F:H2O_classI_1}
\end{center}
\end{figure*}

\begin{figure*}
\begin{center}
\includegraphics[width=0.9\textwidth]{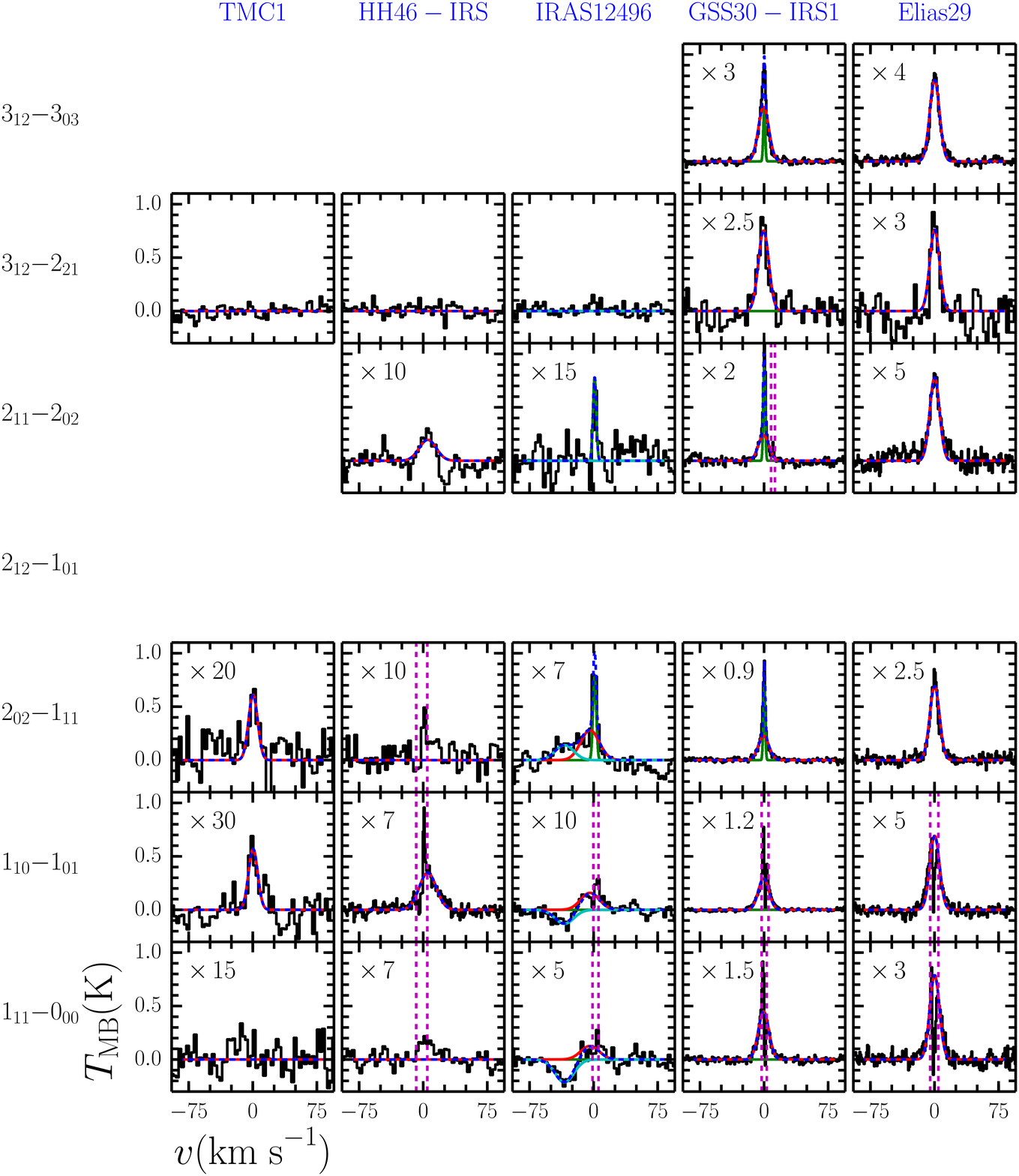}
\caption{As in Figure~\ref{F:H2O_class0_1} but for Class I sources.}
\label{F:H2O_classI_2}
\end{center}
\end{figure*}

\begin{figure*}
\begin{center}
\includegraphics[width=0.9\textwidth]{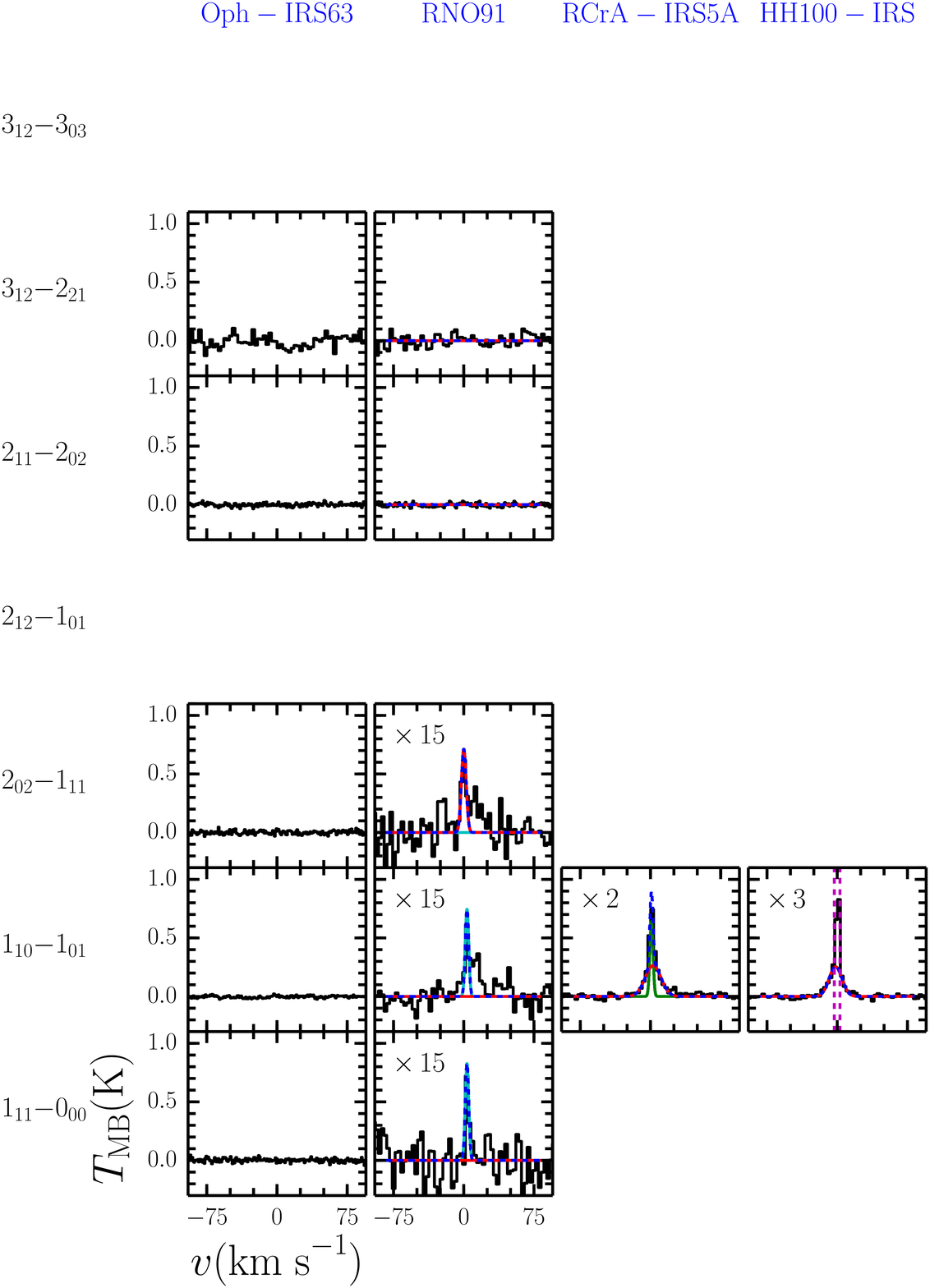}
\caption{As in Figure~\ref{F:H2O_class0_1} but for Class I sources.}
\label{F:H2O_classI_3}
\end{center}
\end{figure*}

\begin{figure*}
\begin{center}
\includegraphics[width=0.9\textwidth]{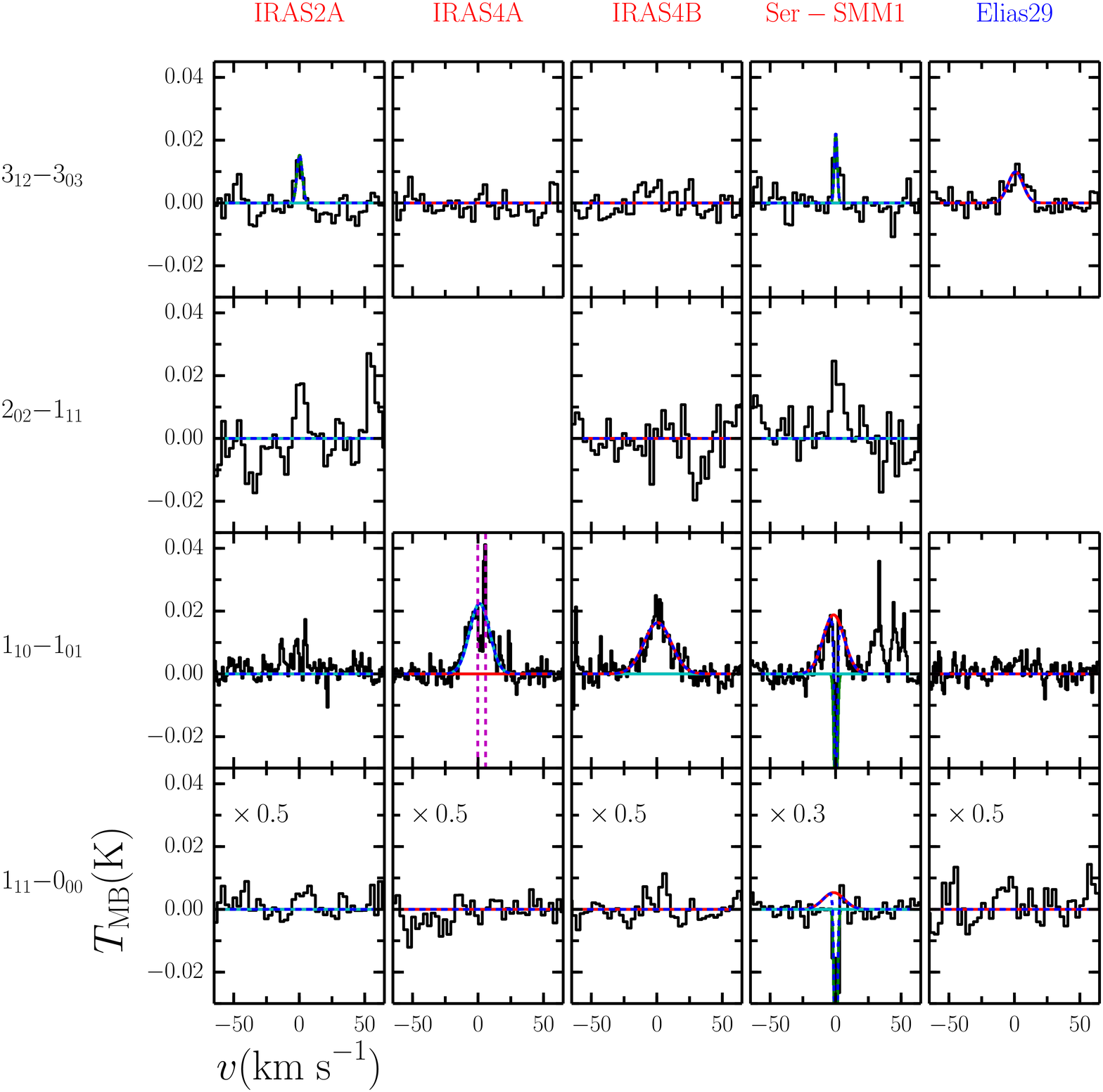}
\caption{Observed H$_{2}^{18}$O spectra for those Class 0 (red) and Class I (blue) sources with detections in at least one of the observed transitions. All spectra are continuum-subtracted and have been recentred so that the source velocity is at 0\kms{} for ease of comparison. Some spectra have also been resampled to a lower velocity resolution for ease of comparison. The red, cyan and green lines show the individual Gaussian components for the cavity shock, spot shocks, and quiescent envelope respectively (see text and Table~\ref{T:results_components_decomposition_previous} for details) while the blue dashed line shows the combined fit for each line. The vertical dashed magenta lines indicate regions which are masked during the fitting. The 1$_{10}-$1$_{01}$ observations also include the CH triplet which is in the other sideband.}
\label{F:H218O_detections}
\end{center}
\end{figure*}

\clearpage

\begin{figure*}
\begin{center}
\includegraphics[width=0.90\textwidth]{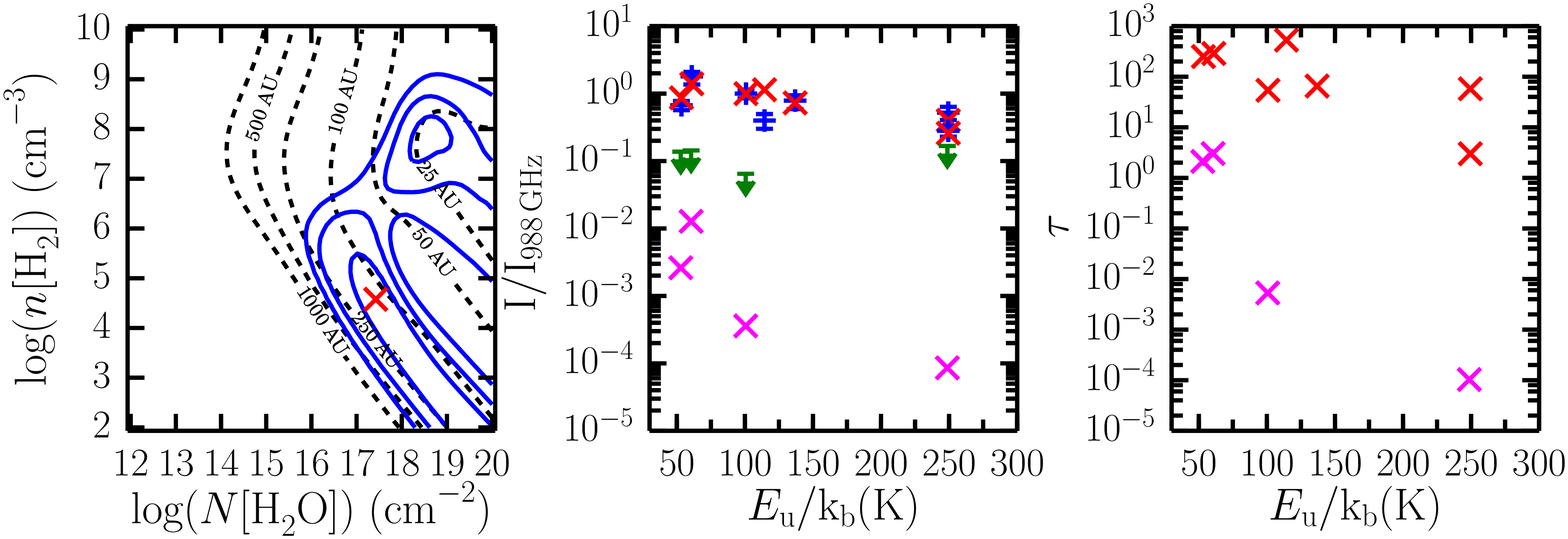}
\includegraphics[width=0.90\textwidth]{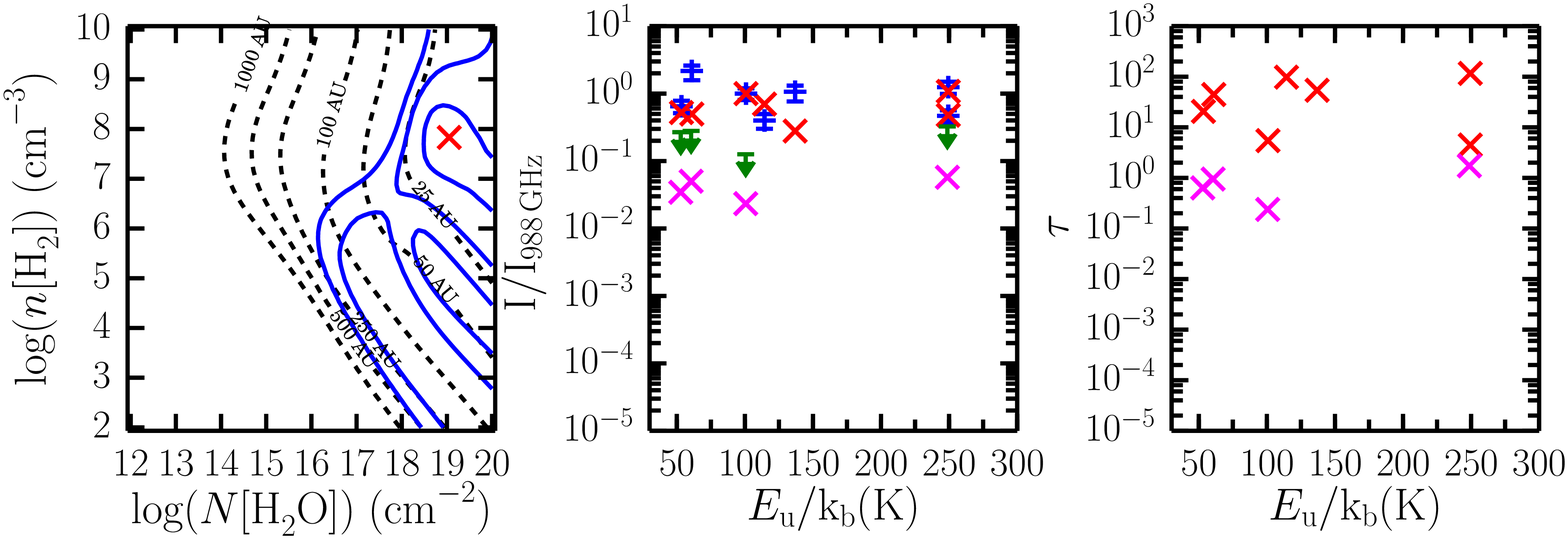}
\includegraphics[width=0.90\textwidth]{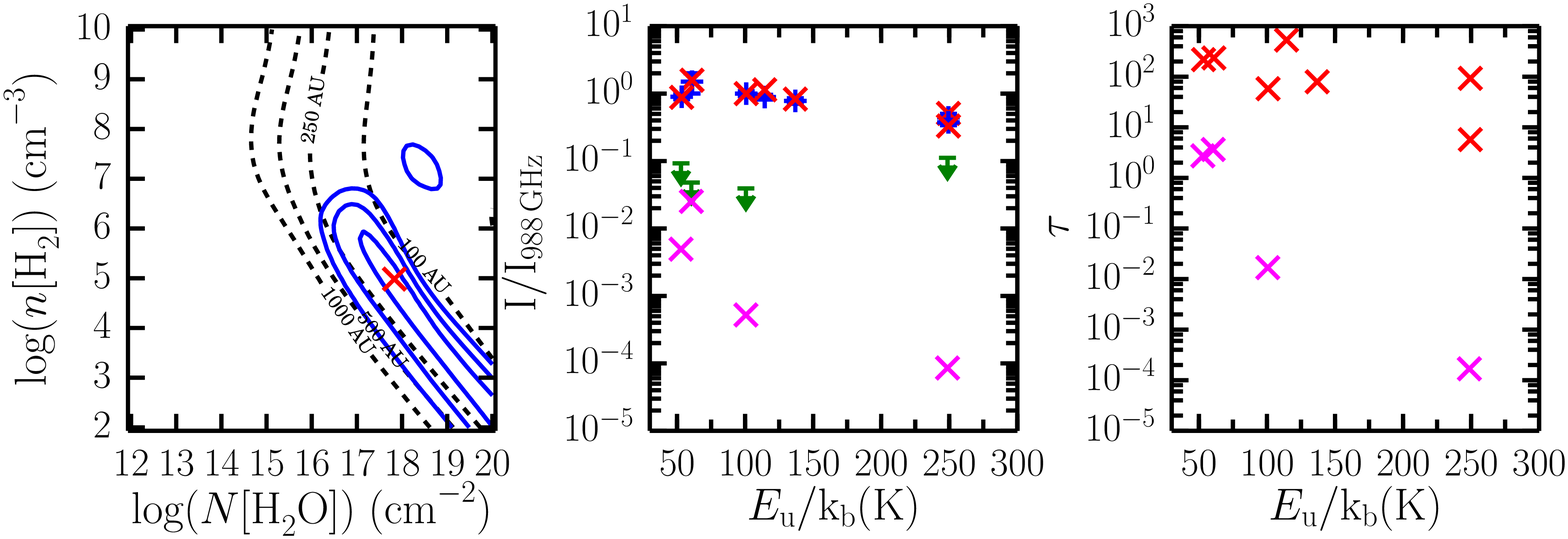}
\includegraphics[width=0.90\textwidth]{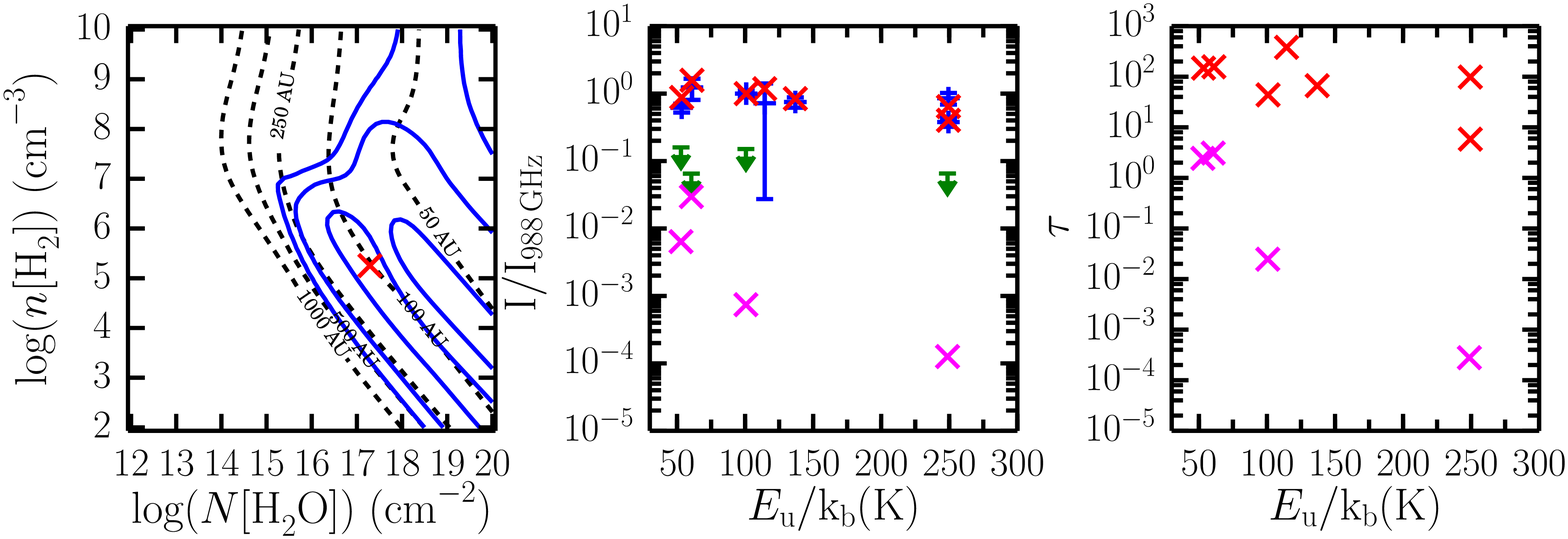}
\caption{\textsc{radex} results for the FWHM=23.0 and 39.8\kms{} spot shocks for L1448-MM (top and upper-middle) and the cavity shock (lower-middle) for L1448-MM and the cavity shock component for IRAS2A (bottom). The left-hand panels show the best-fit (red cross) and 1, 3 and 5$\sigma$ confidence limits (blue contours) for a grid in $n_{\mathrm{H}_{2}}$ and $N_{\mathrm{H}_{2}\mathrm{O}}$. The black dashed contours show the corresponding radius of the emitting region. The middle panels show a spectral line energy distribution comparing the observed (blue for H$_{2}$O, green for H$_{2}^{18}$O) and best-fit model (red for H$_{2}$O, magenta for H$_{2}^{18}$O) results. The right-hand panels show the optical depth for each line for the best-fit model.}
\label{F:analysis_excitation_results_full1}
\end{center}
\end{figure*}

\begin{figure*}
\begin{center}
\includegraphics[width=0.90\textwidth]{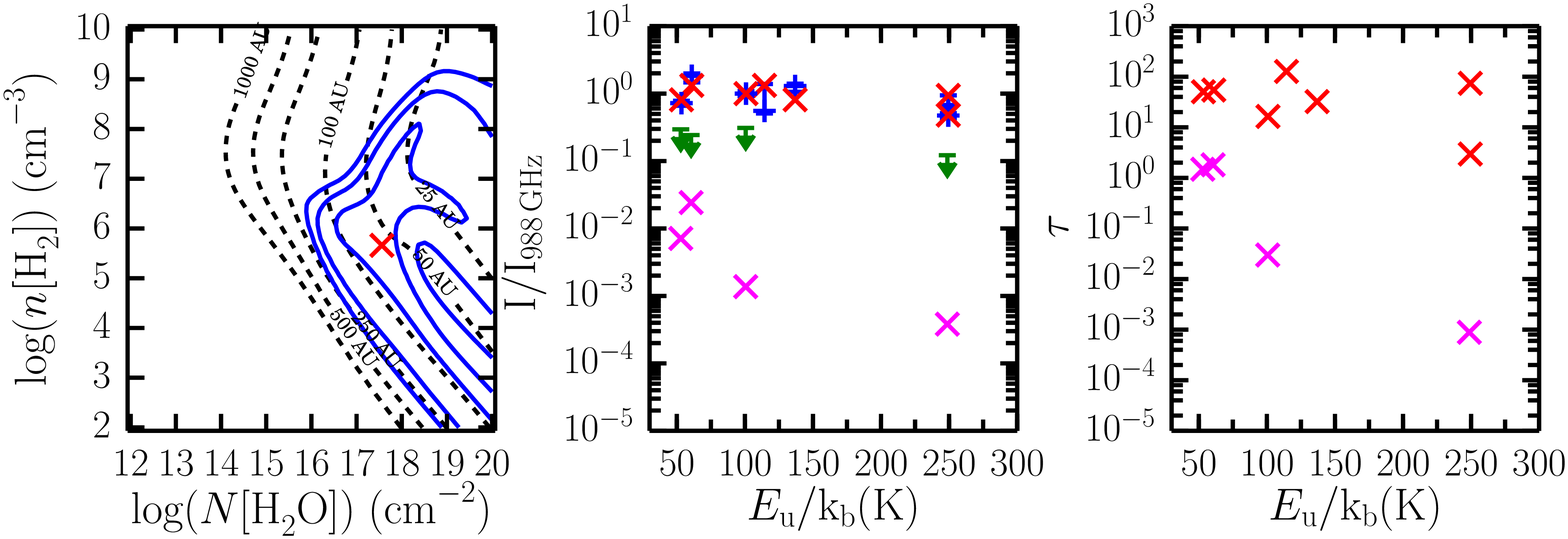}
\includegraphics[width=0.90\textwidth]{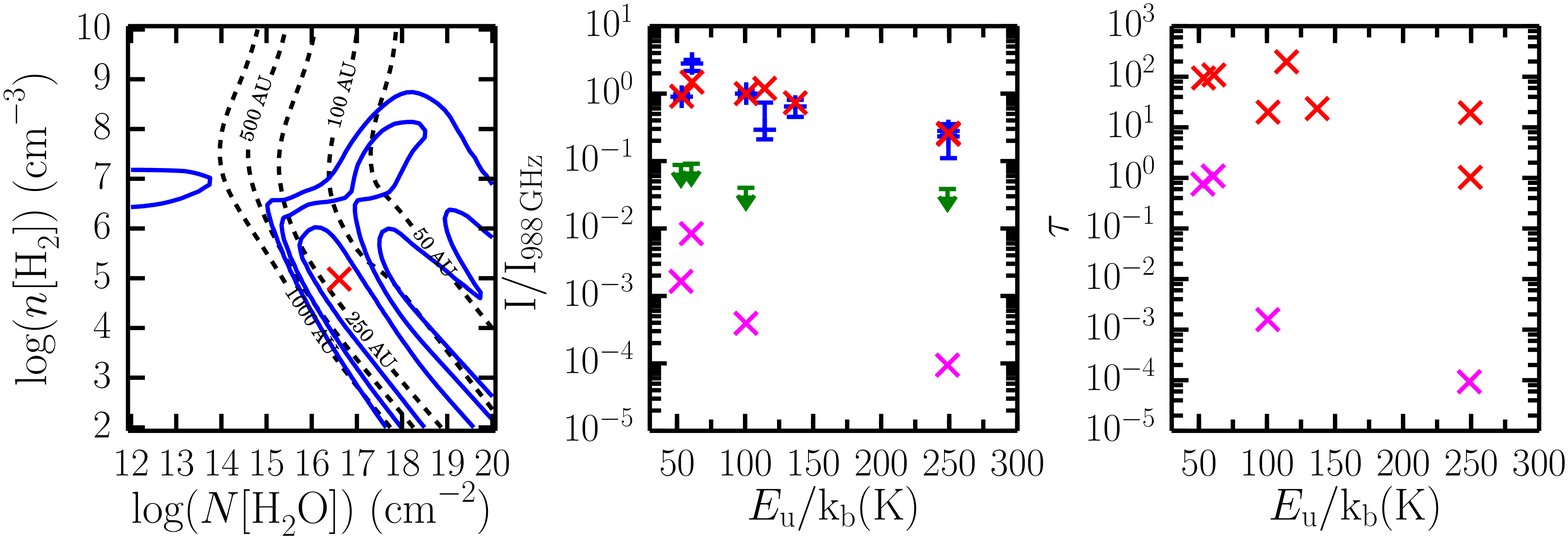}
\includegraphics[width=0.90\textwidth]{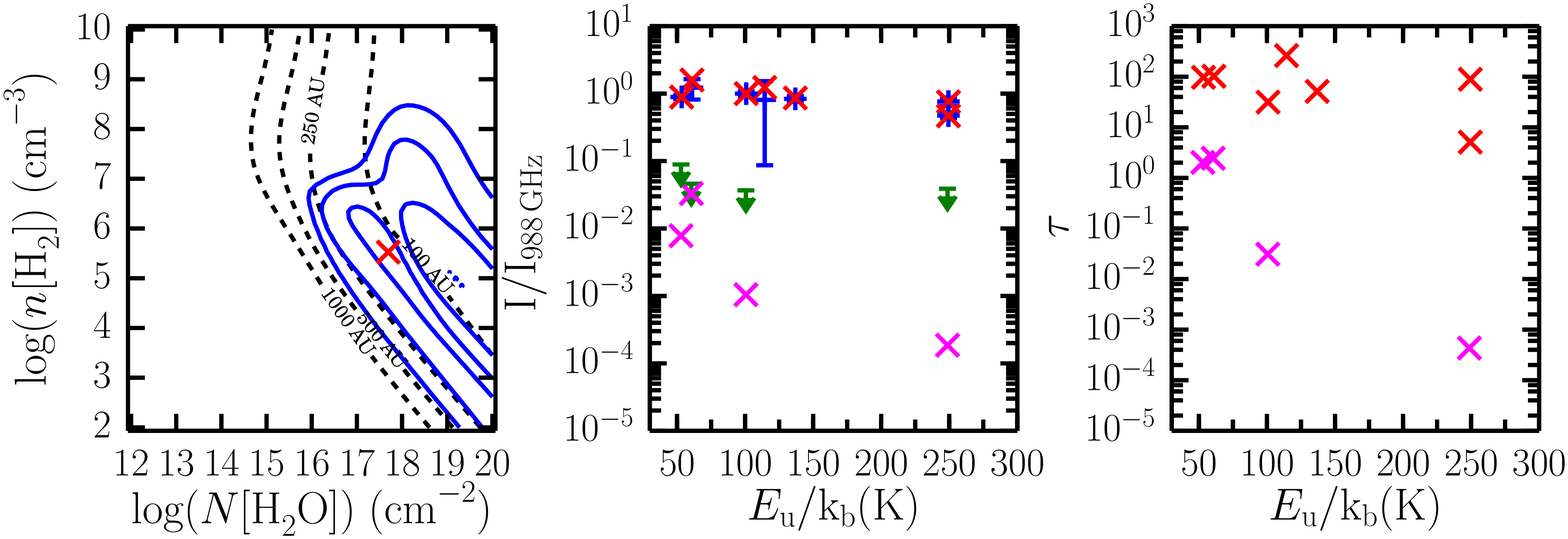}
\includegraphics[width=0.90\textwidth]{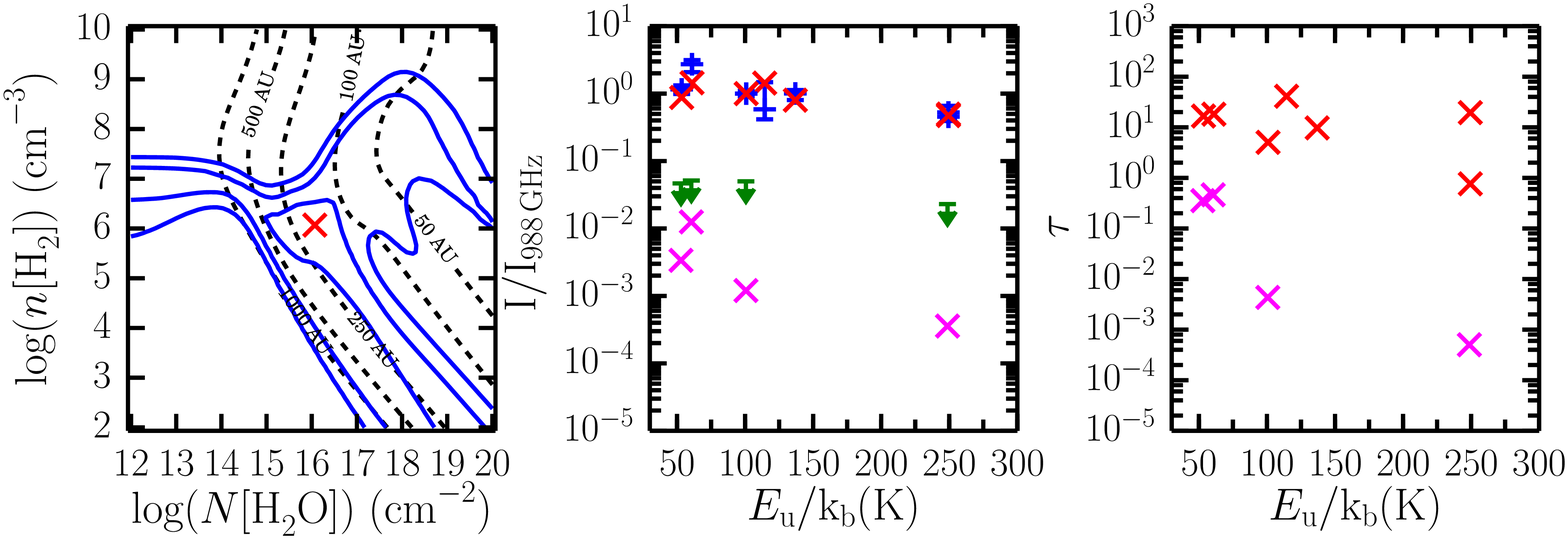}
\caption{As in Fig.~\ref{F:analysis_excitation_results_full1} but for the spot shock component of IRAS2A (top), the FWHM=9.9\kms{} spot shock and cavity shock components of IRAS4A (upper and lower-middle) and the spot shock component of IRAS4B (bottom).}
\label{F:analysis_excitation_results_full2}
\end{center}
\end{figure*}

\begin{figure*}
\begin{center}
\includegraphics[width=0.90\textwidth]{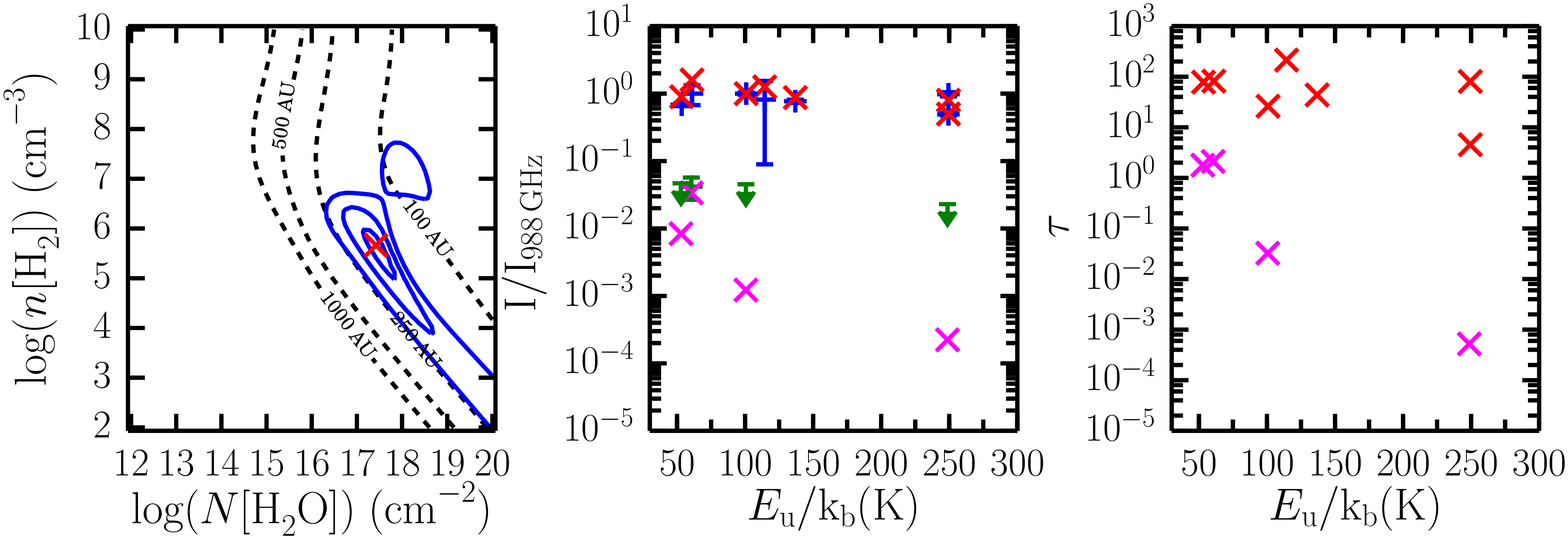}
\includegraphics[width=0.90\textwidth]{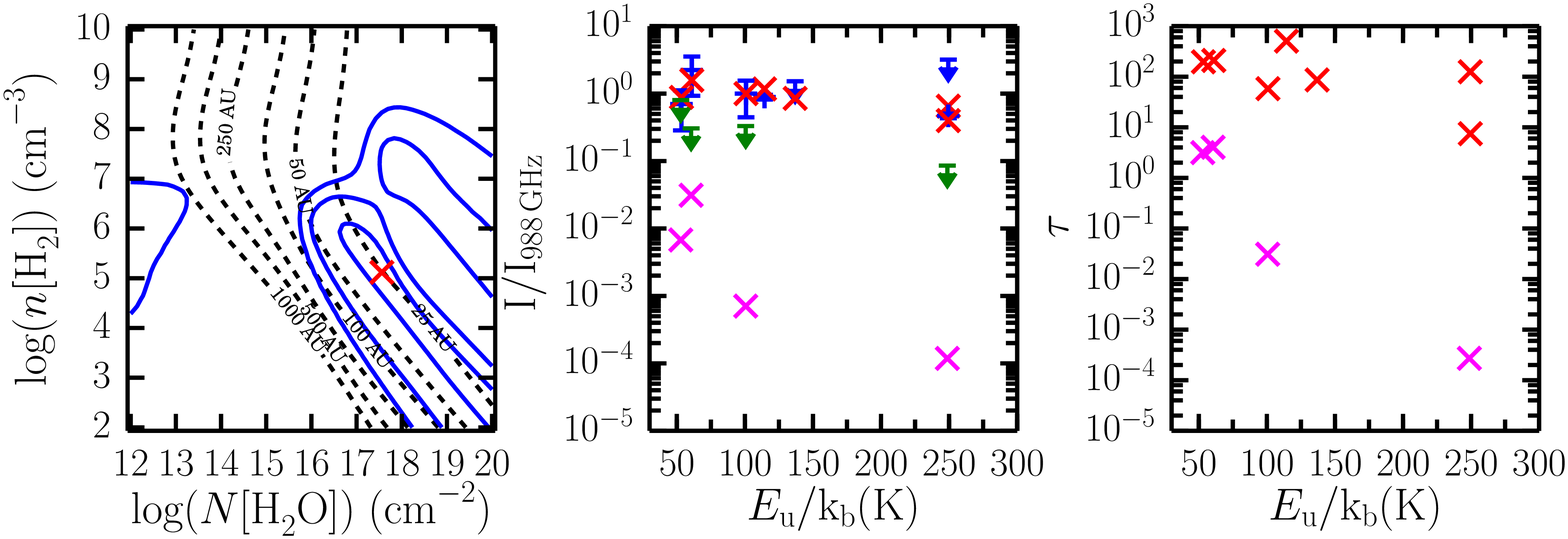}
\includegraphics[width=0.90\textwidth]{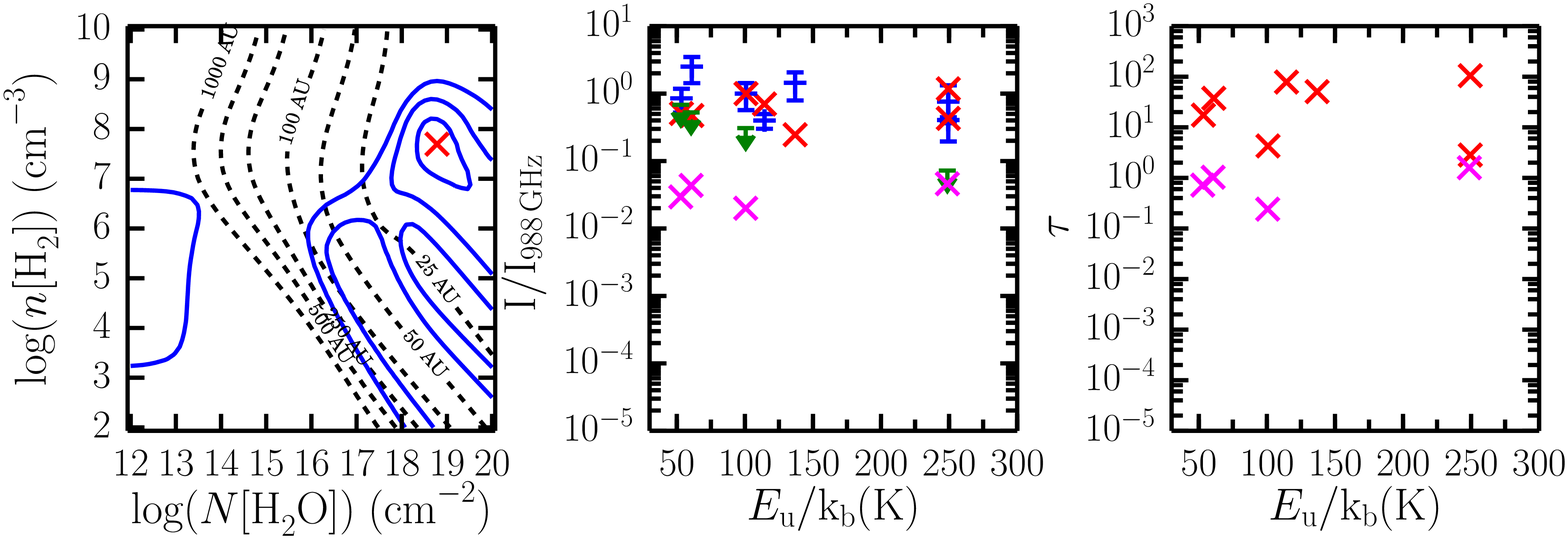}
\includegraphics[width=0.90\textwidth]{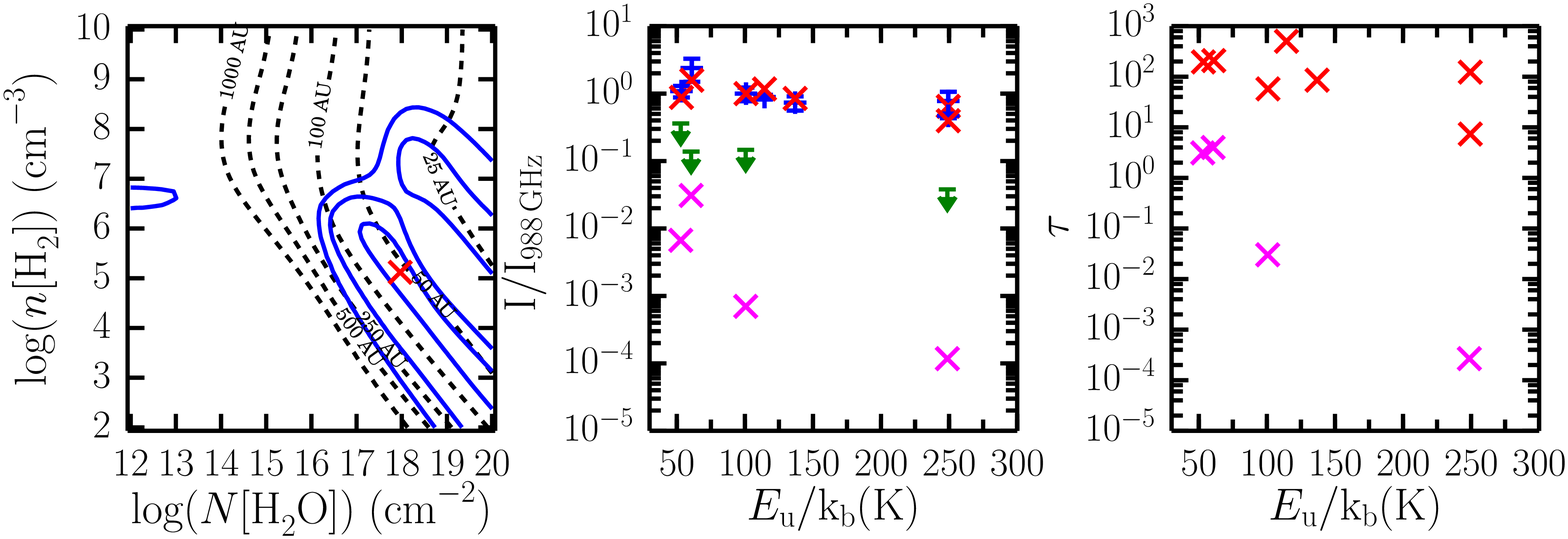}
\caption{As in Fig.~\ref{F:analysis_excitation_results_full1} but for the cavity shock components of IRAS4B (top) L1527 (upper-middle), and the FWHM=28.3\kms{} spot shock and cavity shock components of BHR71.}
\label{F:analysis_excitation_results_full3}
\end{center}
\end{figure*}

\begin{figure*}
\begin{center}
\includegraphics[width=0.90\textwidth]{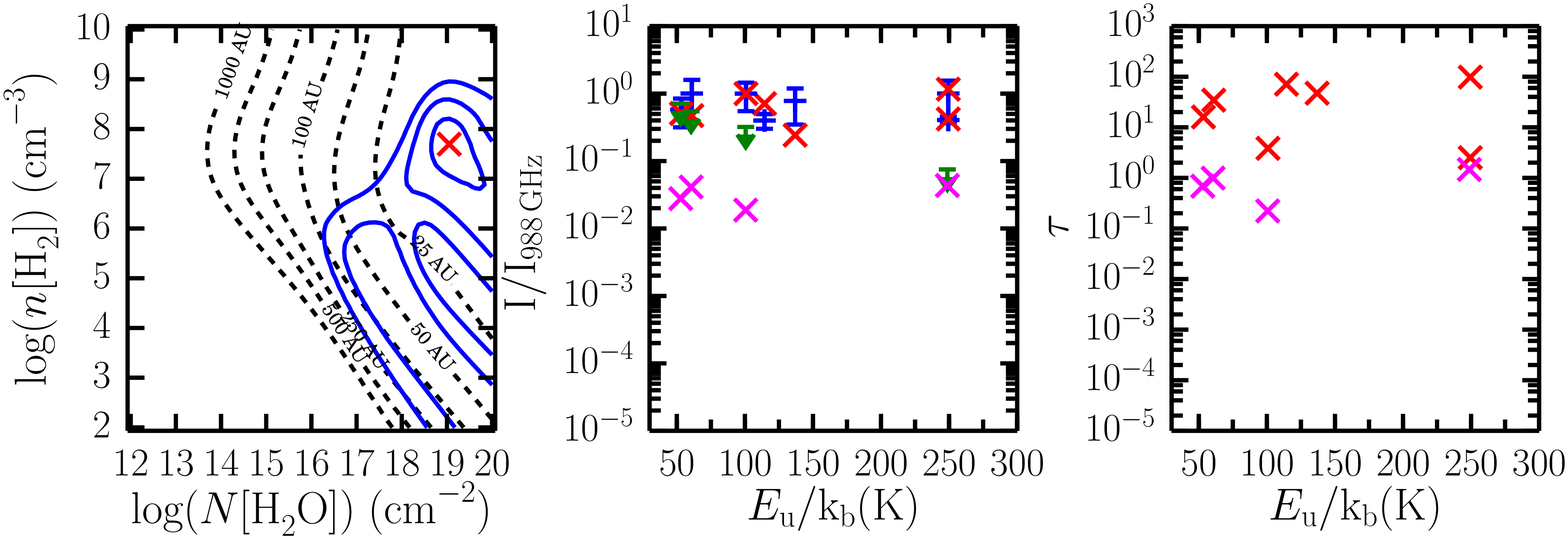}
\includegraphics[width=0.90\textwidth]{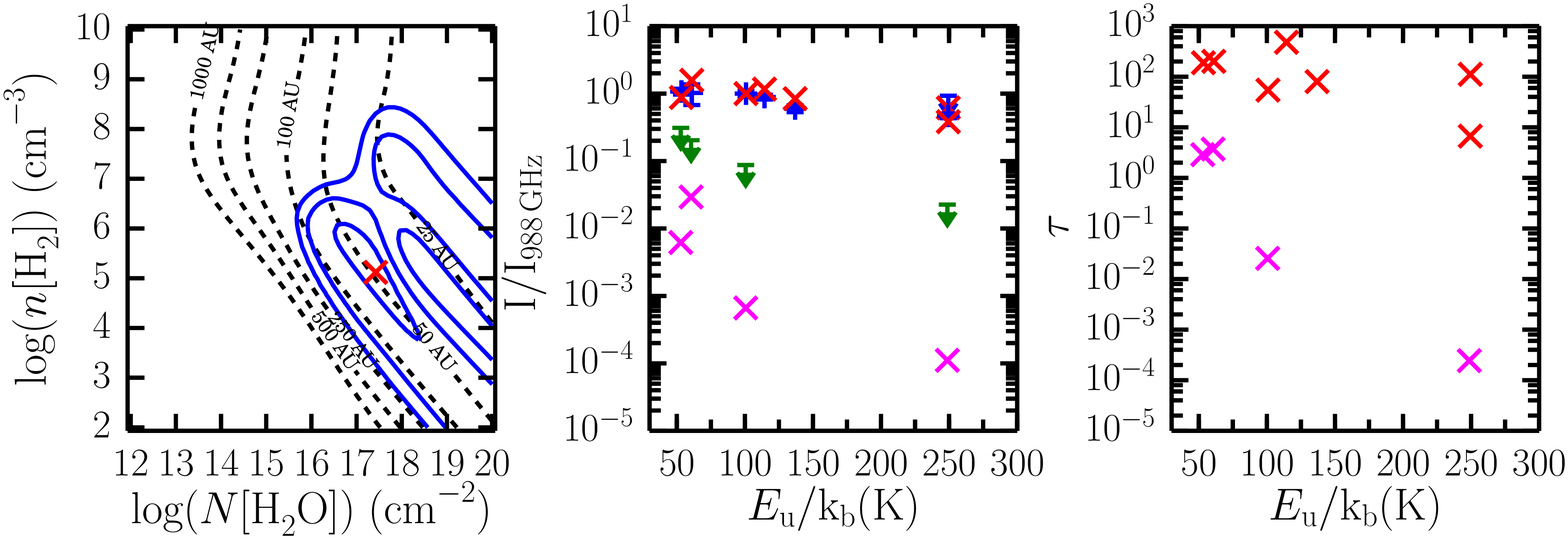}
\includegraphics[width=0.90\textwidth]{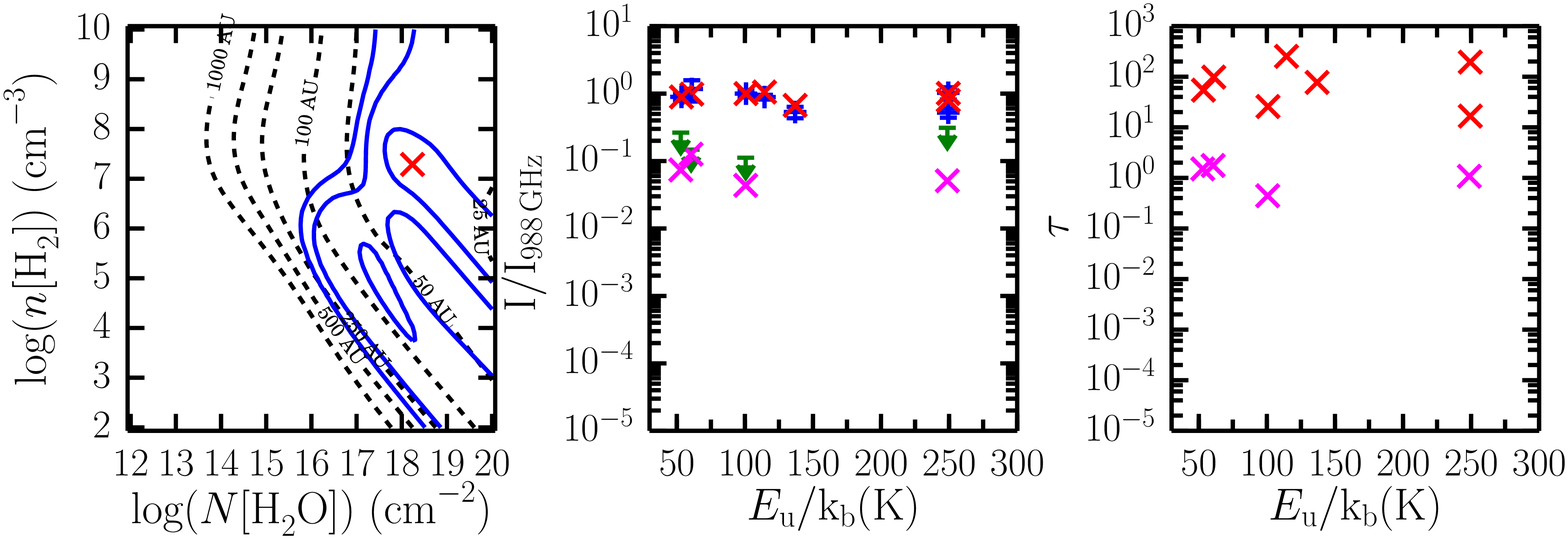}
\includegraphics[width=0.90\textwidth]{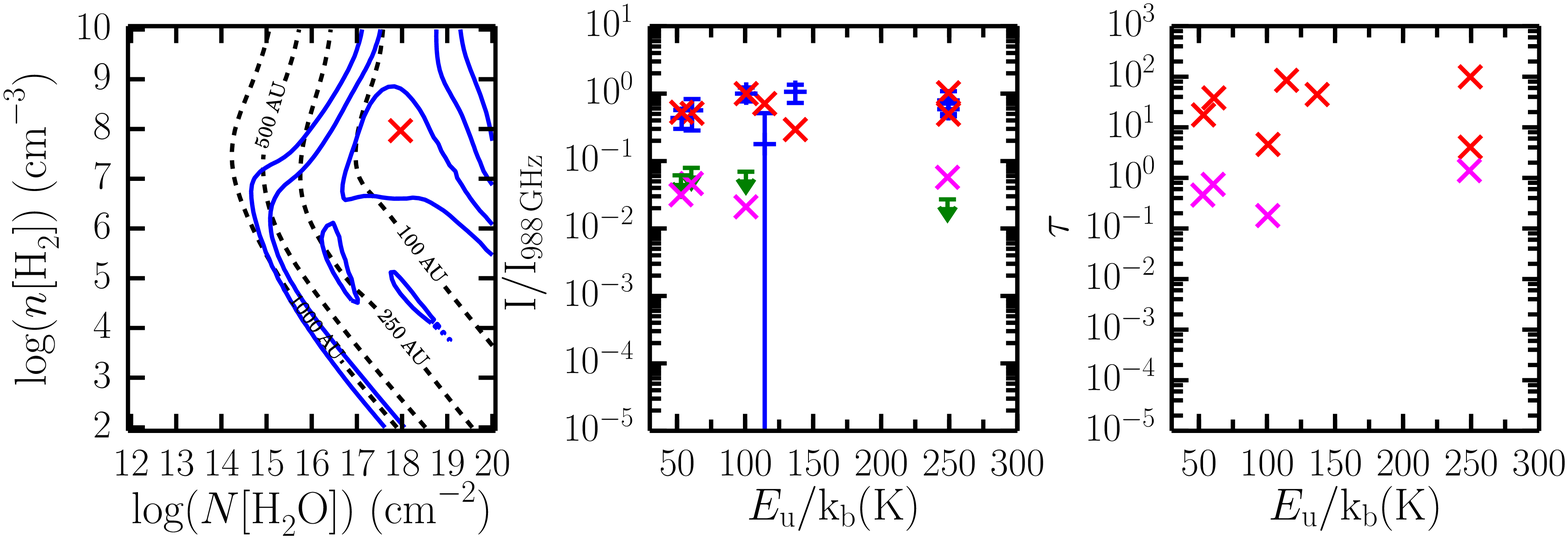}
\caption{As in Fig.~\ref{F:analysis_excitation_results_full1} but for the FWHM=59.0\kms{} spot shock of BHR71 (top) and the cavity shocks of IRAS15398 (upper-middle), L483 (lower-middle) and the FWHM=3.7\kms{} spot shock of Ser-SMM1 (bottom).}
\label{F:analysis_excitation_results_full4}
\end{center}
\end{figure*}

\begin{figure*}
\begin{center}
\includegraphics[width=0.90\textwidth]{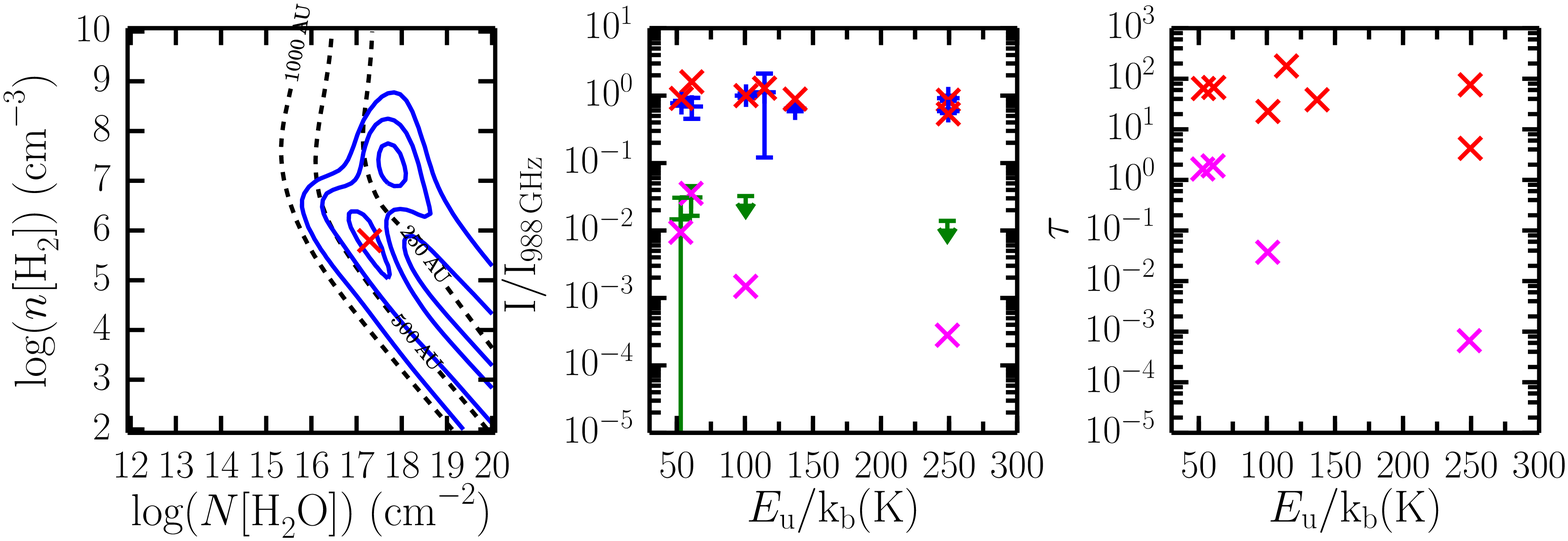}
\includegraphics[width=0.90\textwidth]{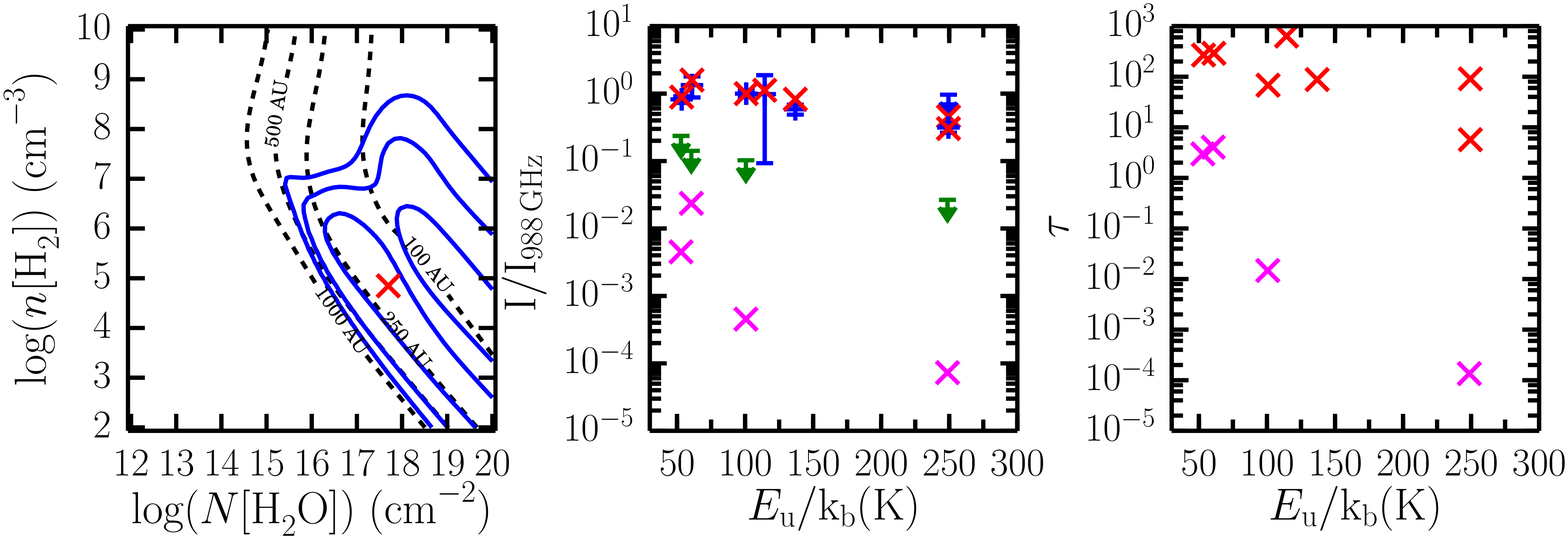}
\includegraphics[width=0.90\textwidth]{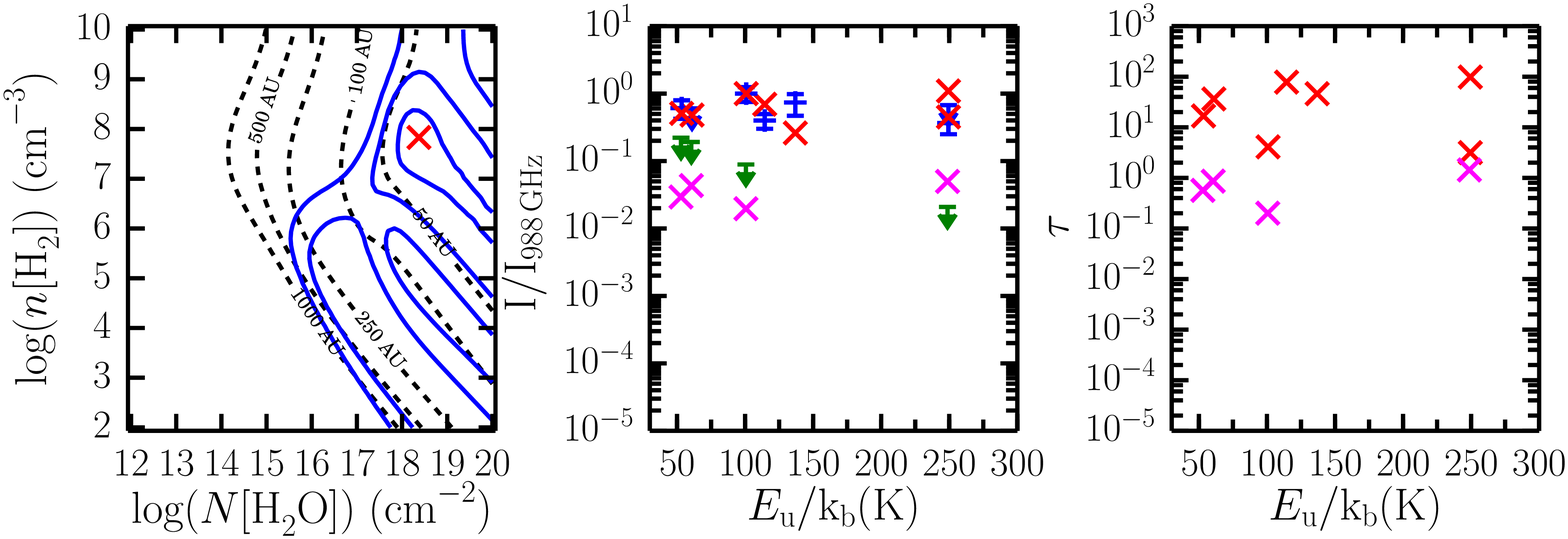}
\includegraphics[width=0.90\textwidth]{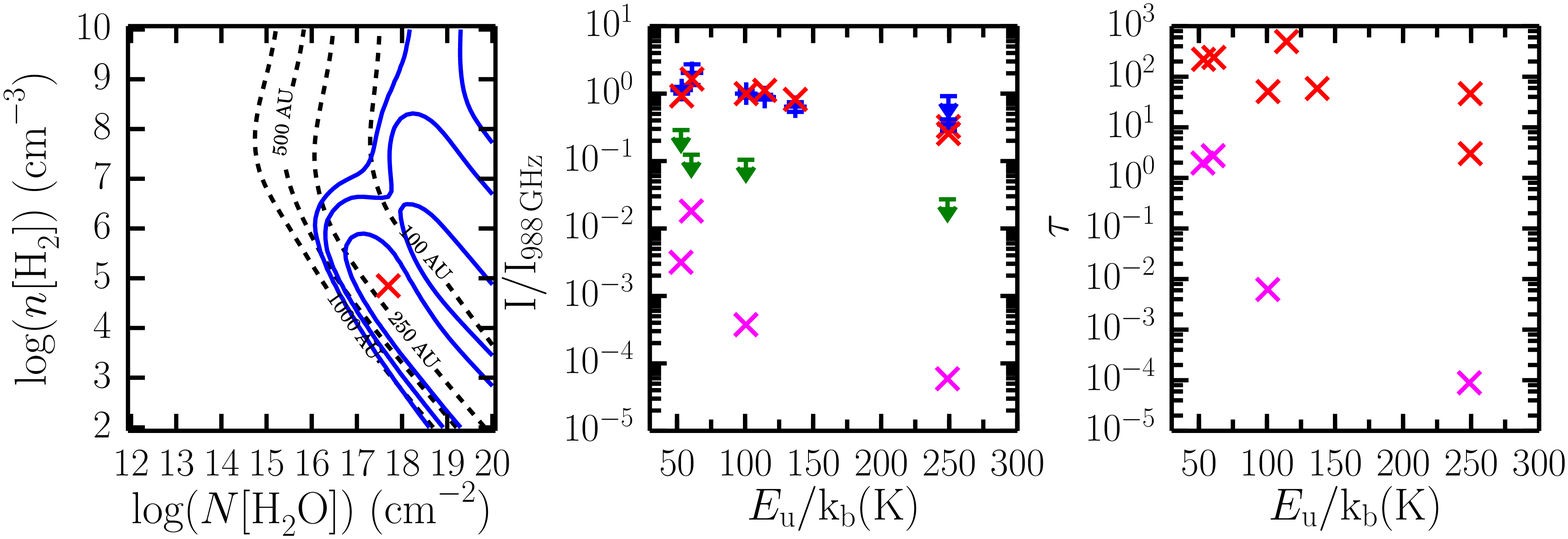}
\caption{As in Fig.~\ref{F:analysis_excitation_results_full1} but for the cavity shock of Ser-SMM1 (top), Ser-SMM3 (upper-middle), the spot shock and cavity shock of Ser-SMM4 (lower-middle and bottom).}
\label{F:analysis_excitation_results_full5}
\end{center}
\end{figure*}

\begin{figure*}
\begin{center}
\includegraphics[width=0.90\textwidth]{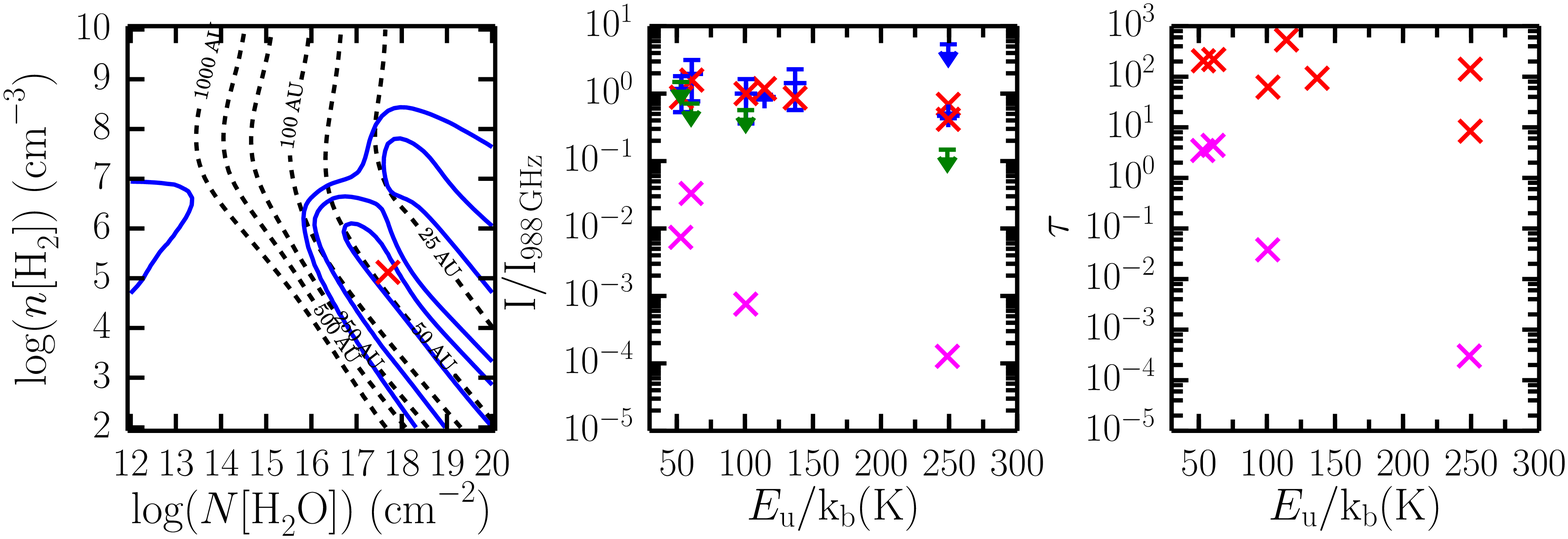}
\includegraphics[width=0.90\textwidth]{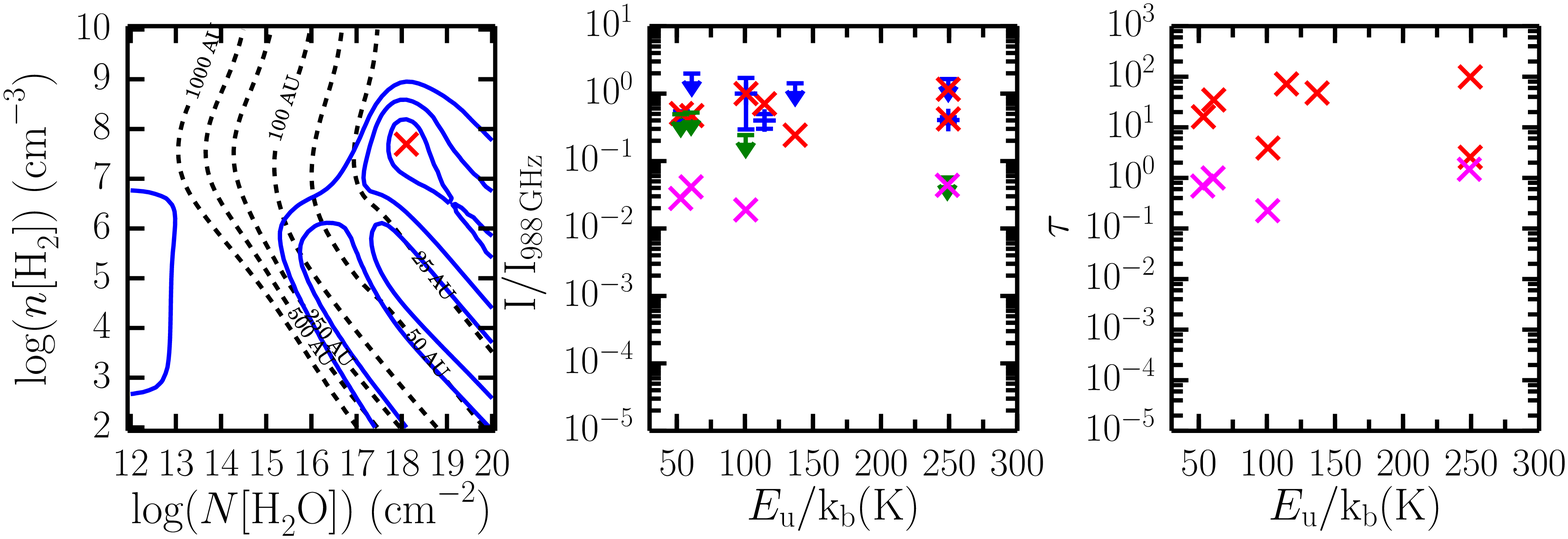}
\includegraphics[width=0.90\textwidth]{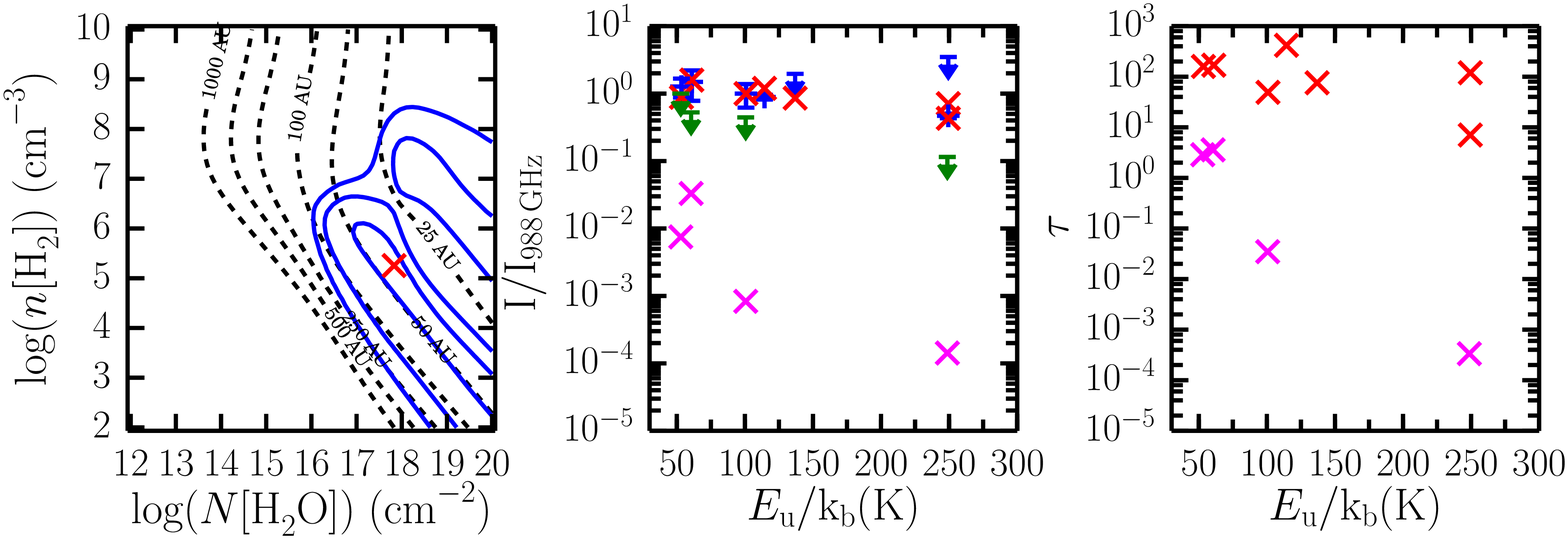}
\includegraphics[width=0.90\textwidth]{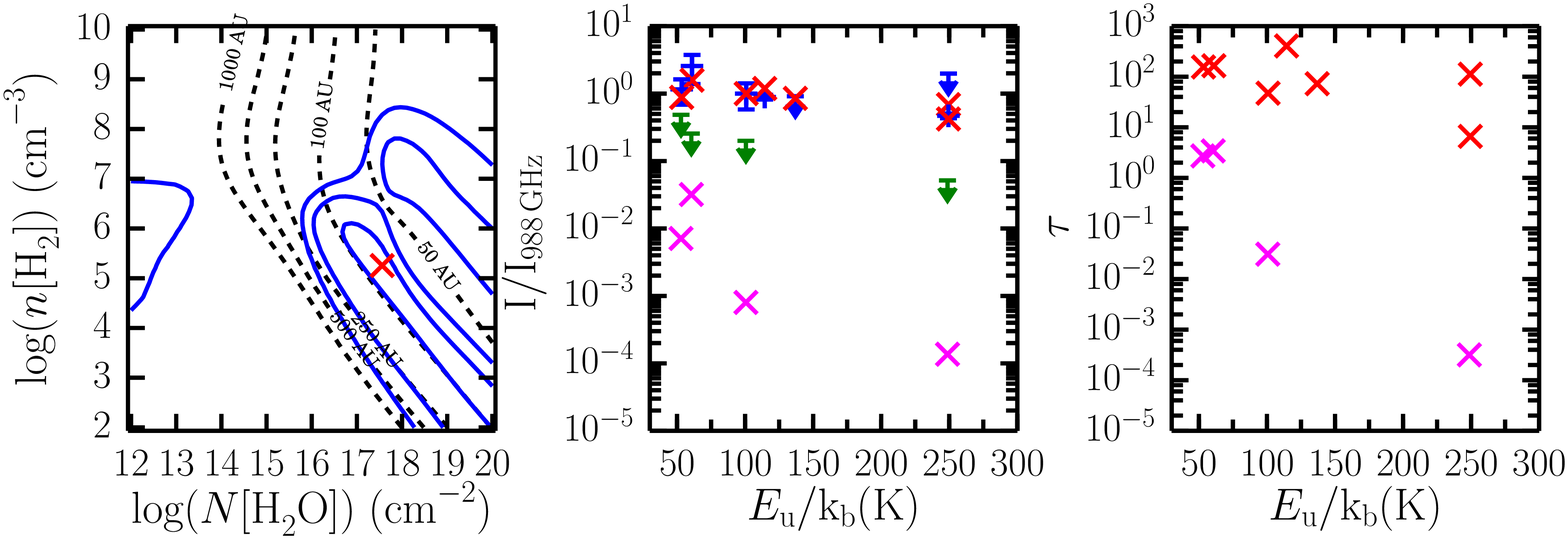}
\caption{As in Fig.~\ref{F:analysis_excitation_results_full1} but for the cavity shock of L723 (top), the spot and cavity shocks of B335 (upper and lower-middle), and the cavity shock of L1157 (bottom).}
\label{F:analysis_excitation_results_full6}
\end{center}
\end{figure*}

\begin{figure*}
\begin{center}
\includegraphics[width=0.90\textwidth]{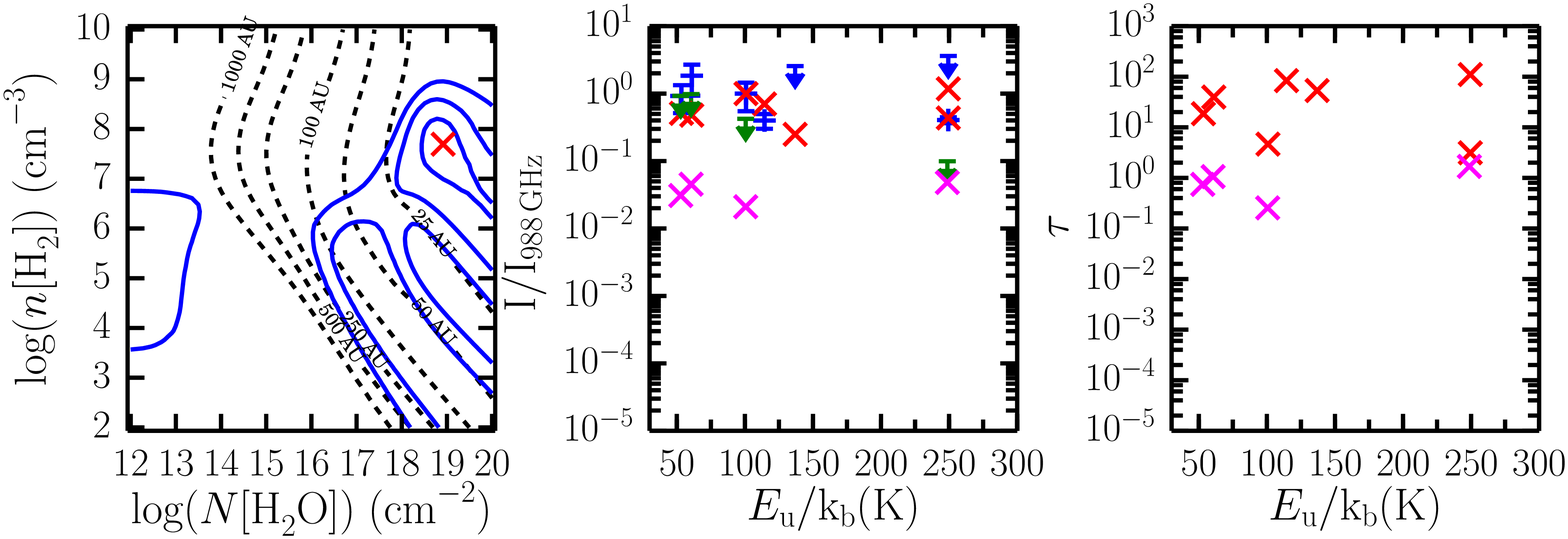}
\includegraphics[width=0.90\textwidth]{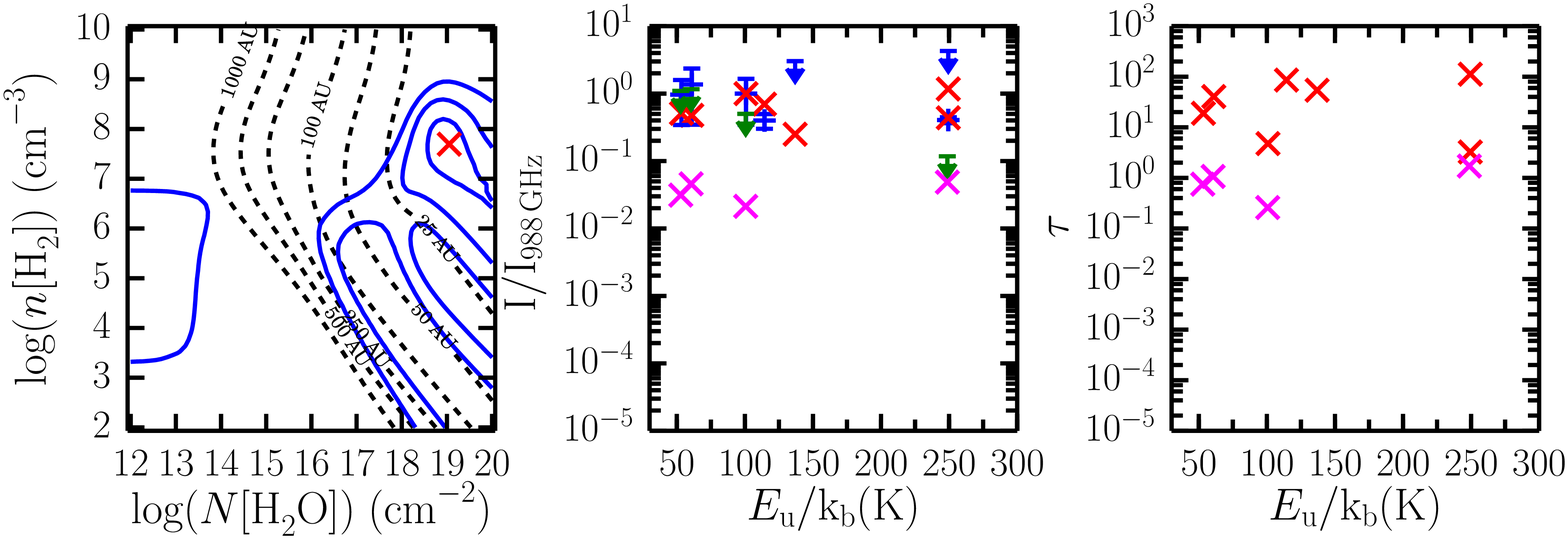}
\includegraphics[width=0.90\textwidth]{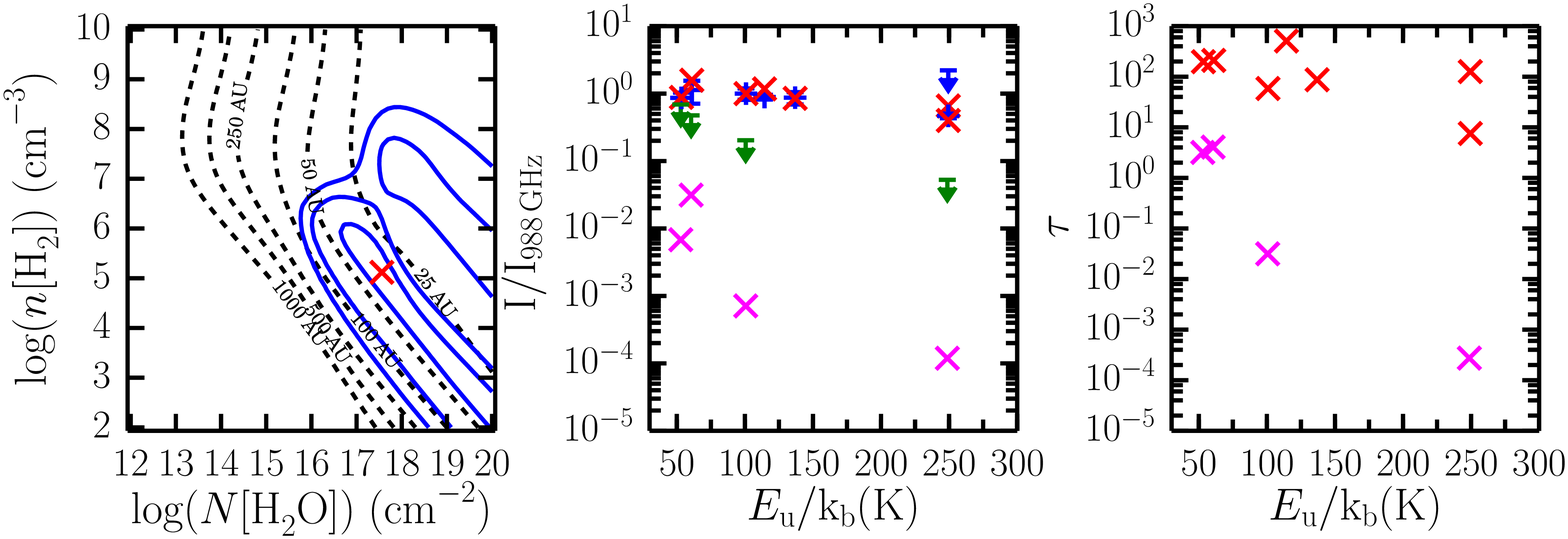}
\includegraphics[width=0.90\textwidth]{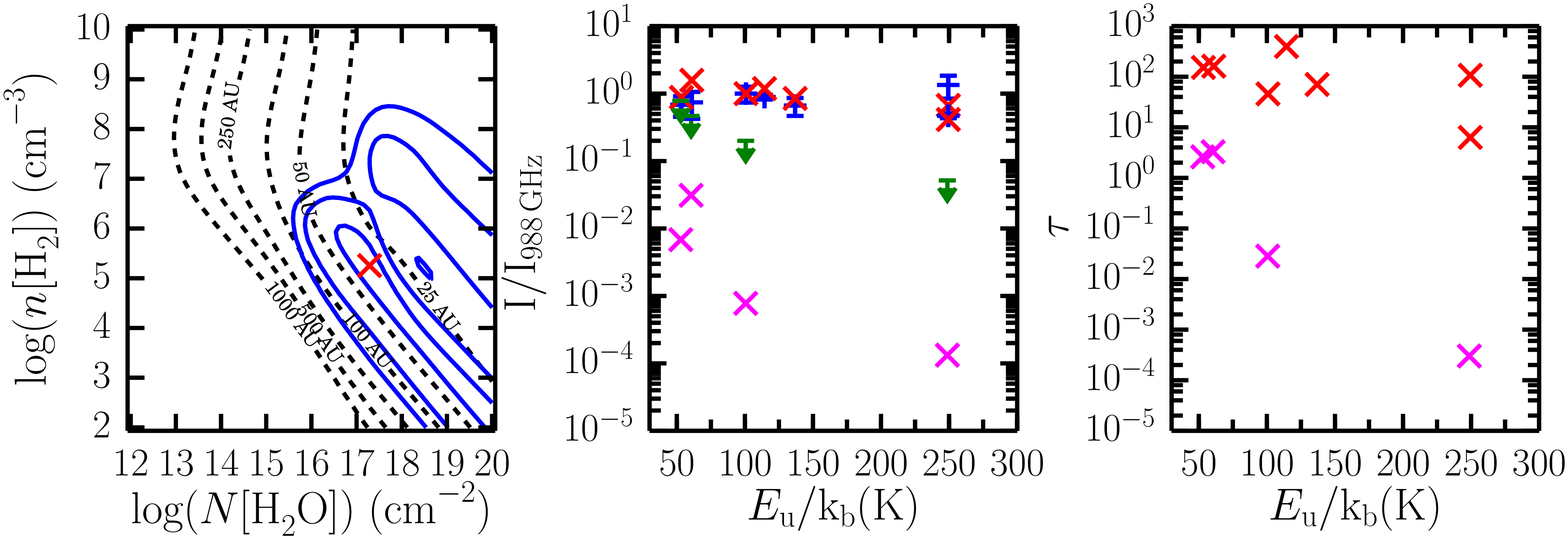}
\caption{As in Fig.~\ref{F:analysis_excitation_results_full1} but for the FWHM=35.7\kms{} and 47.7\kms{} spot shocks of L1157 (top and upper-middle), and the cavity shocks of L1489 (lower-middle) and TMR1 (bottom).}
\label{F:analysis_excitation_results_full7}
\end{center}
\end{figure*}

\begin{figure*}
\begin{center}
\includegraphics[width=0.90\textwidth]{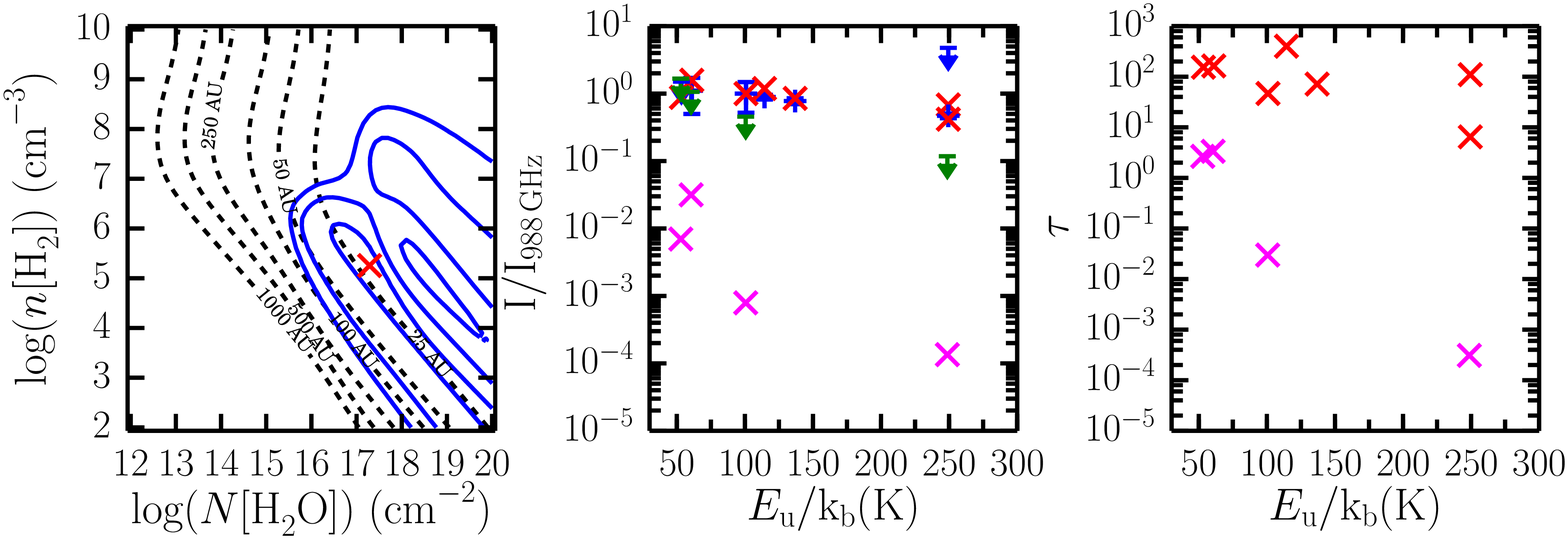}
\includegraphics[width=0.90\textwidth]{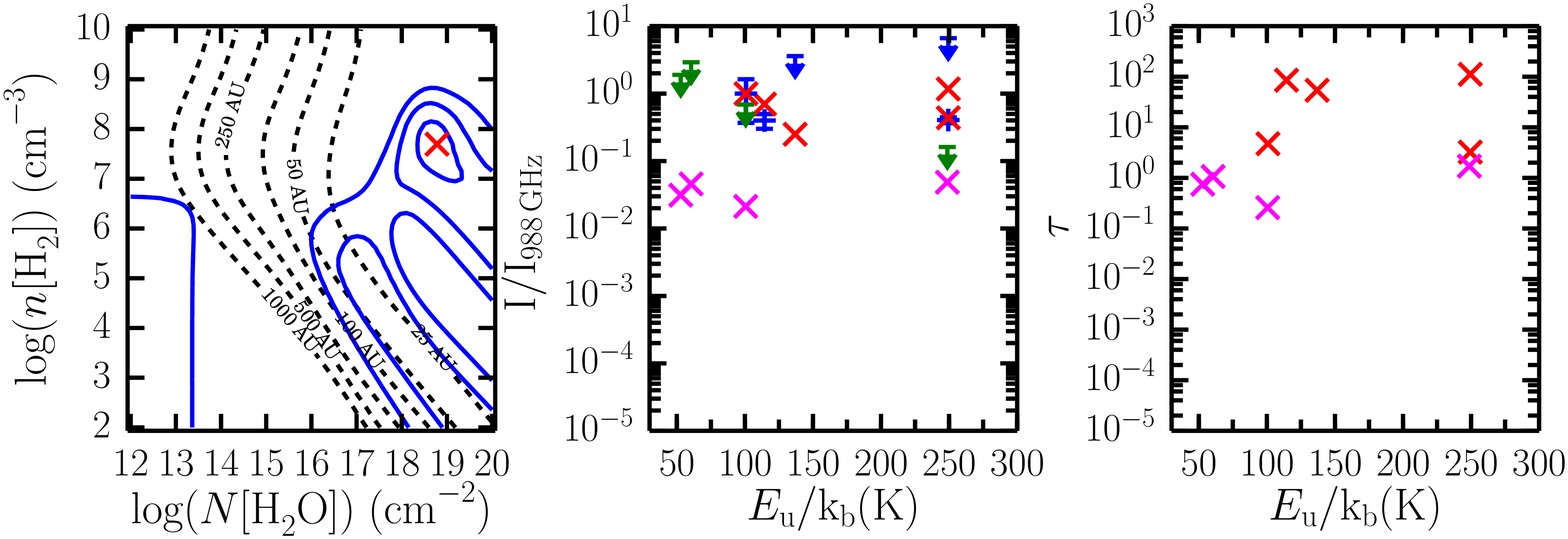}
\includegraphics[width=0.90\textwidth]{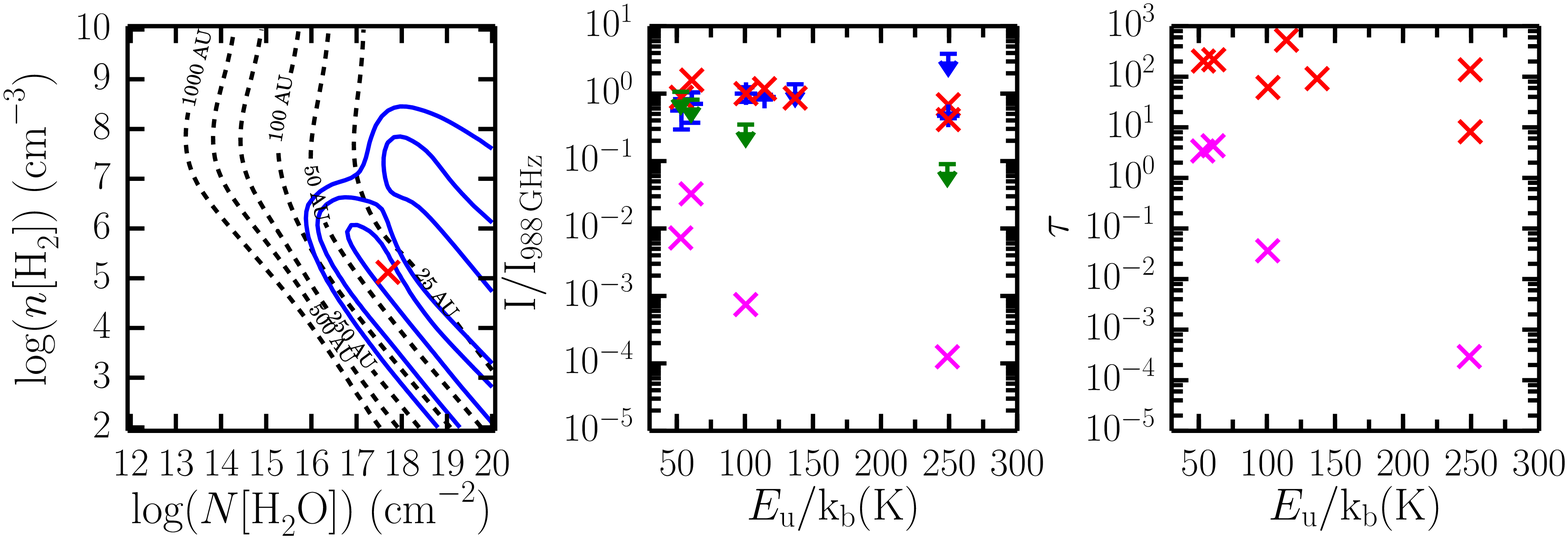}
\includegraphics[width=0.90\textwidth]{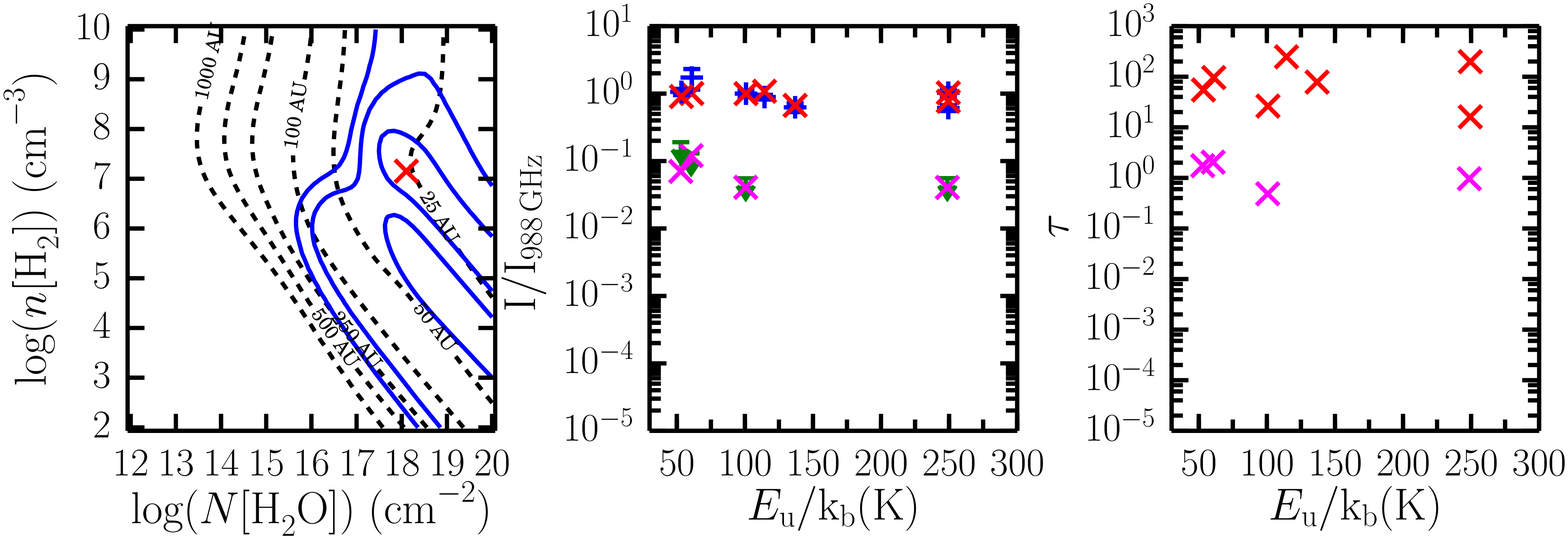}
\caption{As in Fig.~\ref{F:analysis_excitation_results_full1} but for the cavity shocks of TMC1 (top), the spot and cavity shocks of IRAS12496 (upper and lower-middle) and the cavity shock of GSS30-IRS5 (bottom).}
\label{F:analysis_excitation_results_full8}
\end{center}
\end{figure*}

\begin{figure*}
\begin{center}
\includegraphics[width=0.90\textwidth]{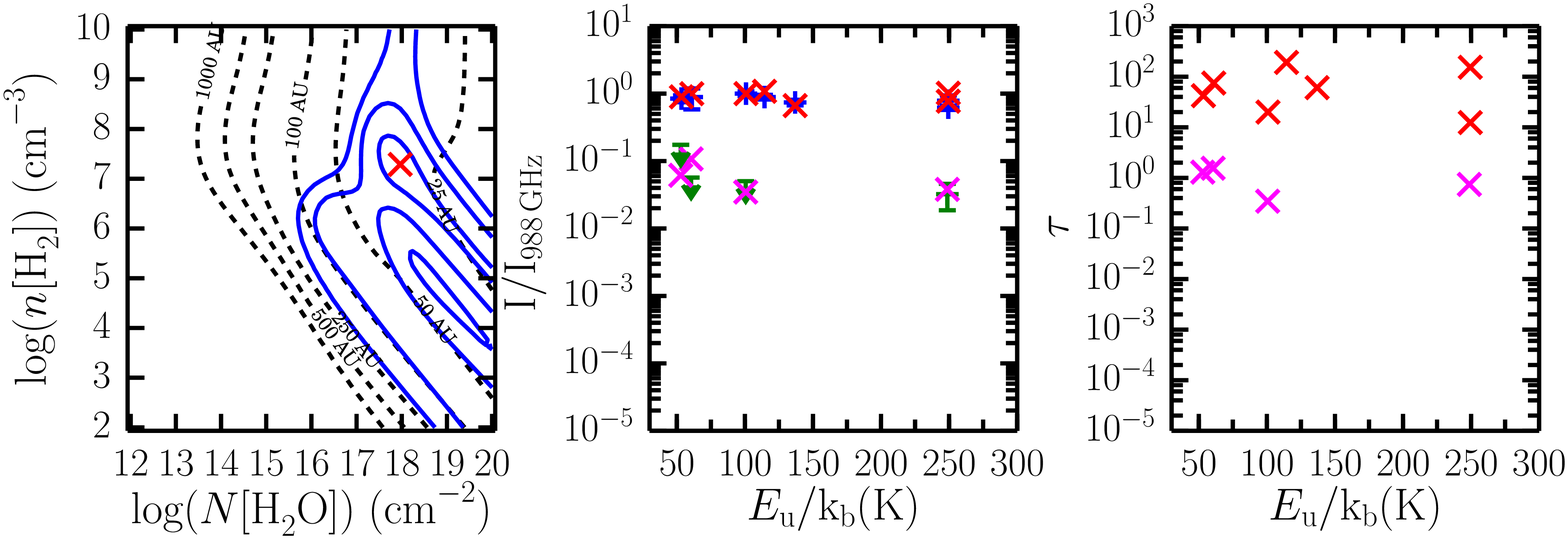}
\includegraphics[width=0.90\textwidth]{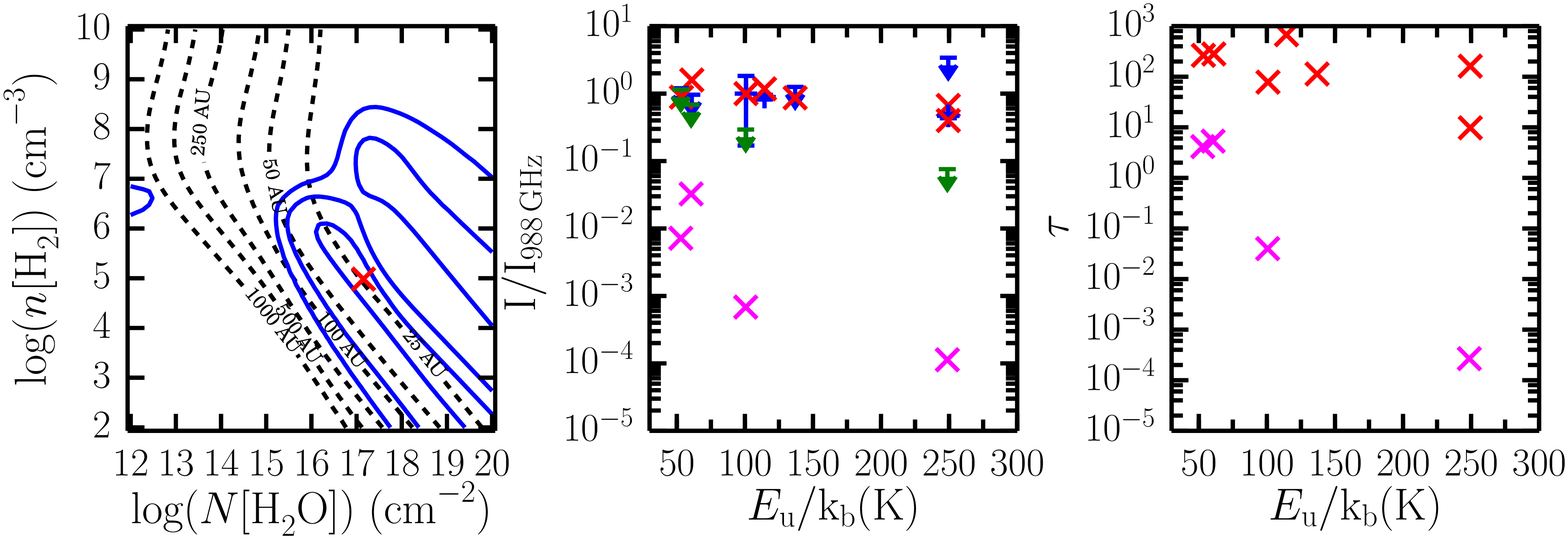}
\caption{As in Fig.~\ref{F:analysis_excitation_results_full1} but for the cavity shocks of Elias29 (top) and RNO91 (bottom).}
\label{F:analysis_excitation_results_full9}
\end{center}
\end{figure*}

\clearpage

\section{Specific Sources}
\label{S:appendix2}

This section discusses in detail the line comparison for three specific sources which require further mention: IRAS15398, Ser-SMM3 and NGC1333-IRAS4A.

\subsection{IRAS15398}
\label{S:appendix2_iras15398}

\begin{figure}
\begin{center}
\includegraphics[width=0.4\textwidth]{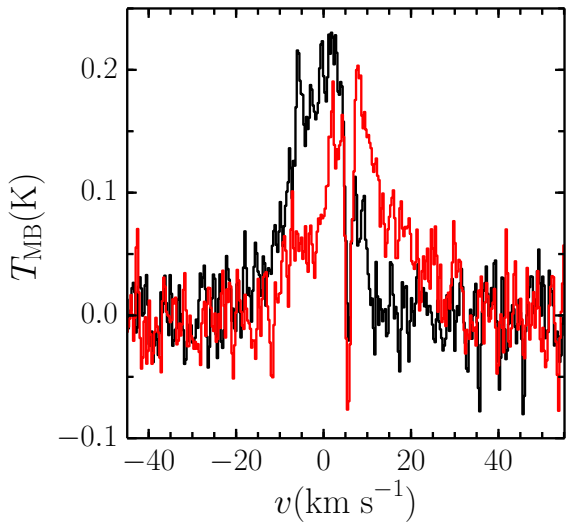}
\caption{Comparison of the H$_{2}$O 1$_{11}-$0$_{00}$ observed as part of WISH towards the coordinates and towards the SMA coordinates as part of OT2\_evandish\_4 (red).}
\label{F:IRAS15398_comp}
\vspace{-6mm}
\end{center}
\end{figure}

As noted in Table~\ref{T:observations_sources}, the position targeted in WISH for a few sources is offset slightly from the source centre as derived by SMA sub-mm interferometric observations. As part of OT2 programme OT2\_evandish\_4 the H$_{2}$O 1$_{11}-$0$_{00}$ line was observed towards the coordinates from \citet{Jorgensen2009}. Figure~\ref{F:IRAS15398_comp} shows a comparison between the WISH observations (black) and those from OT2\_evandish\_4 (red, obsid 1342266006). The central absorption and FWHM of the outflow component are broadly the same, though the outflow is centred closer to the source velocity and has a slightly lower $T_{\mathrm{peak}}$ value in the newer observations. Ultimately this does not impact the conclusions for this source either in this paper or in \citet{Mottram2013}, hence no other lines were re-observed. The H$_{2}^{18}$O 1$_{10}-$1$_{01}$ line was also observed towards IRAS15398 as part of  OT2\_evandish\_4 (obsid 1342266008) with a noise level of 4\,mK, but no line emission or absorption was detected.

\subsection{Ser-SMM3}
\label{S:appendix2_smm3}

Ser-SMM3 is the one source where the line shape seems to vary between the different water transitions. In particular, in Fig.~\ref{F:results_profiles_basic_comparison} the blue wing of the main outflow component seem to be missing in some of the higher frequency transitions, especially for the 2$_{12}-$1$_{01}$ (1670\,GHz) line. This is also one of the few sources to show a shock component in absorption, and it is also blue-shifted with respect to the source velocity. \citet{Kristensen2013} argue that this component is absorbing only against the continuum, which is stronger at higher frequencies. We therefore suggest that the reason the blue wing is decreased or missing in some transitions is due to absorption by the shock component. This source is therefore still consistent with all sources showing no variation in line profile between the different transitions observed with HIFI.

\subsection{NGC1333-IRAS4A}
\label{S:appendix2_iras4A}

\begin{figure}
\begin{center}
\includegraphics[width=0.4\textwidth]{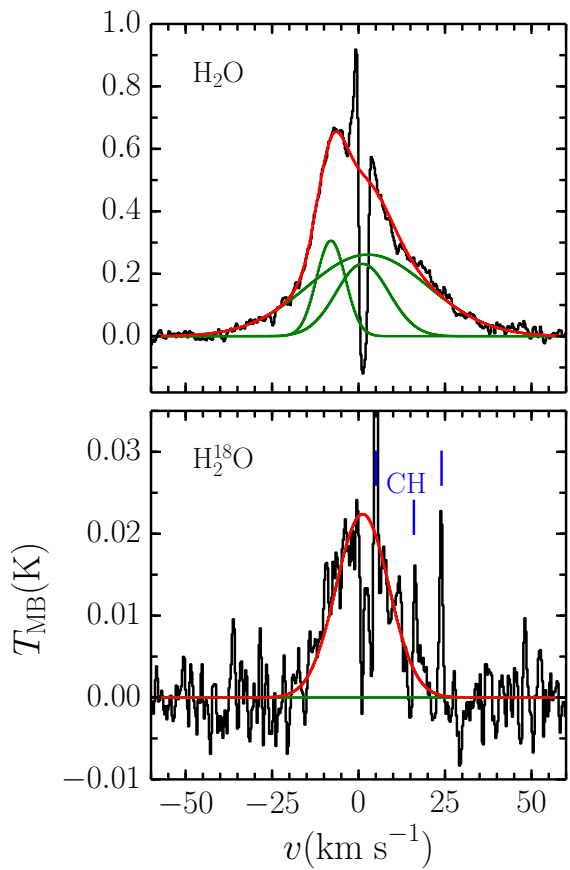}
\caption{Comparison of the 1$_{10}-$1$_{01}$ transitions of H$_{2}$O (top) and H$_{2}^{18}$O (bottom) towards IRAS4A. The black lines show the data, the green show the individual Gaussian components while the red shows the combination of all components present in a given line. The blue lines indicate the approximate velocities of the three CH transitions from the other sideband.}
\label{F:IRAS4A_comp}
\vspace{-6mm}
\end{center}
\end{figure}

Figure~\ref{F:IRAS4A_comp} shows a comparison of the H$_{2}$O and H$_{2}^{18}$O 1$_{10}-$1$_{01}$ observations towards IRAS4A, along with the Gaussian decomposition of those lines. The sharp narrow peaks are due to CH emission in the other side-band and so should be ignored for the following discussion. As discussed in Sec.~\ref{S:results_components} and shown in Fig.~\ref{F:IRAS4A_comp}, the outflow component (broadest Gaussian) is not detected in the H$_{2}^{18}$O observations, but the shock component near the source velocity is. This is also one of the few additional shock components which is approximately at the source velocity and lies in the region occupied by the outflow components in Figure~\ref{F:results_components_comparison_scatter}. Unlike any other shock component, the intensity of this component decreases with decreasing beam-size of the observations and it is not detected at all in the 2$_{12}-$1$_{01}$ (1670\,GHz) transition, which has a beam-size of 12.7\arcsec{}. There are two off-source shocks in the IRAS4A outflow, B1 and R1, which both lie $\sim$10\arcsec{} from the central source. This places at least a part of both B1 and R1 within the beam of most of the on-source observations, with the exception of the 2$_{12}-$1$_{01}$ transition. The R1 position has been observed by \citet{Santangelo2014} and shows similar line-width to the on-source shock component. We therefore conclude that this component is coming from the B1 and/or R1 off-source shocks.

\end{document}